\documentclass[english,reprint,amsmath,amssymb,aps,superscriptaddress,floatfix]{revtex4-1}
\usepackage[T1]{fontenc}
\usepackage[latin9]{inputenc}
\usepackage{babel}
\usepackage{multirow}
\usepackage{amsmath}
\usepackage{amssymb}
\usepackage{graphicx}
\usepackage[unicode=true,
pdfusetitle,
bookmarks=true,bookmarksnumbered=false,bookmarksopen=false,
breaklinks=false,pdfborder={0 0 1},backref=false,
colorlinks=true,
linkcolor={blue},
citecolor={blue},
urlcolor={blue}]{hyperref}
\usepackage{cleveref}
\usepackage{dcolumn}
\usepackage{bm}
\usepackage[caption=false]{subfig}
\usepackage{color}
\makeatletter

\providecommand{\tabularnewline}{\\}

\makeatother

\begin{document}

\title{Orbital loop currents in iron-based superconductors}

\author{Markus Klug}
\email{markus.klug@kit.edu}

\affiliation{Institute for Theoretical Condensed Matter Physics, Karlsruhe Institute
of Technology, D-76131 Karlsruhe, Germany}

\author{Jian Kang}
\author{Rafael M. Fernandes}

\affiliation{School of Physics and Astronomy, University of Minnesota, Minneapolis, Minnesota
55455, USA}

\author{Jörg Schmalian}

\affiliation{Institute for Theoretical Condensed Matter Physics, Karlsruhe Institute
of Technology, D-76131 Karlsruhe, Germany}

\affiliation{Institute of Solid State Physics, Karlsruhe Institute of Technology,
D-76344 Eggenstein-Leopoldshafen, Germany}

\begin{abstract}
We show that the antiferromagnetic state
commonly observed
in the phase diagrams of the iron-based superconductors necessarily
triggers loop currents characterized by charge transfer between different
Fe $ 3d $ orbitals. This effect is rooted on the glide-plane symmetry
of these materials and on the existence of an atomic spin-orbit coupling
that couples states at the $X$ and $Y$ points of the 1-Fe Brillouin
zone. In the particular case in
which the magnetic moments 
are aligned 
parallel to the magnetic ordering vector
direction which is the moment
configuration most commonly
found in the iron-based superconductors these loop currents
involve the $d_{xy}$ orbital and either the $d_{yz}$ orbital (if
the moments point along the $y$ axis) or the $d_{xz}$ orbitals (if
the moments point along the $x$ axis). We show that the two main manifestations
of the orbital loop currents are the emergence of magnetic moments
in the pnictide/chalcogen site and an orbital-selective band splitting
in the magnetically ordered state, both of which could be detected
experimentally. Our results highlight the unique intertwining between
orbital and 
spin 
degrees of freedom in the iron-based superconductors,
and reveal the emergence of an unusual correlated phase that may impact
the normal state and superconducting properties of these materials.
\end{abstract}

\maketitle

\section{Introduction}

One of the hallmarks of the iron-based superconductors is the close
interplay between magnetic and orbital degrees of freedom. Magnetic
order is often observed in the phase diagrams of these materials,
and is characterized by the ordering vectors $\mathbf{Q}_{1}=(\pi,0)$
and $\mathbf{Q}_{2}=(0,\pi)$ of the unfolded 1-Fe Brillouin zone
\cite{Johnston2010,Paglione2010,Stewart2011}. While in most cases
the magnetic ground state corresponds to spin stripes with either
one of these two ordering vectors \cite{Dai2012}, several hole-doped
systems display double-\textbf{Q }magnetic order consisting of a linear
combination of the two possible types of order \cite{AvciChmaissemAllredEtAl2014,BoehmerHardyWangEtAl2015,AllredTaddeiBugarisEtAl2016,MeierDingKreyssigEtAl2017}.
Orbital order is also found in the phase diagram below the nematic
transition temperature, and is characterized by an unequal occupation
between the Fe $d_{xz}$ and $d_{yz}$ orbitals, which breaks the
tetragonal symmetry of the system \cite{Fernandes2014,YiLuChuEtAl2011,ApplegateSinghChenEtAl2012}.
At least in the iron pnictides, the evidence points to a magnetic
origin of this orbital order \cite{Fernandes2013a}, unveiling the
close interaction between these two distinct degrees of freedom.

Microscopically, this interaction is described by the atomic spin-orbit
coupling (SOC), given by the term $\lambda\mathbf{S}\cdot\mathbf{L}$.
ARPES measurements have established that the magnitude of the SOC
$\lambda$ ranges from around $10\text{ meV}$ in the 122-compounds
up to $25\text{ meV}$ in the 11-compounds \cite{Borisenko2015}. Such
an energy scale is comparable to the band splitting caused by orbital
order, to the superconducting gap, and to the spin-density wave gap.
Therefore, the SOC is an integral and necessary ingredient in the
description of the low-energy properties of the iron-based superconductors
\cite{Fernandes2014a}. Besides the coupling between spin-driven nematicity
and orbital order discussed above, SOC also selects the magnetic moment
direction in the stripe phase to be generally parallel to the ordering
vector \cite{Cvetkovic2013,Christensen2015}.

In this work, we demonstrate another important consequence of the
SOC, which has hitherto been unexplored. In particular, we show that
long-range 
spin 
order necessarily generates loop currents in which
charge is transferred between Fe orbitals at different sites,
hence the name orbital loop currents. Although in this paper we focus
on the case of stripe spin magnetic order, the effect is general and should
be present even in double-\textbf{Q} magnetically ordered states, as well as independent of the specific microscopic mechanism for magnetism, e.g., itinerant or localized magnetic moments, since it is solely based on symmetry.
The qualitative argument for this effect is 
general, and can be outlined
as follows. To keep the notation simple, let us introduce the 
intraorbital
stripe magnetic order parameter $m$ with ordering vector $\mathbf{Q}_{i}$
and the interorbital imaginary charge-density wave $\phi$ with ordering
vector $\mathbf{Q}_{j}$. Note that both orders break translational
and tetragonal symmetries, as well as time-reversal symmetry. Because
the latter is broken in spin space by $m$ and in real space by $\phi$,
only SOC can couple both order parameters. In terms of a general Ginzburg-Landau
free-energy expansion, one expects the following bilinear coupling:
\begin{equation}
\delta F=g(\lambda)\phi\, m, \label{eq:couplingSDWICDW}
\end{equation}
where the coupling constant $g(\lambda)$ depends on the strength
of SOC $\lambda$ such that $g\left(\lambda=0\right)=0$\textit{.
}\textit{\emph{At first sight, one may think that this coupling is
only allowed if $\mathbf{Q}_{j}=\mathbf{Q}_{i}$ due to momentum conservation.
}}However, a key property of the iron-based superconductors is the
glide plane symmetry of the FeAs plane \cite{FernandesChubukov2017,TomicJeschkeValenti2014,LinBerlijnWangEtAl2011,ColdeaWatson2017},
which doubles the size of the unit cell to one containing 2 Fe atoms,
making $\mathbf{Q}_{1}$ and $\mathbf{Q}_{2}$ equivalent in the actual
crystallographic unit cell. Thus, depending on the orbital composition
of $\phi$, magnetic order with ordering vector $\mathbf{Q}_{1}$
will trigger imaginary charge-density wave with ordering vector $\mathbf{Q}_{2}$.
An alternative way to interpret this coupling is to note that the
SOC $\lambda$ that couples states at the $X$ and $Y$ points of
the 1-Fe Brillouin zone \cite{Cvetkovic2013,Fernandes2014a} provide
the momentum transfer $\mathbf{Q}_{1}+\mathbf{Q}_{2}$ required by
such a coupling. 

An inter-orbital imaginary charge-density wave corresponds to charge
currents in real space. The inter-orbital character of this order
implies that charge is transferred between different orbitals at different
sites. Charge conservation, however, requires these charge currents
to form closed loops. Thus, hereafter we refer to this type of order
as an orbital loop-current order. Note that the loop-currents discussed
here break the translational symmetry of the lattice by changing sign
between neighboring Fe lattice sites; as a result, they are compatible
with Bloch's theorem, which prohibits a static current carrying ground
state \cite{Bohm49,Kiselev17}.

Interestingly, a similar type of loop-current order, defined in band
basis instead of orbital basis, was introduced by Ref. \cite{Kang2011}
to explain the tetragonal symmetry breaking of the iron-based superconductors.
Ref. \cite{Podolsky2009} proposed the existence of an imaginary charge-density
wave based on an enhanced symmetry associated with nested Fermi pockets.
In Ref. \cite{Chubukov2015}, the combined effects of loop current
fluctuations and magnetic fluctuations was proposed to enhance the
nematic transition temperature of materials with very small Fermi
energies, such as FeSe. Whether the loop-current order can spontaneously
order as a primary order, rather than a secondary order triggered
by magnetic order, remains an open question that is only sharply defined for small $ \lambda $ anyway. Renormalization group
calculations, performed in the band basis, suggest that onsite repulsive
interactions generally select magnetic order over loop-current order
\cite{Chubukov2008}. We also note that loop-current order has also
been investigated in cuprate superconductors, either as a $\mathbf{Q}=0$
intra unit-cell order involving the Cu and the O orbitals \cite{Varma1997},
or as a translational symmetry-breaking order also known as $d$-density
wave \cite{ChakravartyLaughlinMorrEtAl2001}. Note that, in our case,
the loop-current order is necessarily accompanied by magnetic order,
and it also breaks the tetragonal symmetry of the lattice.

In the remainder of the paper, we derive the results outlined above
from a low-energy microscopic model for the iron-based superconductors
that respect all the crystal symmetries of the FeAs plane \textendash{}
including the glide plane symmetry\textbf{ }\cite{Cvetkovic2013,FernandesChubukov2017}.
This low-energy model relies on the smallness of the Fermi energy
of these materials to perform a $\mathbf{k}\cdot\mathbf{p}$ expansion
near high-symmetry points of the lattice. As a result, only three
Fe orbitals are taken into account, namely the $xz$, $yz$, and the
$xy$ $3d$-orbitals, since these orbitals are responsible for the
band dispersions in the vicinity of the Fermi surface. Additional
electronic degrees of freedom, such as electronic states located at
the pnictogen/chalcogen lattice sites, or the two remaining Fe $3d$-orbtials,
are assumed to be integrated out, thus renormalizing the model's parameters.
We stress that our main results are general and independent of the
particular choice of parameters. 

Using this model, we derive the different orbital loop current configurations
that are allowed by symmetry and triggered by the three different
types of long-range magnetic stripe order, providing explicit expressions
for the coupling constant $g(\lambda)$ introduced in Eq. (\ref{eq:couplingSDWICDW})
above. For the most experimentally relevant case of stripe order with
magnetic moments pointing parallel to the wave-vectors $\mathbf{Q}_{1}$
and $\mathbf{Q}_{2}$, we find that the loop currents involve $d_{xy}/d_{yz}$
and $d_{xy}/d_{xz}$ orbitals, respectively. We also discuss two manifestations
of these orbital loop-current orders, which can be detected experimentally.
The first one is the magnetization profile induced by the loop currents
that follows from simple electrodynamic considerations. The main result
is the appearance of magnetic moments on the pnictide/chalcogen sites
pointing out of the plane, in contrast to the magnetic moments on
the Fe sites, which point in the plane. The second manifestation that
we discuss is the spectroscopic signature of the orbital loop current
order parameter on the electronic spectrum. In the particular case
in which the spins in the magnetic stripe state point in the plane,
we show that the magnetic order parameter $M$ and the orbital loop
current order parameter $\phi$ contribute to splittings of different
doublets at the corner of the 2-Fe Brillouin zone. Consequently, we
propose to use ARPES to assess the impact of orbital loop current
order on the normal state electronic spectrum of the pnictides.

The paper is organized as follows. Section II introduces the microscopic
model, whereas Sec. III classifies the different types of orbital
loop current allowed by the different types of stipe magnetic order.
Sec. IV derives the microscopic coupling between these two types
of order, mediated by the SOC, while Sec. V presents the experimental
manifestations of orbital loop-current order. Conclusions are presented
in Sec. VI. To help the reader navigate the paper, Appendix
A describes the various quantities defined
throughout the main text. Appendix B shows the explicit form of the band dispersions, whereas Appendix C presents
explicit expressions for the vertex functions needed in the diagrammatic
calculations.

\section{Microscopic low-energy model \label{sec:Three-orbital-model}}

Instead of the full electronic model which requires 10 Fe orbitals
per unit cell (5 per Fe site), we consider a reduced orbital-projected
band model. This electronic model was derived previously by employing
the symmetry properties of the underlying crystal structure \cite{Cvetkovic2013}
and by expanding the tight binding dispersion of the electronic states
around the high symmetry points of the Brillouin zone \cite{Christensen2015}.
It therefore represents an effective low-energy model of electronic
states in the vicinity of the Fermi surfaces located at these high
symmetry points. Electronic states residing on pnictogen/chalcogenide
lattice sites are energetically well below or above the Fermi surface
and are therefore assumed to effectively renormalize the model's parameters.
Besides its simplicity, the full symmetry properties of the non-symmorphic
$P4/nmm$ space group characterizing a single Fe-pnictogen/chalcogen
plane are encoded in this model, including the glide plane symmetry
resulting from the alternate stacking of the pnictogen/chalcogen
atoms. For the sake of brevity, we review here only the aspects of
this model necessary for our analysis. A more detailed derivation
of this model can be found in Refs. \cite{Cvetkovic2013,Christensen2015}
or in the review \cite{FernandesChubukov2017}.

Before presenting the model, we first comment on the representation
of the electronic states in the 2-Fe and 1-Fe unit cells to clarify
the notation. As discussed above, the 2-Fe unit cell represents the
actual crystallographic unit cell, with the corresponding 2-Fe Brillouin
zone described in terms of the real crystal momenta of the electronic
states following Bloch's theorem. Alternatively, the electronic states
can be represented in the 1-Fe unit cell in terms of the spatial coordinates
$x$ and $y$ connecting nearest-neighbor Fe lattice sites, as well
as in the corresponding 1-Fe Brillouin zone described in terms of
pseudo-crystal momenta $k_{x}$ and $k_{y}$ \cite{Lee2008,LinBerlijnWangEtAl2011}.
For the sake of simplicity, we will express all states in the 1-Fe
Brillouin zone notation, highlighting the difference between real
and pseudo-crystal momentum if necessary. 

The low-energy electronic model consists of two hole-like Fermi pockets
located at the $\Gamma=\left(0,0\right)$ point of the 1-Fe Brillouin
zone and two electron-like Fermi pockets located at $X=\left(\pi,0\right)$
and $Y=\left(0,\pi\right)$. Electronic states around $\Gamma$ transform
as the irreducible $E_{g}$ representation of the tetragonal $D_{4h}$
point group \cite{Cvetkovic2013}, forming therefore an orbital doublet
composed of degenerate $xz$ and $yz$ Fe orbitals (oriented along
the Fe-Fe direction). Thus, it is convenient to introduce the doublet
\begin{align}
\psi_{\Gamma,\mathbf{k}\sigma} & =\begin{pmatrix}d_{yz,\mathbf{k}\sigma}\\
-d_{xz,\mathbf{k}\sigma}
\end{pmatrix}.\label{eq:states-Gamma}
\end{align}

Electronic states near the $X$ and $Y$ points form two electron
pockets centered at momenta $\mathbf{Q}_{1}=(\pi,0)$ and $\mathbf{Q}_{2}=(0,\pi)$,
composed of $yz/xy$ and $xz/xy$ orbitals, respectively. Thus, we
introduce two additional doublets:
\begin{subequations}
\label{eq:states-M}
\begin{align}
\psi_{X,\mathbf{k}+\mathbf{Q}_{1}\sigma} & =\begin{pmatrix}d_{yz,\mathbf{k}+\mathbf{Q}_{1}\sigma}\\
d_{xy,\mathbf{k}+\mathbf{Q}_{1}\sigma}
\end{pmatrix}\quad\text{and}\\
\psi_{Y,\mathbf{k}+\mathbf{Q}_{2}\sigma} & =\begin{pmatrix}d_{xz,\mathbf{k}+\mathbf{Q}_{2}\sigma}\\
d_{xy,\mathbf{k}+\mathbf{Q}_{2}\sigma}
\end{pmatrix}.
\end{align}
\end{subequations}

Because of the glide plane symmetry of the system, which makes the
system invariant under a $\left(\frac{1}{2},\frac{1}{2}\right)$ lattice
translation (in the 2-Fe unit cell) followed by a mirror reflection
with respect to the plane, the elements of the two doublets transform
according to the two-dimensional irreducible representations (denoted
$E_{Mi}$) of the $P4/nmm$ group at the $M=\left(\pi,\pi\right)$
point of the 2-Fe Brillouin zone. In particular, the upper components
form another doublet that transforms as $E_{M1}$ whereas the lower components
form a doublet that transforms as $E_{M3}$. 

The effective tight-binding Hamiltonian in real space is then given
by
\begin{align}
H_{0} & =\sum_{ij\sigma}t_{ij}^{a,b}\left(d_{a,i\sigma}^{\dagger}d_{b,j\sigma}+\mbox{H.c.}\right)\label{eq:tight-binding-H}
\end{align}
with hopping amplitude $t_{ij}^{a,b}$ representing the wave function's
overlap of orbitals $a$ and $b$ at Fe lattice site $i$ and $j$.
Represented in reciprocal space, the Hamiltonian is given by 
\begin{multline}
H_{0}=\sum_{\mathbf{k}\sigma}\psi_{\Gamma,\mathbf{k}\sigma}^{\dagger}\epsilon_{\mathbf{k}}^{\Gamma}\psi_{\Gamma,\mathbf{k}\sigma}\\
+\sum_{\mathbf{k}\sigma}\psi_{X,\mathbf{k}+\mathbf{Q}_{1}\sigma}^{\dagger}\epsilon_{\mathbf{k}}^{X}\psi_{X,\mathbf{k}+\mathbf{Q}_{1}\sigma}\\
+\sum_{\mathbf{k}\sigma}\psi_{Y,\mathbf{k}+\mathbf{Q}_{2}\sigma}^{\dagger}\epsilon_{\mathbf{k}}^{Y}\psi_{Y,\mathbf{k}+\mathbf{Q}_{2}\sigma}\label{H0}
\end{multline}
where $\epsilon^{\Gamma}$, $\epsilon^{X}$ and $\epsilon^{Y}$ are
$2\times2$ matrices acting in the orbital doublet space, whose explicit
expressions are given in Appendix \ref{sec:Orbital-dispersion-relations}.
Diagonalizing these matrices gives the corresponding band dispersions
of the electronic states. 

The atomic spin-orbit coupling, $H_{\text{SOC}}=\lambda\mathbf{S}\cdot\mathbf{L}$,
gives an additional contribution to the non-interacting Hamiltonian
\cite{Cvetkovic2013,Christensen2015}: 
\begin{multline}
H_{\text{SOC}}=i\frac{\lambda}{2}\sum_{\mathbf{k}\alpha\beta}\left(d_{xz,\mathbf{k}\alpha}^{\dagger}\sigma_{\alpha\beta}^{z}d_{yz,\mathbf{k}\beta}-\text{H.c.}\right)\\
+i\frac{\lambda}{2}\sum_{\mathbf{k}\alpha\beta}\left(d_{xz,\mathbf{k}+\mathbf{Q}_{2}\alpha}^{\dagger}\sigma_{\alpha\beta}^{x}d_{xy,\mathbf{k}+\mathbf{Q}_{1}\beta}-\text{H.c.}\right)\\
+i\frac{\lambda}{2}\sum_{\mathbf{k}\alpha\beta}\left(d_{yz,\mathbf{k}+\mathbf{Q}_{1}\alpha}^{\dagger}\sigma_{\alpha\beta}^{y}d_{xy,\mathbf{k}+\mathbf{Q}_{2}\beta}-\text{H.c.}\right)
\end{multline}
with Pauli matrix $\sigma^{a}$ acting on spin space. The first term
shows that, at the center of the Brillouin zone, SOC mixes the $xz$
and $yz$ orbitals, lifting the degeneracy of the orbital doublet.
The second term, on the other hand, couples the electronic states
located at the $X$ and $Y$ points. This term does not violate momentum
conservation because the $xy$ orbital has real crystal momentum $\mathbf{k}+\mathbf{Q}_{i}+\left(\pi,\pi\right)=\mathbf{k}+\mathbf{Q}_{\bar{i}}$
(with $\bar{1}=2$ and $\bar{2}=1$) due to the glide plane symmetry
affecting differently even and odd orbitals \cite{Lee2008}. In terms
of the doublets introduced above, the SOC Hamiltonian is given by:
\begin{multline}
H_{\text{SOC}}=\sum_{\mathbf{k}\sigma}\psi_{\Gamma,\mathbf{k}\alpha}^{\dagger}\Lambda_{\Gamma,\alpha\beta}^{\text{SOC}}\psi_{\Gamma,\mathbf{k\beta}}\\
+\sum_{\mathbf{k}}\left(\psi_{X,\mathbf{k}+\mathbf{Q}_{1}\alpha}^{\dagger}\Lambda_{M,\alpha\beta}^{\text{SOC}}\psi_{Y,\mathbf{k}+\mathbf{Q}_{2}\beta}+\text{H.c.}\right)\label{eq:H-SOC}
\end{multline}
with the vertices:
\begin{subequations}
\begin{eqnarray}
\Lambda_{\Gamma}^{\text{SOC}} & = & i\frac{\lambda}{2}\begin{pmatrix}0 & -1\\
1 & 0
\end{pmatrix}\sigma^{z},\label{eq:VSOC1}\\
\Lambda_{M}^{\text{SOC}} & = & i\frac{\lambda}{2}\left[\begin{pmatrix}0 & 1\\
0 & 0
\end{pmatrix}\sigma^{x}+\begin{pmatrix}0 & 0\\
1 & 0
\end{pmatrix}\sigma^{y}\right]\label{eq:VSOC2}.
\end{eqnarray}
\end{subequations}

	\section{Classification of magnetic and orbital loop-current orders\label{sec:Magnetic-and-orbital}}

Having established the microscopic model, we now classify the possible
types of stripe spin-density wave (SDW) order and orbital loop current
(OLC) order allowed by the symmetries of the system. 

\begin{table}[b]
	\protect\caption{\label{tab:Possible-intra-orbital-SDW} Possible intra-orbital spin
		density-wave and orbital loop-current orders, classified by the two
		dimensional irreducible representations $E_{Mi}:\,\left(E_{Mi}^{+},E_{Mi}^{-}\right)$
		of the $P4/nmm$ space group \cite{Cvetkovic2013}. Also represented
		in the table are the spin composition and orbital composition of each
		order, corresponding to the indices $a$ and $b$ defined in Eqs.
		\ref{eq:order-parameter-M} and \ref{eq:order-poarameter-phi}. In
		the case of SDW order, the orbital composition is intra-orbital $a=b$.
		In the case of orbital loop-current order, the spin composition is
		trivial. Due to the trivial transformation behavior of the free-energy,
		there is only a bilinear coupling between spin magnetic and loop current
		orders that share the same symmetry properties, i.e., that transform
		according to the same irreducible representations. Here, fields
		listed in the same line can couple bilinearly in the free-energy expansion. }
	\begin{ruledtabular}
		\begin{tabular}[b]{>{\raggedright}p{0.2\columnwidth}cccc}
			\noalign{\vskip\doublerulesep}
			\multirow{1}{0.2\columnwidth}{} & \multicolumn{2}{c}{%
				\begin{minipage}[t]{0.35\columnwidth}%
					\begin{center}
						intra-orbital \\
						spin order
						\par\end{center}%
				\end{minipage}} & \multicolumn{2}{c}{%
				\begin{minipage}[t]{0.35\columnwidth}%
					\begin{center}
						orbital loop\\
						current order 
						\par\end{center}%
				\end{minipage}}
				\tabularnewline[0.4cm]
				\cline{2-5} 
				\noalign{\vskip\doublerulesep}
				\noalign{\vskip\doublerulesep}
				irrep of $P4/nmm$ space group & %
				\begin{minipage}[t]{0.15\columnwidth}%
					field%
				\end{minipage}  & %
				\begin{minipage}[t]{0.15\columnwidth}%
					orbital and spin comp.%
				\end{minipage} & %
				\begin{minipage}[t]{0.1\columnwidth}%
					field%
				\end{minipage}  & %
				\begin{minipage}[t]{0.2\columnwidth}%
					orbital composition%
				\end{minipage}
				\tabularnewline[0.4cm]
				\hline 
				\noalign{\vskip\doublerulesep}
				\hline 
				\noalign{\vskip\doublerulesep}
				$\begin{pmatrix}E_{M1}^{+}\\
				E_{M1}^{-}
				\end{pmatrix}$ & $\begin{pmatrix}m_{\mathbf{Q}_{2}}^{y}\\
				m_{\mathbf{Q}_{1}}^{x}
				\end{pmatrix}$ & $\begin{pmatrix}xz,\,\sigma^{y}\\
				yz,\,\sigma^{x}
				\end{pmatrix}$ & $\begin{pmatrix}\phi_{\mathbf{Q}_{1}}^{\left(yz,xy\right)}\\
				\phi_{\mathbf{Q}_{2}}^{\left(xz,xy\right)}
				\end{pmatrix}$ & $\begin{pmatrix}yz,\, xy\\
				xz,\, xy
				\end{pmatrix}$
				\tabularnewline[0.4cm]
				\hline 
				\noalign{\vskip\doublerulesep}
				\noalign{\vskip\doublerulesep}
				$\begin{pmatrix}E_{M2}^{+}\\
				E_{M2}^{-}
				\end{pmatrix}$ & $\begin{pmatrix}m_{\mathbf{Q}_{2}}^{x}\\
				m_{\mathbf{Q}_{1}}^{y}
				\end{pmatrix}$ & $\begin{pmatrix}xz,\,\sigma^{x}\\
				yz,\,\sigma^{y}
				\end{pmatrix}$ & $\begin{pmatrix}\phi_{\mathbf{Q}_{1}}^{\left(xz,xy\right)}\\
				\phi_{\mathbf{Q}_{2}}^{\left(yz,xy\right)}
				\end{pmatrix}$ & $\begin{pmatrix}xz,\, xy\\
				yz,\, xy
				\end{pmatrix}$
				\tabularnewline[0.4cm]
				\hline 
				\noalign{\vskip\doublerulesep}
				\noalign{\vskip\doublerulesep}
				$\begin{pmatrix}E_{M3}^{+}\\
				E_{M3}^{-}
				\end{pmatrix}$ & $\begin{pmatrix}m_{\mathbf{Q}_{1}}^{z}\\
				m_{\mathbf{Q}_{2}}^{z}
				\end{pmatrix}$ & $\begin{pmatrix}yz,\,\sigma^{z}\\
				xz,\,\sigma^{z}
				\end{pmatrix}$ & $\begin{pmatrix}\phi_{\mathbf{Q}_{1}}^{\left(xz,yz\right)}\\
				\phi_{\mathbf{Q}_{2}}^{\left(yz,xz\right)}
				\end{pmatrix}$ & $\begin{pmatrix}xz,\, yz\\
				yz,\, xz
				\end{pmatrix}$\tabularnewline[0.4cm]
				\hline 
				\noalign{\vskip\doublerulesep}
				\noalign{\vskip\doublerulesep}
				$\begin{pmatrix}E_{M4}^{+}\\
				E_{M4}^{-}
				\end{pmatrix}$ & - & - & $\begin{pmatrix}\phi_{\mathbf{Q}_{1}}^{\left(yz,yz\right)}\\
				\phi_{\mathbf{Q}_{2}}^{\left(xz,xz\right)}
				\end{pmatrix}$ & $\begin{pmatrix}yz,\, yz\\
				xz,\, xz
				\end{pmatrix}$\tabularnewline[0.4cm]
			\end{tabular}
		\end{ruledtabular}
	\end{table}

\subsection{Magnetic Order}

The SDW order realized in the iron pnictides is parametrized in terms
of magnetic order parameters $\mathbf{m}_{\mathbf{Q}_{j}}$ associated
with each of the two magnetic ordering vectors $\mathbf{Q}_{1}=(\pi,0)$
and $\mathbf{Q}_{2}=(0,\pi)$. Generally, these order parameters are
$2\times2$ tensors in orbital space and vectors in spin space. They
are defined in terms of the electronic operators according to (summation
over spin indices is left implicit):
\begin{equation}
\mathbf{m}_{\mathbf{Q}_{j}}^{\left(a,b\right)}\propto\sum_{\mathbf{k}}\langle d_{a,\mathbf{k}\alpha}^{\dagger}\boldsymbol{\sigma}_{\alpha\beta}d_{b,\mathbf{k}+\mathbf{Q}_{j}\beta}+\text{H.c.}\rangle\label{eq:order-parameter-M}
\end{equation}
with $\alpha,\beta$ corresponding to spin indices and $a,b$ to orbital
indices. Thus, we can write down the SDW Hamiltonian:
\begin{equation}
H_{\text{SDW}}=\sum_{\mathbf{k}j,ab}\mathbf{m}_{\mathbf{Q}_{j}}^{\left(a,b\right)}\cdot\left(d_{a,\mathbf{k}\alpha}^{\dagger}\boldsymbol{\sigma}_{\alpha\beta}d_{b,\mathbf{k}+\mathbf{Q}_{j}\beta}+\text{H.c.}\right).
\end{equation}

Hereafter, we focus on intra-orbital SDW, $a=b$, which is believed
to be the leading magnetic instability in the iron-based superconductors
\cite{Christensen2015,GastiasoroAndersen2015}. Now, SDW order involves
combinations of electronic states at the $\Gamma$ and $X/Y$ points
of the 1-Fe Brillouin zone. Because in our low-energy model the electronic
states at $\Gamma$ contain only contributions from $xz/yz$ orbitals,
whereas the states at $X$ have contributions from $yz/xy$ and at
$Y$, from $xz/xy$, there is only one intra-orbital component $\mathbf{m}_{\mathbf{Q}_{j}}^{\left(a,a\right)}$
relevant for each ordering vector $\mathbf{Q}_{j}$, namely, $a=yz$
for $\mathbf{Q}_{1}=\left(\pi,0\right)$ and $a=xz$ for $\mathbf{Q}_{2}=\left(0,\pi\right)$.
Therefore, to simplify the notation, hereafter we set:
\begin{align}
\mathbf{m}_{\mathbf{Q}_{1}} & \equiv\mathbf{m}_{\mathbf{Q}_{1}}^{\left(yz,yz\right)}\nonumber \\
\mathbf{m}_{\mathbf{Q}_{2}} & \equiv\mathbf{m}_{\mathbf{Q}_{2}}^{\left(xz,xz\right)}.
\end{align}

Note that, since only the $xz$ and $yz$ orbitals are involved, the
pseudo-crystal momentum for the SDW order parameters coincides with
the real crystal momentum. In the case of stripe SDW, which is our
interest here, it follows that either $\mathbf{m}_{\mathbf{Q}_{1}}=0$
or $\mathbf{m}_{\mathbf{Q}_{2}}=0$. 

As pointed out in Ref. \cite{Cvetkovic2013}, the possible long-range
magnetic orders are classified by the irreducible representations
of the $P4/nmm$ space group at the $M$ point of the 2-Fe Brillouin
zone. In particular, the six different vector components of the two
$\mathbf{m}_{\mathbf{Q}_{j}}$ order parameters are grouped in three
doublets, which transform according to the three two-dimensional irreducible
representations $E_{Mi}$ with $i\in\{1,2,3\}$. The full classification
is presented in Table \ref{tab:Possible-intra-orbital-SDW}. Note
that with the constraint of intra-orbital order only, there is no
magnetic order that transforms as the irreducible representation
$E_{M4}$. 

Thus, this classification allows us to write the SDW order parameters
not as two three-dimensional vector order parameters $m_{\mathbf{Q}_{j}}^{\alpha}$
(with $\alpha=x,y,z$ and $j=1,2$), but rather as three doublets
$M_{E_{M_{i}}}^{\mu}$, with $i\in\{1,2,3\}$ denoting the appropriate
irreducible representation, and $\mu\in\{+,-\}$ denoting the component
of the doublet. The relationship between these two representations
are those already presented in Table \ref{tab:Possible-intra-orbital-SDW}.
In what follows, we will use the $M_{E_{Mi}}^{\mu}$ notation for
convenience. Their physical interpretation is straightforward: the
doublet component $\mu$ corresponds to either $\mathbf{Q}_{1}$ or
$\mathbf{Q}_{2}$ order. Moreover, $M_{E_{M1}}^{\mu}$ corresponds
to stripes with in-plane moments parallel to $\mathbf{Q}_{j}$, i.e., the one frequently observed in iron-based systems; $M_{E_{M2}}^{\mu}$ corresponds to stripes with in-plane moments perpendicular to $\mathbf{Q}_{j}$;
and $M_{E_{M3}}^{\mu}$ corresponds to stripes with out-of-plane moments.

In our weak-coupling approach, magnetism arises from low-energy states near the Fermi level. In some iron-based systems, there is also another hole pocket made out of $d_{xy}$ orbitals centered at the point $ M=(\pi,\pi) $ of the 1-Fe Brillouin zone (equivalent to the $ \Gamma $ point of the 2-Fe Brillouin zone). As a result, it is possible to form intra-orbital magnetic order parameters in Eq. \ref{eq:order-parameter-M} with $a=b=xy$. 
These $xy$ SDW order parameters are classified in identical manner as the ones involving the $xz$ and $yz$ orbitals, via the two-dimensional irreducible representations $E_{Mi}:\,\left(E_{Mi}^{+},E_{Mi}^{-}\right)$ of the $P4/nmm$ space group. This classification is presented in the Appendix in Table \ref{tab:Possible-intra-orbital-SDW-XY}. Clearly, by symmetry, the same types of OLC order are induced by $(\pi,0)$ or $(0,\pi)$ magnetic order parameters, regardless of their orbital composition. Thus, while in the remainder of the paper we will focus for simplicity on the microscopic model that does not have intra-orbital $xy$ magnetism, the qualitative results remain
the same even if one includes this type of SDW order parameters. 
\\
\begin{figure*}
\subfloat[\label{fig:Orbital-loop-current a}Loop current order belonging to
the irrep $E_{M1}^{-}$: $\phi_{\mathbf{Q}_{2}}^{\left(xz,xy\right)}$ ]{\includegraphics[width=0.25\paperwidth]{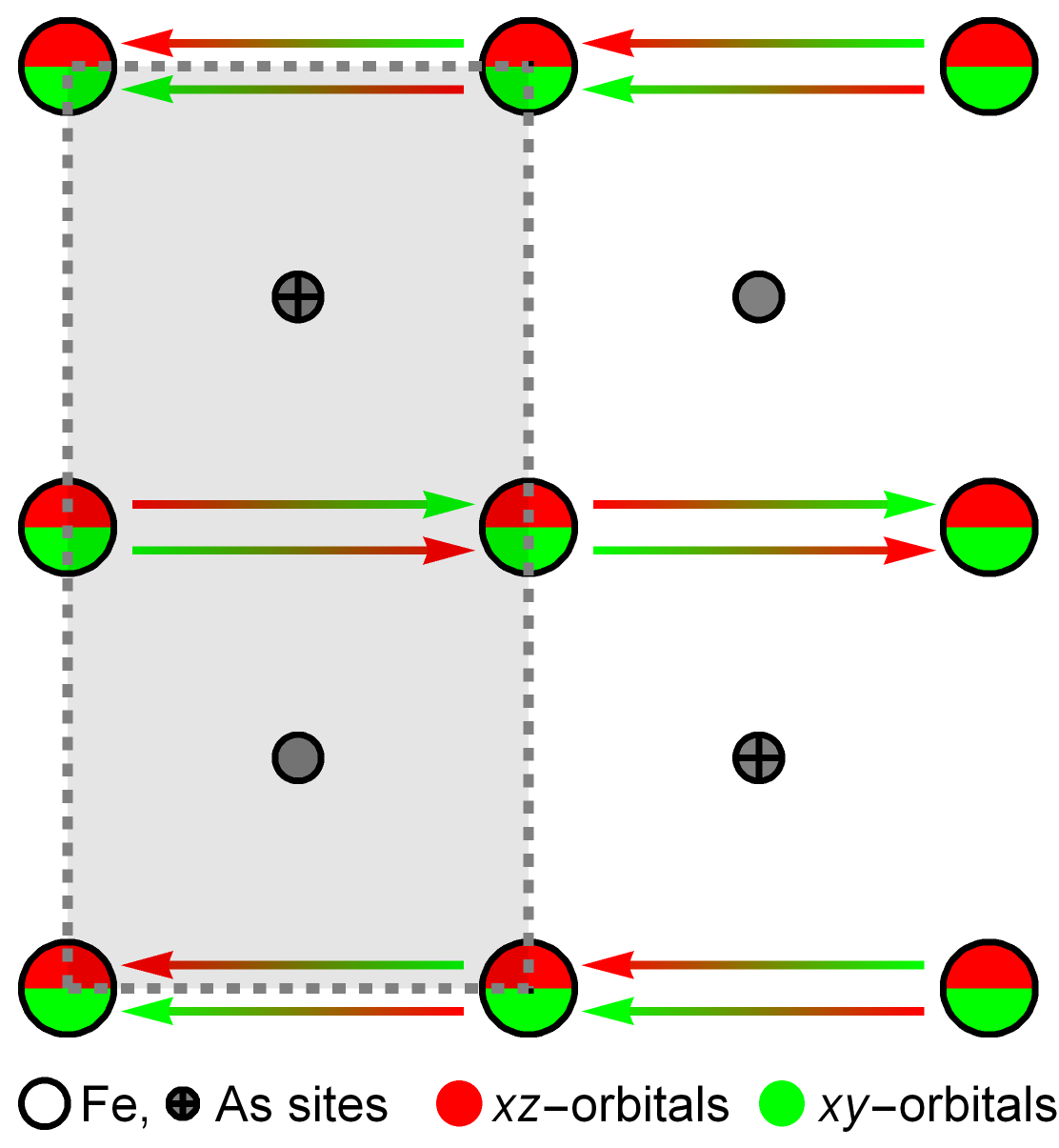}
}\hfill{}\subfloat[\label{fig:Orbital-loop-current b}Loop current order belonging to
the irrep $E_{M2}^{-}$: $\phi_{\mathbf{Q}_{2}}^{\left(yz,xy\right)}$ ]{\includegraphics[width=0.25\paperwidth]{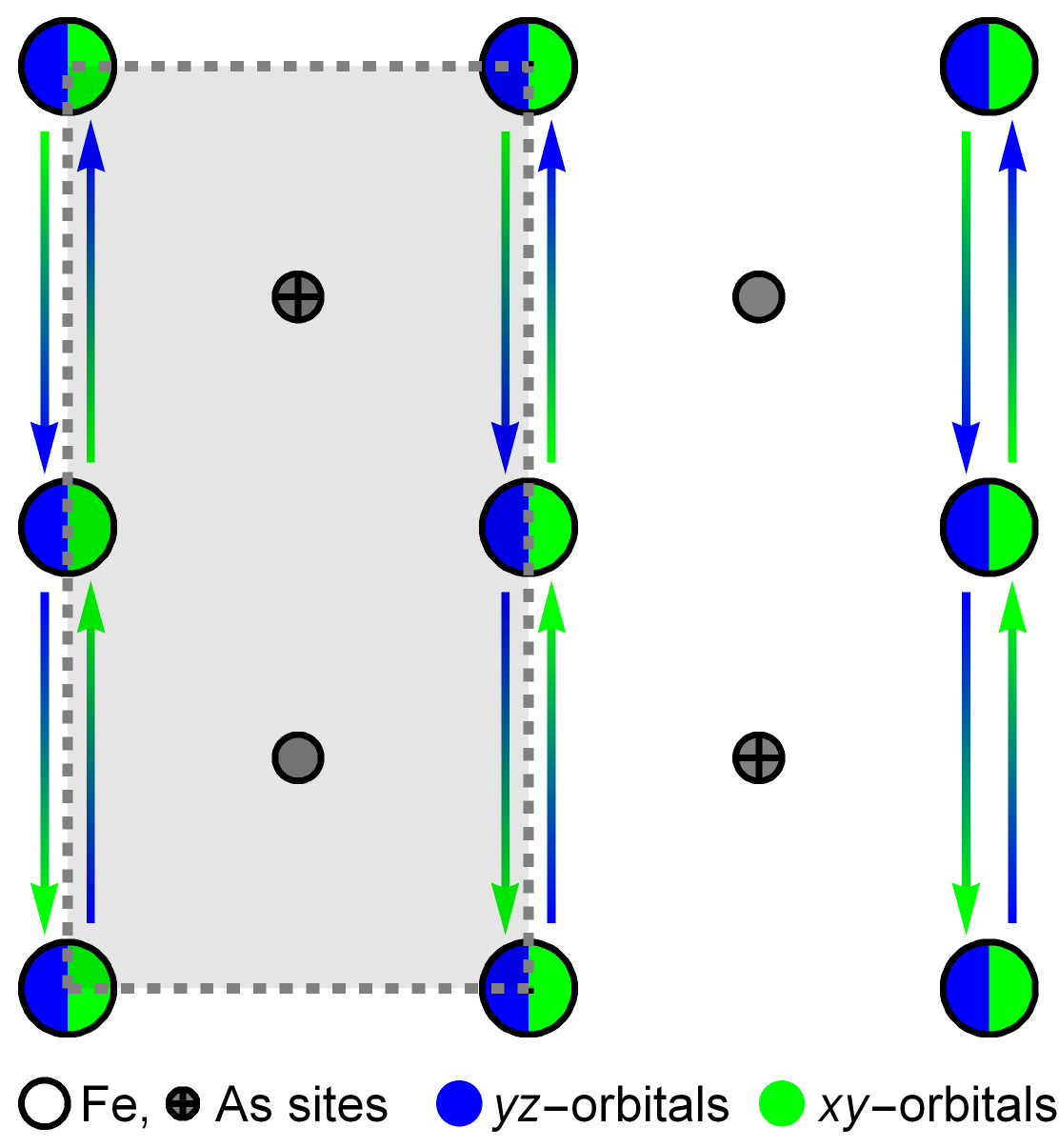}}\hfill{}\subfloat[\label{fig:Orbital-loop-current c}Loop current order belonging to
the irrep $E_{M3}^{+}$: $\,\phi_{\mathbf{Q}_{1}}^{\left(xz,yz\right)}$ ]{\includegraphics[width=0.25\paperwidth]{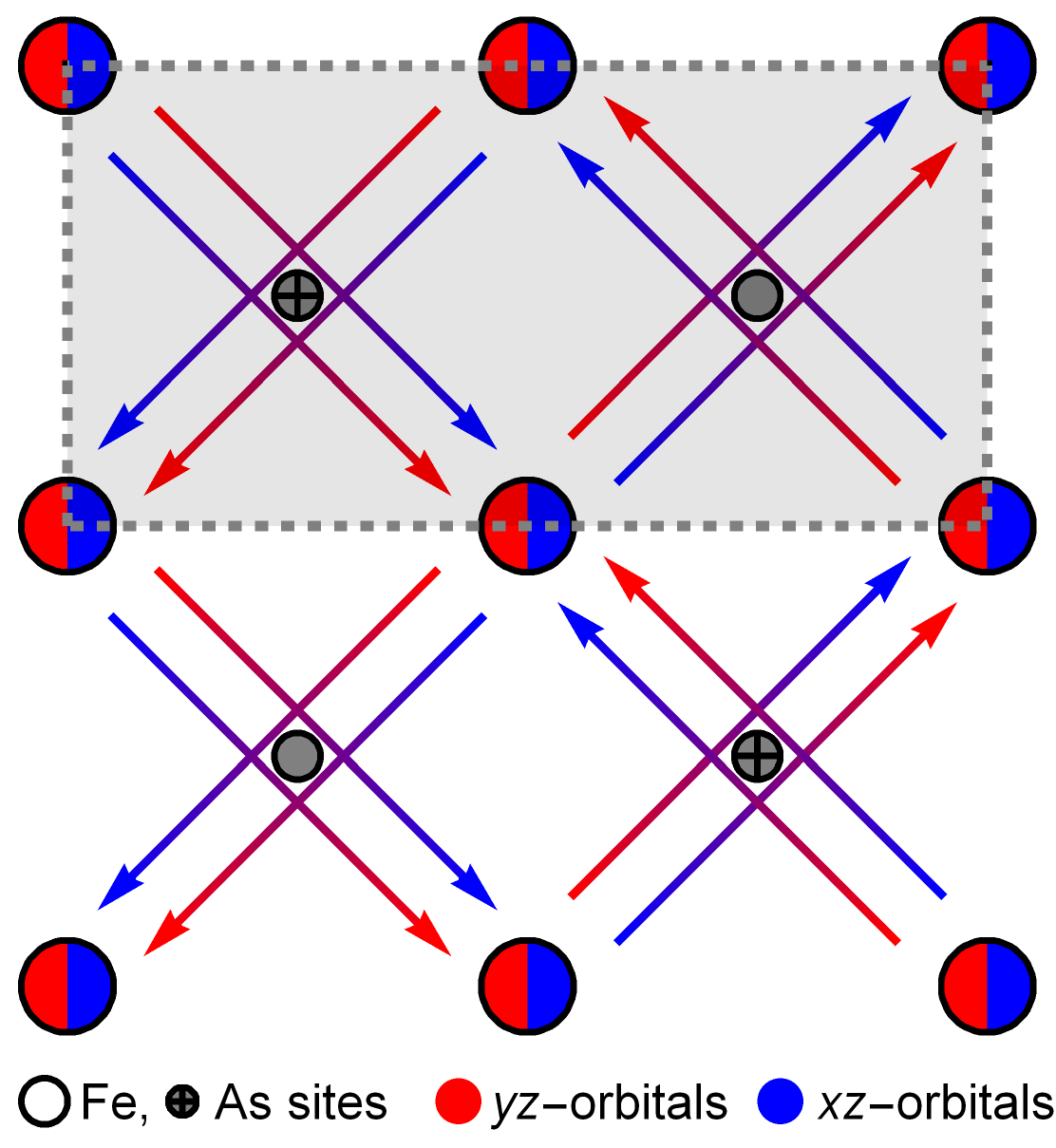}

}

\protect\caption{\label{fig:Orbital-current-textures} 
Loop-current patterns of three
different OLC ground states, one for each possible irreducible representation
$E_{Mi}^{\mu}$ where periodic boundary conditions are assumed. The representation of the $\mu$ component not depicted
in each panel is obtained by performing a mirror reflection with respect
to planes whose normals are along the diagonals of the 1-Fe unit cell.
To create these configurations, only leading order Fe-Fe hopping parameters
are considered. The coloring of the arrows representing the electronic
flows refers to their orbital composition. Small gray circles refer
to pnictogen/chalcogen sites lying above (crossed circles) or below
(non-crossed circles) the Fe plane; the large circles represent Fe
lattice sites. Shaded rectangles indicate the enlarged unit cell due
to the translational symmetry breaking imposed by the OLC order. 
}
\end{figure*}

\subsection{Orbital Loop Current Order}
To classify the different types of translational symmetry-breaking
orbital loop current (OLC) order, we look for bilinear combinations
of fermionic field operators with momentum transfer $\mathbf{Q}_{1}$
and $\mathbf{Q}_{2}$ that are trivial in spin space while breaking
time reversal symmetry. Therefore, in analogy to the SDW order parameters,
the most general OLC order parameter is given in terms of the orbital
electronic operators by:

\begin{eqnarray}
\phi_{\mathbf{Q}_{j}}^{(a,b)} & \propto & \sum_{\mathbf{k}}i\langle d_{a,\mathbf{k}\sigma}^{\dagger}d_{b,\mathbf{k}+\mathbf{Q}_{j}\sigma}-\text{H.c.}\rangle
\end{eqnarray}
yielding the following OLC Hamiltonian:

\begin{equation}
H_{\text{OLC}}=\sum_{\mathbf{k}j,ab}i\,\phi_{\mathbf{Q}_{j}}^{\left(a,b\right)}\left(d_{a,\mathbf{k}\sigma}^{\dagger}d_{b,\mathbf{k}+\mathbf{Q}_{j}\sigma}-\text{H.c.}\right)\label{eq:order-poarameter-phi}.
\end{equation}

Similarly to the case of the SDW orders, the different types of OLC
orders must transform according to one of the three two-dimensional
irreducible representations $E_{M_{i}}$. Consequently, this restricts
the possible orbital indices $\phi_{\mathbf{Q}_{j}}^{(a,b)}$. The
orbital composition of all possible orbital loop-current orders and
their classification in terms of the $E_{M_{i}}$ irreducible representation
of the $P4/nmm$ space group is given in Table \ref{tab:Possible-intra-orbital-SDW}.
In analogy to the notation introduced in the SDW case, we will express
the OLC order parameters also in terms of their corresponding doublets,
denoted by $\Phi_{E_{Mi}}^{\mu}$. 

Note that, in contrast to the SDW orders, we can construct a loop
current order which transforms as the irreducible representation $E_{M4}$.
This is the only type of OLC order that is intra-orbital; all other
ones are inter-orbital order. Because this OLC order cannot couple
to any intra-orbital SDW order, this component will not be considered hereafter.
We also point out that the two orbital loop-current orders that couple
$xz$/$yz$ orbitals with $xy$ orbitals (corresponding to the $E_{M1}$
and $E_{M2}$ irreducible representations) actually carry real crystal
momentum $\mathbf{Q}_{\text{Real}}=\mathbf{Q}_{i}+\left(\pi,\pi\right)=\mathbf{Q}_{\bar{i}}$.
Thus, these OLC order parameters $\phi_{\mathbf{Q}_{i}}$ must couple
to an SDW order parameter with momentum $\mathbf{Q}_{\bar{i}}$, instead
of momentum $\mathbf{Q}_{i}$, as one would naively expect. The difference
in pseudo-crystal momentum is compensated by the out-of-plane position
of the pnictogen/chalcogenide lattice sites as we will see later in
the discussion of loop current induced magnetic moments in Sec. \ref{sec:Experimental-consequences}.

Due to charge conservation, the OLC orders are represented in real
space by closed current loops. The link between the order parameter
$\phi_{\mathbf{Q}_{j}}^{(a,b)}$ and the physical electronic currents
is established by considering the current operator $\hat{\mathbf{j}}_{ij}$
between lattice sites $i$ and $j$ with distance $\mathbf{R}_{ij}\equiv\mathbf{R}_{i}-\mathbf{R}_{j}$:

\begin{equation}
\hat{\mathbf{j}}_{ij}=\sum_{ab}\hat{\mathbf{j}}_{ij}^{ab}=\frac{ie}{\hbar}\frac{\mathbf{R}_{ij}}{R_{ij}}\sum_{ab\sigma}t_{ij}^{a,b}\left(d_{j,a\sigma}^{\dagger}d_{i,b\sigma}-\mbox{H.c.}\right),
\end{equation}
where $t_{ij}^{a,b}$ refers to the hopping amplitudes introduced
in context of the tight-binding Hamiltonian Eq. (\ref{eq:tight-binding-H}),
corresponding to the the overlap of localized Wannier states.\\
\begin{table}[b]
	\protect\caption{\label{tab:Leading-order-hopping}Leading order hopping processes
		between Fe lattice sites \cite{Graser2009} that are used to construct
		the loop current patterns depicted in Fig. \ref{fig:Orbital-current-textures}.
		NN and NNN refer to \textit{\emph{nearest-}}\emph{ }and\emph{ }\textit{\emph{next-nearest-neighbor}} hopping, respectively. }
		\begin{ruledtabular}
\begin{tabular}{lr}
\begin{minipage}[t]{0.2\columnwidth}%
\begin{flushleft}
coupled Fe\\
orbitals
\par\end{flushleft}%
\end{minipage} & %
\begin{minipage}[t]{0.3\columnwidth}%
\begin{flushright}
leading order \\
hopping processes
\par\end{flushright}%
\end{minipage}
\tabularnewline[0.4cm]
\hline 
\hline 
\noalign{\vskip\doublerulesep}
$xz$ and $xy$ & NN along $x$\tabularnewline
\hline 
\noalign{\vskip\doublerulesep}
$yz$ and $xy$ & NN along $y$\tabularnewline
\hline 
\noalign{\vskip\doublerulesep}
$xz$ and $yz$ & NNN along $x\pm y$ 
\end{tabular}
	\end{ruledtabular}
\end{table}
The key point is that a non-vanishing OLC order parameter causes a
finite expectation value for the current operator. Consequently, the
resulting loop current patterns depend on the hopping matrix elements
$t_{ij}^{a,b}$. Depending on the amplitude of the different elements
$t_{ij}^{a,b}$ connecting different sites $i$ and $j$ and orbitals
$a$ and $b$, arbitrary complex current patterns can be constructed
in principle. However, the symmetry properties of the resulting current
pattern do not depend on the actual set and strength of possible hopping
processes. For the sake of simplicity, we therefore consider the leading
order hopping parameters involving either nearest- or next-nearest
neighbors for each combination of orbitals, as summarized in Table
\ref{tab:Leading-order-hopping}. To construct the appropriate loop
current patterns for each orbital loop-current order parameter $\phi_{\mathbf{Q}_{j}}^{(a,b)}$,
we take into account the corresponding hopping parameters of Table
\ref{tab:Leading-order-hopping}, combined with the additional constraint
for local charge conservation 

\begin{equation}
\sum_{j}|\langle\hat{\mathbf{j}}_{ij}\rangle|=0,\quad\forall i\label{eq:charge-neutrality}
\end{equation}

As an example, let us show explicitly how to obtain the loop current
pattern corresponding to $\phi_{\mathbf{Q}_{2}}^{(xz,xy)}$ ($E_{M1}^{-}$
irreducible representation). The dominating hopping process connecting
$xz$ and $xy$ orbitals is the nearest-neighbor hopping along the
$x$-direction \cite{Graser2009}. Thus, the dominating contribution
to the current is given by:

\begin{equation}
\langle\hat{\mathbf{j}}\rangle_{ij'}^{xz,xy}=\frac{ie}{\hbar}\hat{\mathbf{e}}_{x}\, t_{ij'}^{xz,xy}\langle d_{j',xz\sigma}^{\dagger}d_{i,xy\sigma}-\mbox{H.c.}\rangle
\end{equation}

for all nearest neighbors $j'$ which fulfill $\mathbf{R}_{j'}=\mathbf{R}_{i}\pm\mathbf{\hat{e}}_{x}$.
Here, $\hat{\mathbf{e}}_{x}$ represents the unit vector along the
$x$ direction. Taking the Fourier transform, we find:

\begin{multline}
\langle\hat{\mathbf{j}}\rangle_{ij'}^{xz,xy}= 
\\ \frac{ie}{\hbar}\hat{\mathbf{e}}_{x}\, t_{ij'}^{xz,xy}\sum_{\mathbf{k},\mathbf{p}}e^{-i\left(\mathbf{k}-\mathbf{p}\right)\cdot\mathbf{R}_{i}}e^{-i\mathbf{k}\cdot\mathrm{\hat{\mathbf{e}}}_{x}}\langle d_{\mathbf{k},xz\sigma}^{\dagger}d_{\mathbf{p},xy\sigma}-\mbox{H.c.}\rangle.
\end{multline}

In the case of $\phi_{\mathbf{Q}_{2}}^{(xz,xy)}$ order, only $\mathbf{p}=\mathbf{k}+\mathbf{Q}_{2}$
survives in the sum above, yielding:
\begin{equation}
\langle\hat{\mathbf{j}}\rangle_{ij'}^{xz,xy}\propto\frac{e}{\hbar}\hat{\mathbf{e}}_{x}\, t_{ij'}^{xz,xy}\,\phi_{\mathbf{Q}_{2}}^{xz,xy}\, e^{i\mathbf{Q}_{2}\cdot\mathbf{R}_{i}},
\end{equation}
where, in the last step, we used the fact that $e^{-i\mathbf{k}\cdot\mathrm{\hat{\mathbf{e}}}_{x}}\approx1$
since the magnitude of $\mathbf{k}$ is restricted to small values.
Thus, there is a constant flow along the $x$ direction, which changes
sign as one changes the spatial position along the $y$-direction
due to the oscillating factor $e^{i\mathbf{Q}_{2}\cdot\mathbf{R}_{i}}$,
thus preserving local charge conservation as demanded. 

The analysis for other types of loop currents orders works analogously,
resulting in the loop-current patterns for $\phi_{\mathbf{Q}_{2}}^{\left(xz,xy\right)}$,
$\phi_{\mathbf{Q}_{2}}^{\left(yz,xy\right)}$ and $\phi_{\mathbf{Q}_{1}}^{\left(xz,yz\right)}$
depicted in Fig. \ref{fig:Orbital-current-textures}, where we assume periodic boundary conditions. As explained
above, the current patterns displayed are one contribution to the
total current flow present. However, they posses the full symmetry
properties imposed by their classification into the three irreducible
representations, providing thus insight onto the relationship between
the OLC order parameters to microscopic charge flows. Moreover, we
emphasize that the inter-site flow between the Fe lattice sites is
actually mediated by the pnictogen/chalcogen lattice sites, though
their electronic degrees of freedom are integrated out in our microscopic
model. This will be relevant to discuss the experimental manifestations
on OLC orders below.

\section{Microscopic coupling between magnetic and orbital loop-current orders}

\begin{figure*}
\subfloat[Coupling between the $E_{M1}^{-}$ fields $m_{\mathbf{Q}_{1}}^{x}$
and $\phi_{\mathbf{Q}_{2}}^{\left(xz,xy\right)}$ ]{\includegraphics[width=0.22\paperwidth]{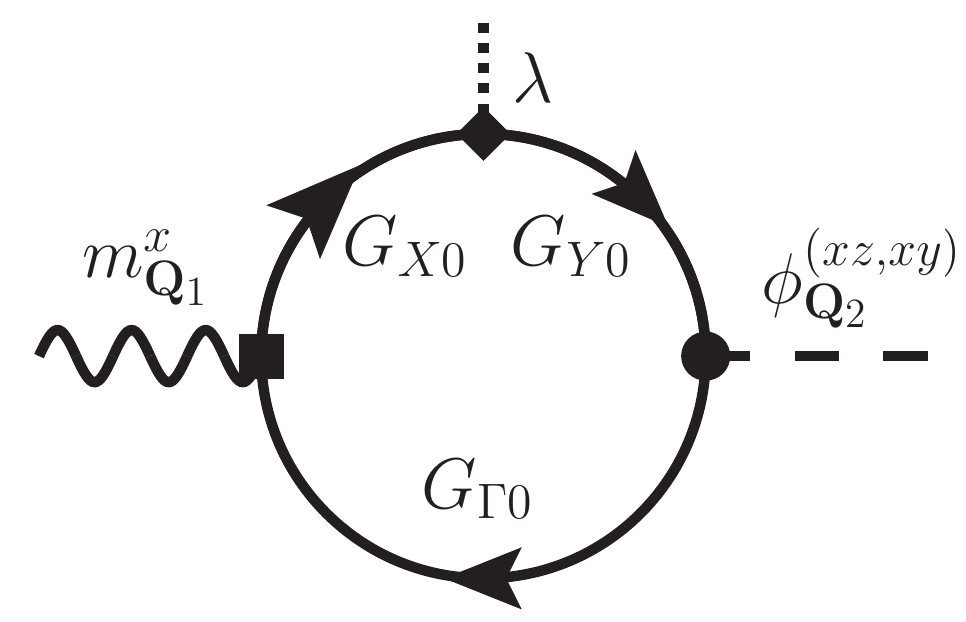}

}\hfill{}\subfloat[Coupling between the $E_{M2}^{-}$ fields $m_{\mathbf{Q}_{1}}^{y}$
and $\phi_{\mathbf{Q}_{2}}^{\left(yz,xy\right)}$]{\includegraphics[width=0.22\paperwidth]{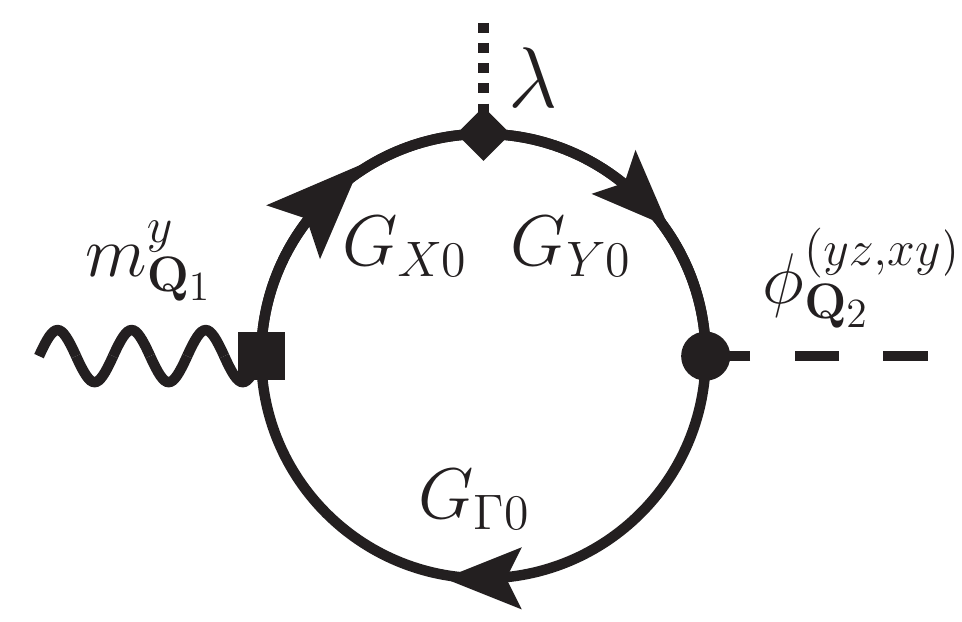}}\hfill{}\subfloat[Coupling between the $E_{M3}^{+}$ fields $m_{\mathbf{Q}_{1}}^{z}$
and $\phi_{\mathbf{Q}_{1}}^{\left(xz,yz\right)}$]{\includegraphics[width=0.22\paperwidth]{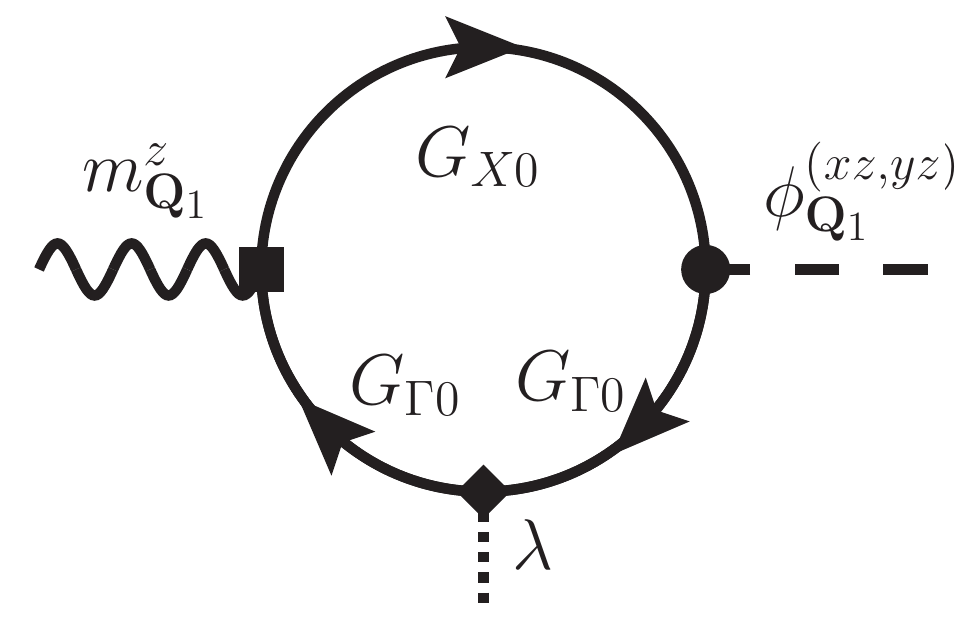}

\label{sub:Induced-Magnetic-moments}}

\protect\caption{\label{fig:microscopical-coupling-M-phi}Microscopic representation
of the coupling coefficient $g_{E_{Mi}}^{\mu}(\lambda)$ in terms of Feynman diagrams
coupling spin density-wave and orbital loop-current orders. Wavy and
dashed lines represent SDW and OLC current fields, respectively, whereas
dotted lines refer to the spin-orbit interaction. $G_{\Gamma,0}$,
$G_{X0},\, G_{Y0}$ represent fermionic propagators of electronic
states residing at the three different pockets of the Brillouin zone.
The vertices for the coupling between electronic and bosonic degrees
of freedom and the SOC are represented by a square, a circle, and
a diamond, respectively. }
\end{figure*}

The analysis of the previous section reveals a set of three magnetic
order parameters $M_{E_{M_{i}}}^{\mu}$ and three orbital loop current
order parameters $\Phi_{E_{Mi}}^{\mu}$ that transform according to
the same two-dimensional irreducible representations $E_{M_{i}}$,
with $i=1,\,2,\,3$ and $\mu\in\{+,-\}$ denoting the two elements
of each doublet. As a result, symmetry dictates that if $M_{E_{Mi}}^{\mu}\neq0$,
the corresponding $\Phi_{E_{Mi}}^{\mu}$ must also become non-zero.
In other words, stripe SDW order induces OLC in the iron-based superconductors.
Mathematically, this implies that the Ginzburg-Landau free-energy
expansion of this model must contain bilinear terms of the form $g_{E_{Mi}}^{\mu} M_{E_{Mi}}^{\mu}\Phi_{E_{Mi}}^{\mu}$.
In this section, we derive microscopically these coupling constants
$g_{E_{Mi}}^{\mu}$. Since $\Phi$ transforms trivially in spin space, it is clear
that the coupling constants must be non-zero only in the presence of
the SOC, i.e. $g_{E_{Mi}}^{\mu}(\lambda=0)=0$. 

The free-energy functional is obtained by following the standard procedure
of integrating out the electronic degrees of freedom \cite{Fernandes2012a}.
The non-interacting Hamiltonian was already defined in Sec. II.
Introducing the 12-dimensional enlarged electronic spinor $\Psi_{\mathbf{k}}=\left(\psi_{\Gamma,\mathbf{k}},\psi_{X,\mathbf{k}+\mathbf{Q}_{1}},\psi_{Y,\mathbf{k}+\mathbf{Q}_{2}}\right)^{\text{T}}$
associated with the low-energy electronic states, we can write it
in the compact form:
\begin{align}
H_{0} & =\sum_{\mathbf{k}}\Psi_{\mathbf{k}}^{\dagger}\mathcal{H}_{0}\Psi_{\mathbf{k}}\nonumber \\
H_{\mathrm{SOC}} & =\sum_{\mathbf{k}}\Psi_{\mathbf{k}}^{\dagger}\mathcal{H}_{\mathrm{SOC}}\Psi_{\mathbf{k}}
\end{align}
where the matrices $\mathcal{H}_{0}$ and $\mathcal{H}_{\mathrm{SOC}}$
can be directly read off Eqs. (\ref{H0}) and (\ref{eq:H-SOC}). As
for the interacting part, we start by projecting all possible electron-electron
interactions in the spin and orbital loop-current channels:
\begin{multline}
H_{\text{int}}=\Bigg[\sum_{i\mathbf{q}\mu}\left(U_{E_{Mi}}^{M}\hat{M}_{E_{Mi}}^{\mu}(\mathbf{q})\hat{M}_{E_{Mi}}^{\mu}(-\mathbf{q})\right)\\
+U_{E_{Mi}}^{\Phi}\hat{\Phi}_{E_{Mi}}^{\mu}(\mathbf{q})\hat{\Phi}_{E_{Mi}}^{\mu}(-\mathbf{q})\Bigg]
\end{multline}
where the SDW and OLC bosonic operators are given by 
\begin{eqnarray}
\hat{M}_{E_{Mi}}^{\mu}(\mathbf{q}) & = & \sum_{\mathbf{k}}\Psi_{\mathbf{k}+\mathbf{q}}^{\dagger}\varUpsilon_{E_{Mi},\mu}^{M}\Psi_{\mathbf{k}},\label{eq:Mvertex}\\
\hat{\Phi}_{E_{Mi}}^{\mu}(\mathbf{q}) & = & \sum_{\mathbf{k}}\Psi_{\mathbf{k}+\mathbf{q}}^{\dagger}\varUpsilon_{E_{Mi},\mu}^{\Phi}\Psi_{\mathbf{k}}.\label{eq:phiVertex}
\end{eqnarray}
Here, the hat indicates that the quantity is an operator. The bosonic
vertices $\varUpsilon_{E_{Mi},\mu}^{M}$ and $\varUpsilon_{E_{Mi},\mu}^{\Phi}$
for each channel indicate how each order couples to the low-energy
electrons. They are represented in terms of $12\times12$ matrices
acting in the enlarged spinor space and can be read off using Table
\ref{tab:Possible-intra-orbital-SDW}. For instance, for the order
parameters belonging to the irreducible representations $E_{M1}^{-}$,
the vertices are given by 
\begin{subequations}
\begin{align}
\varUpsilon_{E_{M1},-}^{M} & =\begin{pmatrix}0 & \tau^{\uparrow}\sigma^{x} & 0\\
\tau^{\uparrow}\sigma^{x} & 0 & 0\\
0 & 0 & 0
\end{pmatrix}\label{eq:vertexM}\\
\varUpsilon_{E_{M1},-}^{\Phi} & =\begin{pmatrix}0 & 0 & -i\tau^{\downarrow}\sigma^{0}\\
0 & 0 & 0\\
i\tau^{\downarrow}\sigma^{0} & 0 & 0
\end{pmatrix}\label{eq:vertexPhi}
\end{align}
\end{subequations}
with $\tau^{\uparrow}=\begin{pmatrix}1 & 0\\
0 & 0
\end{pmatrix}$ and $\tau^{\downarrow}=\begin{pmatrix}0 & 0\\
0 & 1
\end{pmatrix}$ acting in orbital doublet space and Pauli matrices $\sigma^{a}$
acting in spin space. Note that each entry of the matrices above is
itself a $4\times4$ matrix. The vertices of the remaining channels
are listed in Appendix \ref{sec:Vertex-Parts} for convenience. The
effective interactions $U_{E_{Mi}}^{M}$ and $U_{E_{Mi}}^{\Phi}$
have contributions from interorbital and intraorbital on-site interactions,
and will not be discussed here. Previous results have established
that, for most of the parameter space, only the SDW state develops
spontaneously (see for instance Ref. \cite{FernandesChubukov2017}). 

We now introduced the bosonic fields $M_{E_{Mi}}^{\mu}$ and $\Phi_{E_{Mi}}^{\mu}$
via Hubbard-Stratonovich transformations of the quartic interaction
terms. This decoupling makes the Hamiltonian quadratic in the fermions,
which can now be integrated out, resulting in an effective free-energy
functional for the two bosonic fields:
\begin{multline}
F[M,\Phi]=F_{0}-\beta^{-1}\text{tr}\ln[1-\mathcal{G}V]+\\
+\sum_{i\mu q}\left[\frac{2}{U_{E_{Mi}}^{M}}|M_{E_{Mi}}^{\mu}(q)|^{2}+\frac{2}{U_{E_{Mi}}^{\Phi}}|\Phi_{E_{Mi}}^{\mu}(q)|^{2}\right]\label{eq:free-energy-functional}
\end{multline}
with $q=(i\nu_{n},\mathbf{q})$ denoting bosonic Matsubara frequency
$\nu_{n}=2n\pi T$ and momentum $\mathbf{q}$. The non-interacting
fermionic propagator, defined in the 12-dimensional enlarged spinor
space, is given by: 
\begin{equation}
\mathcal{G}^{-1}=i\omega_{n}\mathbb{I}-\left(\mathcal{H}_{0}+\mathcal{H}_{\mathrm{SOC}}\right)
\end{equation}
with $\omega_{n}=\left(2n+1\right)\pi T$ denoting a fermionic Matsubara
frequency. The interaction term $V$ is given in terms of the vertex
functions by
\begin{equation}
V=\sum_{i\mu}\left(\varUpsilon_{E_{Mi},\mu}^{M}M_{E_{Mi}}^{\mu}+\varUpsilon_{E_{Mi},\mu}^{\Phi}\Phi_{E_{Mi}}^{\mu}\right)\label{eq:V}
\end{equation}
The trace runs over spin and orbital indices as well as momenta and
frequencies of the electronic degrees of freedom. 

Since the electronic states at the $\Gamma$ and $X/Y$ points fully
decouple, even in the presence of spin-orbit coupling, we rewrite
the non-interacting propagator $\mathcal{G}$ in the more convenient
block-diagonal form:
\begin{equation}
\mathcal{G}=\left(\begin{array}{ccc}
G_{\Gamma\Gamma} & 0 & 0\\
0 & G_{XX} & G_{XY}\\
0 & G_{YX} & G_{YY}
\end{array}\right).
\end{equation}
Near the $\Gamma$ point, we have
\begin{equation}
G_{\Gamma\Gamma}=\left(G_{\Gamma,0}^{-1}-\Lambda{}_{\Gamma}^{\text{SOC}}\right)^{-1}\label{eq:GGamma}
\end{equation}
with:
\begin{equation}
G_{\Gamma,0}^{-1}=i\omega_{n}\mathbb{I}-\epsilon_{\mathbf{k}}^{\Gamma}
\end{equation}
and $\Lambda{}_{\Gamma}^{\text{SOC}}$ as defined in Eq. (\ref{eq:VSOC1}).
Note that the momentum and spin indices are dropped for the sake of
simplicity. 

At the $X$ and $Y$ points, the electronic states of both Fermi pockets
are coupled to each other by the spin-orbit interaction. Defining:
\begin{align}
G_{X,0}^{-1}  =i\omega_{n}\mathbb{I}-\epsilon_{\mathbf{k}}^{X}, 
\quad  
G_{Y,0}^{-1}  =i\omega_{n}\mathbb{I}-\epsilon_{\mathbf{k}}^{Y},
\end{align}
we have the block-diagonal matrix:
\begin{multline}
\begin{pmatrix}G_{XX} & G_{XY}\\
G_{YX} & G_{YY}
\end{pmatrix}=\\
\left[\begin{pmatrix}G_{X,0}^{-1} & 0\\
0 & G_{Y,0}^{-1}
\end{pmatrix}-\begin{pmatrix}0 & \Lambda_{M}^{\text{SOC}}\\
(\Lambda_{M}^{\text{SOC}})^{\dagger} & 0
\end{pmatrix}\right]^{-1}\label{eq:GM}
\end{multline}
with $\Lambda_{M}^{\text{SOC}}$ given in Eq. (\ref{eq:VSOC2}). 

We are now in position to expand the logarithm in Eq. (\ref{eq:free-energy-functional})
in powers of the bosonic fields using the identity $\ln(1-GV)=-\sum_{n=1}^{\infty}\frac{1}{2n}(GV)^{2n}$.
Focusing on the terms that are second order in the bosonic fields
$M_{E_{Mi}}^{\mu}$ and $\Phi_{E_{Mi}}^{\mu}$, we find the bilinear
coupling

\begin{equation}
\delta F=\sum_{i\mu q}g_{E_{Mi}}^{\mu}(\lambda)\,\Phi_{E_{Mi}}^{\mu}(-q)M_{E_{Mi}}^{\mu}(q)+\text{c.c.}
\end{equation}
with 
\begin{equation}
g_{E_{Mi}}^{\mu}(\lambda)=\frac{1}{2}\mbox{tr}\left[\mathcal{G}\varUpsilon_{E_{Mi},\mu}^{\Phi}\mathcal{G}\varUpsilon_{E_{Mi},\mu}^{M}\right],\label{eq:g}
\end{equation}
where the trace runs over orbital doublet and spin space indices,
as well as momentum and frequency. It is clear why this term vanishes
in the absence of SOC: the SDW vertex $\varUpsilon_{E_{Mi},\mu}^{M}$
has a Pauli matrix, whereas the OLC vertex $\varUpsilon_{E_{Mi},\mu}^{\Phi}$
does not. If there is no SOC, then the non-interacting Green's functions
$\mathcal{G}$ are also spin independent, implying that the trace
over spin indices gives zero.

To better highlight the effects of the SOC, we perform an expansion
in powers of $\lambda$. A graphical representation of the microscopic
process generating $g_{E_{Mi}}^{\mu}$ in terms of Feynman diagrams
of three of the six introduced orders is depicted in Fig. \ref{fig:microscopical-coupling-M-phi}.
Note that for a coupling between static and spatially uniform orders,
which is of interest in this work, the external momentum and frequency
in the evaluation of Eq. (\ref{eq:g}) are set to zero. The calculation
of the coupling constants is now straightforward. For $E_{M1}^{-}$,
where $m_{\mathbf{Q}_{1}}^{x}$ couples to $\phi_{\mathbf{Q}_{2}}^{\left(xz,xy\right)}$,
we obtain:
\begin{widetext}
\begin{eqnarray}
g_{E_{M1}}^{-}(\lambda)&=&\frac{1}{2}\text{tr}\left[G_{XY}\left(i\tau^{\downarrow}\sigma^{0}\right)G_{\Gamma}\left(\tau^{\uparrow}\sigma^{x}\right)-G_{\Gamma}\left(i\tau^{\downarrow}\sigma^{0}\right)G_{YX}\left(\tau^{\uparrow}\sigma^{x}\right)\right]
\notag \\&\approx&\frac{1}{2}\text{tr}\left[G_{X,0}\Lambda_{M}^{\text{SOC}}G_{Y,0}\left(i\tau^{\downarrow}\sigma^{0}\right)G_{\Gamma,0}\left(\tau^{\uparrow}\sigma^{x}\right)-G_{\Gamma}\left(i\tau^{\downarrow}\sigma^{0}\right)G_{Y,0}\left(\Lambda_{M}^{\text{SOC}}\right)^{\dagger}G_{X,0}\left(\tau^{\uparrow}\sigma^{x}\right)\right] \notag \\&=&-\lambda\int_{k}\left[G_{X,0}\right]_{11}\left[G_{Y,0}\right]_{22}\left[G_{\Gamma,0}\right]_{21} \label{eq:g-EM1}
\end{eqnarray}
\end{widetext}
where we approximated $ G_{XY}\approx G_{X,0}\Lambda_{M}^{\text{SOC}}G_{Y,0} $
and $ G_{YX}\approx G_{Y,0}\left(\Lambda_{M}^{\text{SOC}}\right)^{\dagger}G_{X,0} $.
In the last line we performed the trace over orbital doublet and spin
space and left the remaining integration over momentum and frequency
symbolized by$\int_{k}=T\sum_{\omega_{n}}\int\frac{d\mathbf{k}}{\left(2\pi\right)^{2}}$. 

The evaluations of the other coefficients are analogous. For $E_{M2}^{-}$ we find:
\begin{equation}
g_{E_{M2}}^{-}(\lambda)\approx \lambda \int_{k}\left[G_{X,0}\right]_{12}\left[G_{Y,0}\right]_{12}\left[G_{\Gamma,0}\right]_{11}, \label{eq:g-EM2}
\end{equation}
whereas for $E_{M3}^{+}$
\begin{multline}
g_{E_{M3}}^{+}(\lambda) \\ \approx
\lambda \int_{k}\left[G_{X,0}\right]_{11}\left(\left[G_{\Gamma,0}\right]_{21}\left[G_{\Gamma,0}\right]_{21}-\left[G_{\Gamma,0}\right]_{11}\left[G_{\Gamma,0}\right]_{22}\right). \label{eq:g-EM3}
\end{multline}
In all cases, we find that $g_{E_{Mi}}^{\mu}(\lambda=0)=0$, as expected.
The bilinear coupling in the free-energy ensures that $\Phi_{E_{Mi}}^{\mu}$
becomes non-zero once $M_{E_{Mi}}^{\mu}$ orders. Indeed, the full
free-energy expansion becomes (assuming real fields):
\begin{align}
F[M,\Phi] & =g(\lambda)\Phi M\nonumber \\
 & +\frac{r_{M}}{2}M^{2}+\frac{r_{\Phi}}{2}\Phi^{2}+\frac{u_{M}}{4}M^{4} \label{aux_F_M},
\end{align}
where we dropped the indices $E_{Mi},\mu$ for simplicity of notation.
The terms $r_{M}$ and $r_{\Phi}$ trigger the onset of long-range
order in the usual manner, and $u_{M}>0$. Because $\Phi$ does not
usually order on its own, $r_{\Phi}$ remains positive and we can
keep the free-energy expansion to quadratic order in $\Phi$. Minimization
of the free-energy gives:
\begin{equation}
\Phi=-\frac{g(\lambda)}{r_{\Phi}}\, M\label{eq:phi2}
\end{equation}
showing that $M\neq0$ gives $\Phi\neq0$. Note also that, once $\Phi$
is integrated out, the magnetic free-energy changes to:
\begin{equation}
F[M]=\left(\frac{r_{M}}{2}-\frac{g^{2}(\lambda)}{r_{\Phi}}\right)M^{2}+\frac{u_{M}}{4}M^{4}
\end{equation}
which demonstrates that the SDW transition temperature is enhanced.

To give a rough estimate of the magnitude of the OLC order parameter induced by the SDW order parameter, Eq. (\ref{eq:phi2}), we consider how the relevant energy scales of the problem enter in the combination $g(\lambda)/r_{\Phi}$ that determines the ratio $\Phi/M$. Although numerical value may be obtained by fitting the microscopic parameters such as the electronic dispersion and interaction parameters to experiments or first-principle calculations, this is not the scope of this work.  To estimate the microscopic coupling $ g $, we note that it is generically given by the SOC times a product of three fermionic Green's functions, see Eqs. (\ref{eq:g-EM1})--(\ref{eq:g-EM3}) of type
\begin{align}
g(\lambda) = \lambda T \sum_{\omega_{n}}\int_{\textbf{k}} G_{i,0} G_{j,0} G_{k,0} . 
\end{align}
Using the general form $ G^{-1}_{i} = i(2n+1)\pi T-\epsilon_{i,\textbf{k}} $ and changing $ \int \frac{d\textbf{k}}{(2\pi)^{2}} = \rho_{F}\int d\epsilon $, we obtain   
\begin{align}
g(\lambda) \approx A \frac{\lambda}{T \epsilon_{F}},
\end{align}
where the numerical factor $ A $ depends on the details of band structure. Here, the density of states at the Fermi level was approximated by the inverse Fermi energy $ \rho_{F} \sim \epsilon_{F}^{-1} $. 
To estimate the parameter $r_{\Phi}$, we compare it to the quadratic Ginzburg-Landau coefficient of the SDW degrees of freedom, $r_M$, see Eq. (\ref{aux_F_M}). The difference between them is expected to be dominated, in the weak-coupling regime, by the difference in the effective electron-electron interaction strengths projected in the corresponding channels, denoted by $U_M$ (for SDW) and $U_\Phi$ (for OLC):
\begin{equation}
r_{\Phi}-r_{M} \approx \frac{1}{U^{\Phi}}-\frac{1}{U^{M} }.
\end{equation}
Near the SDW transition temperature $T_M$, $ r_{M}\approx 0 $. Thus, the estimate for the ratio between the OLC and SDW order parameters is:
\begin{equation}
	\frac{g(\lambda)}{r_{\Phi}} \approx A \frac{\lambda}{T_{M}}\frac{U^{\Phi}}{\epsilon_{F}}\frac{U^{M}}{U^{M}-U^{\Phi}}. 
\end{equation}
Therefore, the ratio $\Phi/M$ depends on four factors: There is a dimensionless factor $A $ depending on the details of the band structure. 
The second factor is the ratio between the SOC energy scale and SDW transition temperature. 
A rough estimate for the iron-based systems with $ \lambda \approx 10 \text{meV} $ and $ T_{M} \approx  100\text{K} $ yields $ \frac{\lambda}{T_{M}} \approx 1$. 
The third factor is the ratio between the OLC interaction strength and Fermi energy, which we expect to be small in a weak-coupling expansion $ \frac{U^{\Phi}}{\epsilon_{F}} \ll 1$. 
The fourth factor depends on the difference between the interaction strengths in the OLC and SDW channels: For $ U^{\Phi} \ll U^{M} $, $ \frac{U^{M}}{U^{M}-U^{\Phi}} \approx 1 $ and the induced OLC is expected to be small $ \Phi \ll M $. 
However, if the 
system is sufficiently close to an OLC instability $ U_{\Phi} \lessapprox U_{M} $, the induced OLC can be comparable with the SDW, $ \Phi \sim M $. 

\section{Experimental manifestations of orbital loop-current order \label{sec:Experimental-consequences}}

The previous sections show that OLC is induced by SDW.
An important issue is about the impact of OLC to the physics of the
iron-based superconductors. In this section, we discuss two direct
experimental manifestations of OLC that can in principle be detected
with appropriate probes. Detection of these effects and of their amplitudes
will allow one to assess the significance of these degrees of freedom
to the properties of these systems.

\begin{figure*}
\subfloat[\label{fig:Loop-current-order a}Loop current order $E_{M1}^{-}:\,\phi_{\mathbf{Q}_{2}}^{\left(xz,xy\right)}$]{\includegraphics[width=0.25\paperwidth]{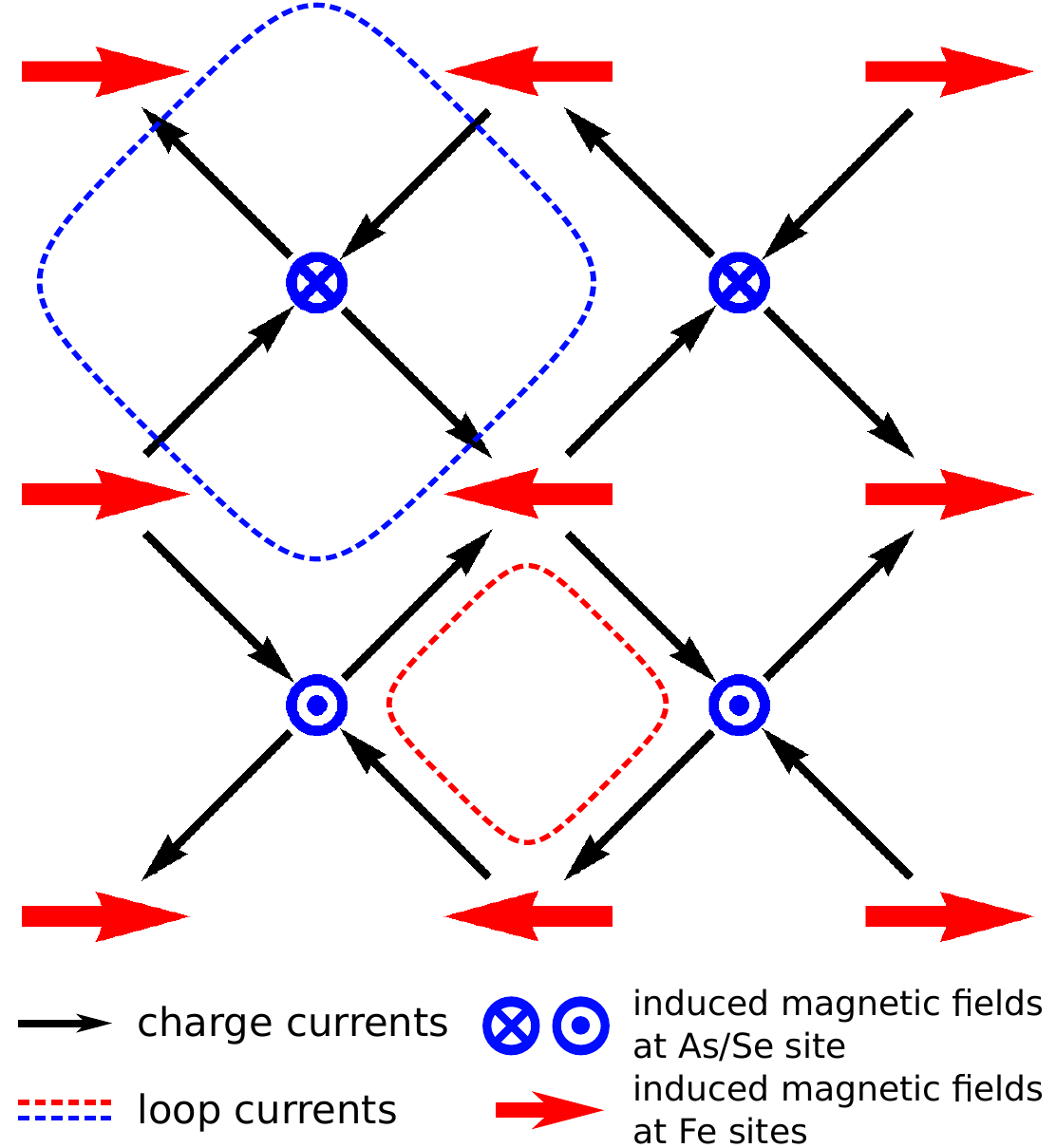}

}\hfill{}\subfloat[\label{fig:Loop-current-order b}Loop current order $E_{M2}^{-}:\,\phi_{\mathbf{Q}_{2}}^{\left(yz,xy\right)}$]{\includegraphics[width=0.25\paperwidth]{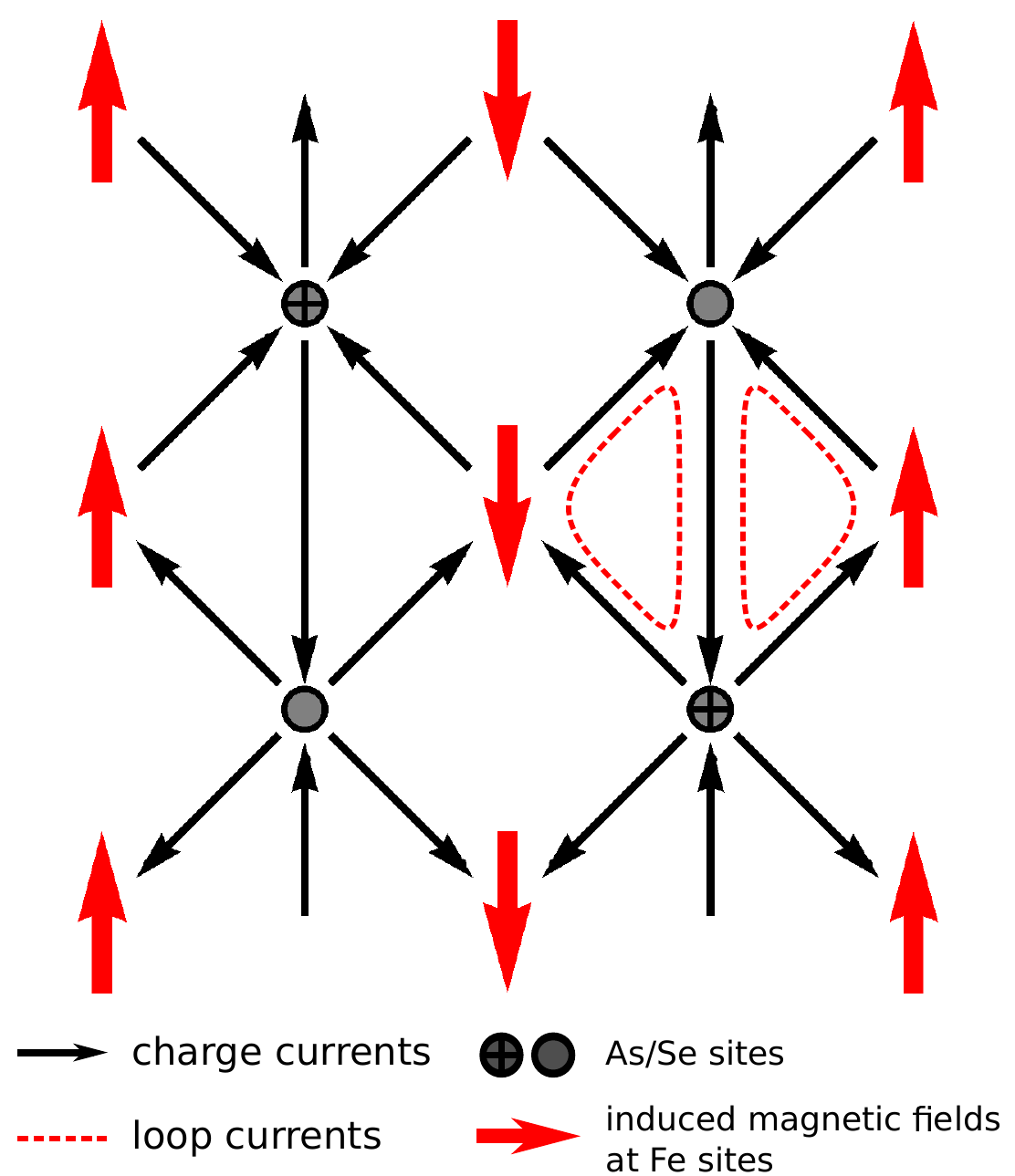}}\hfill{}\subfloat[\label{fig:Loop-current-order c}Loop current order $E_{M3}^{+}:\,\phi_{\mathbf{Q}_{1}}^{\left(xz,yz\right)}$]{\includegraphics[width=0.25\paperwidth]{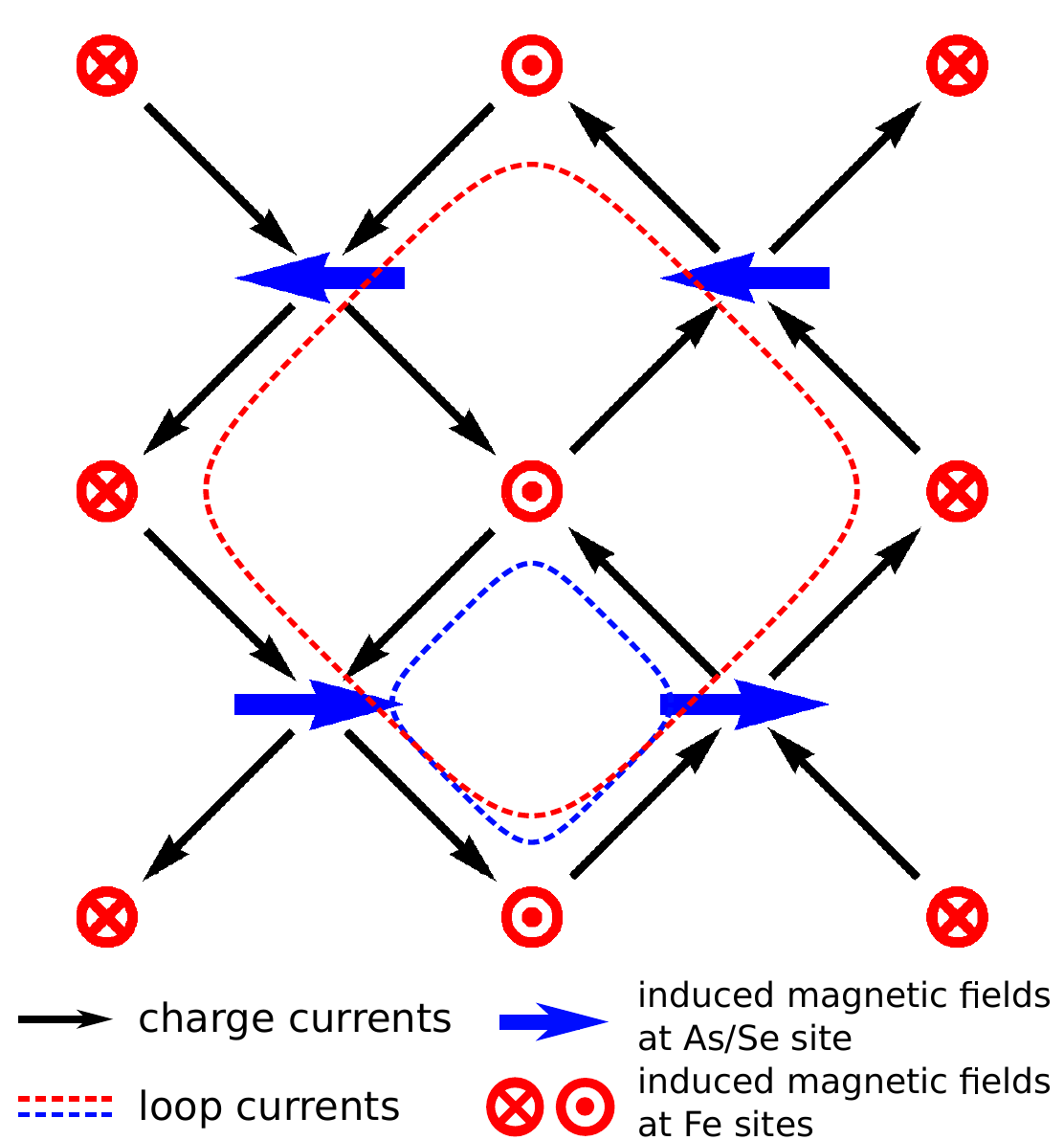}
}

\protect\caption{\label{fig:Orbital-current-textures-1} Magnetic moments induced by
different loop current configurations. Note that, in contrast to Fig.
\ref{fig:Orbital-current-textures}, here we show explicitly the charge
current path going through the As/Se site (black arrows). The corresponding
closed loops that induce magnetic moments at the Fe sites and at the
As/Se sites are shown by dashed red and blue curves, respectively.
Accordingly, the induced magnetic moments at the Fe and As/Se sites
are represented by red and blue arrows, respectively. Recall that
the As/Se lattice sites are positioned above and below the horizontal
Fe plane in a staggering pattern. This feature of the crystal lattice
is what gives rise to in-plane induced magnetic moments at the Fe/Se
sites. }
\end{figure*}

\subsection{Induced Magnetic Moments on the Pnictogen/Chalcogen Sites}

\begin{table}[b]
	\begin{ruledtabular}
	\protect\caption{\label{tab:Character-table-of}Transformation behavior of the irreducible
		representations $E_{Mi}$ for the point symmetry operations discussed
		in the main text, $\sigma_{x}^{\text{As/Se}}$, $\sigma_{y}^{\text{As/Se}}$,
		$\sigma_{z}\sigma_{x}^{\text{Fe}}$, and $\sigma_{z}\sigma_{y}^{\text{Fe}}$. }
\begin{tabular}{ccccc}
irrep & $\sigma_{x}^{\text{As/Se}}$ & $\sigma_{y}^{\text{As/Se}}$ & $\sigma_{z}\sigma_{x}^{\text{Fe}}$ & $\sigma_{z}\sigma_{y}^{\text{Fe}}$
\tabularnewline
\hline 
\hline 
\noalign{\vskip\doublerulesep}
\noalign{\vskip\doublerulesep}

$E_{M1}$ & $\begin{bmatrix}-1 & 0\\
0 & -1
\end{bmatrix}$ & $\begin{bmatrix}-1 & 0\\
0 & -1
\end{bmatrix}$ & $\begin{bmatrix}1 & 0\\
0 & -1
\end{bmatrix}$  & $\begin{bmatrix}-1 & 0\\
0 & 1
\end{bmatrix}$
\tabularnewline
\noalign{\vskip\doublerulesep}
\noalign{\vskip\doublerulesep}
\hline 
\noalign{\vskip\doublerulesep}
\noalign{\vskip\doublerulesep}
$E_{M2}$ & $\begin{bmatrix}1 & 0\\
0 & 1
\end{bmatrix}$ & $\begin{bmatrix}1 & 0\\
0 & 1
\end{bmatrix}$ & $\begin{bmatrix}-1 & 0\\
0 & 1
\end{bmatrix}$ & $\begin{bmatrix}1 & 0\\
0 & -1
\end{bmatrix}$\tabularnewline
\noalign{\vskip\doublerulesep}
\noalign{\vskip\doublerulesep}
\hline 
\noalign{\vskip\doublerulesep}
\noalign{\vskip\doublerulesep}
$E_{M3}$ & $\begin{bmatrix}1 & 0\\
0 & -1
\end{bmatrix}$ & $\begin{bmatrix}-1 & 0\\
0 & 1
\end{bmatrix}$ & $\begin{bmatrix}-1 & 0\\
0 & -1
\end{bmatrix}$ & $\begin{bmatrix}-1 & 0\\
0 & -1
\end{bmatrix}$\tabularnewline
\noalign{\vskip\doublerulesep}
\noalign{\vskip\doublerulesep}
\hline 
\noalign{\vskip\doublerulesep}
\noalign{\vskip\doublerulesep}
$E_{M4}$ & $\begin{bmatrix}-1 & 0\\
0 & 1
\end{bmatrix}$ & $\begin{bmatrix}1 & 0\\
0 & -1
\end{bmatrix}$ & $\begin{bmatrix}1 & 0\\
0 & 1
\end{bmatrix}$ & $\begin{bmatrix}1 & 0\\
0 & 1
\end{bmatrix}$\tabularnewline
\end{tabular}
	\end{ruledtabular}
\end{table}

\begin{table}[b]
	\protect\caption{\label{tab:induced-mag-fields} Direction of the loop-current induced
		magnetic moments $\mathbf{B}_{\text{ind}}$ at Fe and pnictogen/chalcogen
		lattice sites as well as the spin magnetic moments $\boldsymbol{\mu}$
		at the Fe lattice sites. }
\begin{ruledtabular}
\begin{tabular}{lccc}
\multirow{2}{*}{%
\begin{minipage}[t]{0.2\columnwidth}%
irrep%
\end{minipage}} & \multicolumn{2}{c}{%
\begin{minipage}[t]{0.4\columnwidth}%
Loop current induced moments $\mathbf{B}_{\text{ind}}$%
\end{minipage}} & %
\begin{minipage}[t]{0.22\columnwidth}%
Spin moments $\boldsymbol{\mu}$%
\end{minipage}\tabularnewline[0.1cm]
\cline{2-4} 
\noalign{\vskip\doublerulesep}
 & at Fe & at As/Se & at Fe\tabularnewline[0.1cm]
\hline 
\noalign{\vskip\doublerulesep}
\hline 
\noalign{\vskip\doublerulesep}
\noalign{\vskip\doublerulesep}
$\begin{pmatrix}E_{M1}^{+}\\
E_{M1}^{-}
\end{pmatrix}$ & $\begin{pmatrix}\mathbf{\hat{B}}_{y}^{\text{ }}\\
\mathbf{\hat{B}}_{x}^{\text{ }}
\end{pmatrix}_{\text{Fe}}$ & $\begin{pmatrix}\mathbf{\hat{B}}_{z}\\
\mathbf{\hat{B}}_{z}
\end{pmatrix}_{\text{As}}$ & $\begin{pmatrix}\mathbf{\hat{\boldsymbol{\mu}}}_{y}\\
\mathbf{\hat{\boldsymbol{\mu}}}_{x}
\end{pmatrix}_{\text{Fe}}$\tabularnewline[0.4cm]
\hline 
\noalign{\vskip\doublerulesep}
$\begin{pmatrix}E_{M2}^{+}\\
E_{M2}^{-}
\end{pmatrix}$ & $\begin{pmatrix}\mathbf{\hat{B}}_{x}\\
\mathbf{\hat{B}}_{y}
\end{pmatrix}_{\text{Fe}}$ & - & $\begin{pmatrix}\mathbf{\hat{\boldsymbol{\mu}}}_{x}\\
\mathbf{\hat{\boldsymbol{\mu}}}_{y}
\end{pmatrix}_{\text{Fe}}$\tabularnewline[0.4cm]
\hline 
\noalign{\vskip\doublerulesep}
$\begin{pmatrix}E_{M3}^{+}\\
E_{M3}^{-}
\end{pmatrix}$ & $\begin{pmatrix}\mathbf{\hat{B}}_{z}\\
\mathbf{\hat{B}}_{z}
\end{pmatrix}_{\text{Fe}}$ & $\begin{pmatrix}\mathbf{\hat{B}}_{x}\\
\mathbf{\hat{B}}_{y}
\end{pmatrix}_{\text{As}}$ & $\begin{pmatrix}\mathbf{\hat{\boldsymbol{\mu}}}_{z}\\
\mathbf{\hat{\boldsymbol{\mu}}}_{z}
\end{pmatrix}_{\text{Fe}}$
\end{tabular}
\end{ruledtabular}
\end{table}

An obvious manifestation of static loop currents is the generation of
magnetic fields, according to the Biot-Savart law of classical electrodynamics.
To determine the magnetic field distributions corresponding to each
loop current configuration, we employ symmetry-based arguments, i.e.
the induced magnetic fields must transform as the underlying loop
current pattern under the symmetry operations of the lattice. We follow a procedure similar to Ref.  \cite{Lederer2012}.

The point symmetry operations which will be used in the following
analysis are given by mirror reflections cutting Fe and As/Se sites
combined with horizontal reflections. At the As/Se lattice sites,
there are mirror planes whose normals point along the $x$ and $y$
directions, denoted respectively by $\sigma_{x}^{\text{As/Se}}$ and
$\sigma_{y}^{\text{As/Se}}$. At the Fe lattice sites, these mirror
reflections are combined with reflections with respect to the horizontal
Fe plane yielding the point symmetry operations $\sigma_{z}\sigma_{x}^{\text{Fe}}$
and $\sigma_{z}\sigma_{y}^{\text{Fe}}$. The transformation behavior
of the four distinct irreducible representations $E_{Mi}$ under these
point symmetry operations are listed in Table \ref{tab:Character-table-of}.
The latter can be verified using the current pattern representations
in Fig. \ref{fig:Orbital-current-textures}, and were also given in
Ref. \cite{Cvetkovic2013}. For the subsequent analysis, note that
whereas a current transforms as a vector, an induced magnetic field
transforms as a pseudo-vector, meaning that the normal component remains
invariant whereas in-plane components change sign under mirror reflections. 

To illustrate the logic of the symmetry-based argument for the existence
of induced magnetic moments introduced above, we discuss in detail
loop currents, which transform as the irreducible representation $E_{M1}$.
In the following, we denote the direction of induced magnetic moments
by $\hat{\mathbf{B}}_{\alpha}=\frac{\mathbf{B}_{\text{ind}}}{|\mathbf{B}_{\text{ind}}|}$
with $\alpha=x,y,z$ representing the particular directions along
the Fe-Fe bonds $(x,y)$ or pointing out of plane $(z)$. We are looking
for a doublet of directions of induced magnetic moments $(\hat{\mathbf{B}}_{\alpha},\hat{\mathbf{B}}_{\alpha'})^{\text{T}}$
that corresponds to loop currents of one irreducible representation.
We find at the Fe and As/Se lattice, respectively, 
\begin{subequations}
\begin{eqnarray}
\sigma_{z}\sigma_{x}^{\text{Fe}}\begin{pmatrix}\mathbf{\hat{B}}_{y}\\
\mathbf{\hat{B}}_{x}
\end{pmatrix}_{\text{Fe}} & = & \begin{pmatrix}\mathbf{\hat{B}}_{y}\\
-\mathbf{\hat{B}}_{x}
\end{pmatrix}_{\text{Fe}},\\
\sigma_{z}\sigma_{y}^{\text{Fe}}\begin{pmatrix}\mathbf{\hat{B}}_{y}\\
\mathbf{\hat{B}}_{x}
\end{pmatrix}_{\text{Fe}} & = & \begin{pmatrix}-\mathbf{\hat{B}}_{y}\\
\mathbf{\hat{B}}_{x}
\end{pmatrix}_{\text{Fe}},\\
\sigma_{x}^{\text{As/Se}}\begin{pmatrix}\mathbf{\hat{B}}_{z}\\
\mathbf{\hat{B}}_{z}
\end{pmatrix}_{\text{As}} & = & -\begin{pmatrix}\mathbf{\hat{B}}_{z}\\
\mathbf{\hat{B}}_{z}
\end{pmatrix}_{\text{As}},\\
\sigma_{y}^{\text{As/Se}}\begin{pmatrix}\mathbf{\hat{B}}_{z}\\
\mathbf{\hat{B}}_{z}
\end{pmatrix}_{\text{As}} & = & -\begin{pmatrix}\mathbf{\hat{B}}_{z}\\
\mathbf{\hat{B}}_{z}
\end{pmatrix}_{\text{As}}
\end{eqnarray}
\end{subequations}
which transform consistently as the $E_{M1}$ loop currents as indicated
in Table \ref{tab:Character-table-of}. We therefore conclude that
these are the directions of local magnetic moments at the Fe and As/Se
lattice sites which are induced by these particular orbital loop currents.
It is straightforward to verify that this is the only possible
distribution of moments that is allowed by symmetry. 

Our results for the directions of possible induced magnetic moments
for all types of loop-current orders are presented in Table \ref{tab:induced-mag-fields}.
Note that they agree with the As/Se magnetic moments discussed in
Ref. \cite{Cvetkovic2013} derived solely on symmetry considerations.
We find that at Fe lattice sites the induced magnetic fields are pointing
in the same direction as the magnetic moments of the corresponding
SDW orders. This is not surprising since the SDW and OLC orders share
the same symmetry properties. It can also be traced back to the physical
fact that the two orders couple via these moments. However, there
are also induced magnetic fields located at the pnictogen/chalcogenide
(As/Se) lattice sites which represent a unique signature of the distinct
loop-current patterns. 

To establish a link between the derived schematic current pattern
shown in Fig. \ref{fig:Orbital-current-textures} and the direction
of the induced magnetic fields, one has to include the actual current
flow via the As/Se lattice sites. This immediately gives rise to orbital
independent current pattern for $E_{M1}^{\mu}$ and $E_{M3}^{\mu}$
loop-current orders as depicted in Fig. \ref{fig:Loop-current-order a}
and \ref{fig:Loop-current-order c}, respectively. In case of $E_{M2}^{\mu}$
current order we encounter a problem: The net charge flow between
Fe lattice sites sums up to zero as seen in Fig. \ref{fig:Loop-current-order b}.
This can traced back to the fact that the $E_{M2}^{\mu}$ current
pattern transform evenly under point symmetry operations with respect
to the Fe lattice sites (see Table \ref{tab:Character-table-of}).
Thus, a refined analysis, which takes higher-order Fe-Fe hopping processes
into account as conducted in the previous section, would not yield
any solution and only an explicit consideration of electronic degrees
of freedom of pnictogen / chalcogenide atoms would remedy this problem.
Here, we will use the symmetry properties of the $E_{M2}^{\mu}$ irreducible
representation to derive an effective current pattern: Besides the
transformation behaviors under point symmetry operations, we require
loop currents that preserve local charge conservation and posses
ordering momentum $\mathbf{Q}_{j}$. The result is depicted in Fig.
\ref{fig:Loop-current-order b}. 

Besides the effective current flow pattern, the induced magnetic fields
in combination with one realization of the responsible loop currents
are visualized in Fig. \ref{fig:Orbital-current-textures-1}. The
out-of plane position and the alternating stacking of As/Se lattice
sites are crucial features and allow for in-plane magnetic fields.
They also provide a physical understanding of the fact that loop-current
orders, which are composed out of evenly and oddly transforming orbitals,
couple to magnetic orders with different pseudo-crystal momentum as
discussed in Sec. \ref{sec:Three-orbital-model}: The momentum of
the induced magnetic fields is shifted by $(\pi,\pi)$ due to the
alternate stacking of As/Se lattice sites. 

We now estimate the magnitude of the induced magnetic fields at the
As/Se sites. Given the area of the loop current and the magnitude
of the current, we can obtain the corresponding magnetic moment in
a straightforward way:
\begin{equation}
\mu_{\mathrm{OLC}}=A_{\perp}|\langle\hat{\mathbf{j}}\rangle|\sim\frac{A_{\perp}et}{\hbar}\,\Phi
\end{equation}
with $A_{\perp}$ being the area enclosed by the loop current $\langle\hat{\mathbf{j}}\rangle$
projected onto the direction of the magnetic moment. For magnetic
moments lying out of plane, $A_{\perp}\sim a_{\text{Fe}}^{2}$, while
for moments lying in-plane, $A_{\perp}\sim a_{\text{Fe}}a_{\text{Pn}}$
where $a_{\text{Fe}}$ is the Fe-Fe lattice spacing and $a_{\text{Pn}}$
denotes the out-of-plane position of the pnictogen/chalcogen lattice
sites. In the last step, we used the relationship between the average
current and the OLC order parameter $\left|\langle\hat{\mathbf{j}}\rangle\right|\sim\frac{et}{\hbar}\,\Phi$
as discussed in the previous section, with $t$ being some dominant
kinetic energy scale of the electronic system. With the expression
for the magnetic moments we approximate the induced magnetic field
strength to 
\begin{equation}
B_{\text{ind}}=\frac{2\mu_{\mathrm{OLC}}}{cr^{3}}\sim\frac{A_{\perp}et}{c\hbar a_{\text{Fe}}^{3}}\,\Phi.
\end{equation}

For a quantitative statement we compare the system of the Fe-based
superconductors to that of the cuprates superconductors. In those
systems the induced magnetic field strength due to loop currents was
estimated to be $B_{\text{Cu}}\sim100-1000\mbox{ G}$ \cite{Lederer2012}.
In comparison, the typical energy scales in the Fe-based systems are
one order of magnitude smaller $\frac{t_{\text{Fe}}}{t_{\text{Cu}}}\sim0.1$.
Thus, we expect the induced magnetic field strength to be of order
$B_{\text{ind}}\sim10-100\mbox{ G}$. 

An important issue is about how to detect these induced fields $\mathbf{B}_{\text{ind}}$
on the As/Se sites. NMR is a natural candidate to detect local fields,
particularly since it can be done on the As and Se nuclei. However,
the internal field measured by NMR also has contributions from the
hyperfine field $\mathbf{B}_{\mathrm{hf}}$ arising due to the coupling
between the As/Se nuclei spins and the electronic Fe spins. Not surprisingly,
the directions of these hyperfine fields $\mathbf{B}_{\mathrm{hf}}$
are exactly the same as the directions of the fields induced by the
loop currents, $\mathbf{B}_{\text{ind}}$ (see, for instance, Ref. \cite{KissikovSarkarLawsonEtAl2017}).
Thus, unambiguously distinguishing the fields generated by the loop
currents from the fields generated by the hyperfine nuclei-spins coupling
is a difficult task. Instead of NMR, x-ray scattering may provide
more unambiguous signatures of the fields induced by the loop currents.
First, because the loop currents are extended objects, they will likely
produce form factors that are rather distinct from those arising from
point-like moments. Second, the x-rays can be tuned to the As absorption
edge, thus providing information about the electronic magnetization
of the As atoms only. Recently, x-ray absorption measurements tuned
to the As $K$-edge were performed to unveil ferromagnetic order in
the As sites of materials closely related to the 122 iron pnictides
\cite{UelandPandeyLeeEtAl2015}.
\vspace{-.15cm}

\subsection{Orbitally-Selective Band Splittings}
\vspace{-.1cm}
\begin{figure}
	\includegraphics[width=\columnwidth]{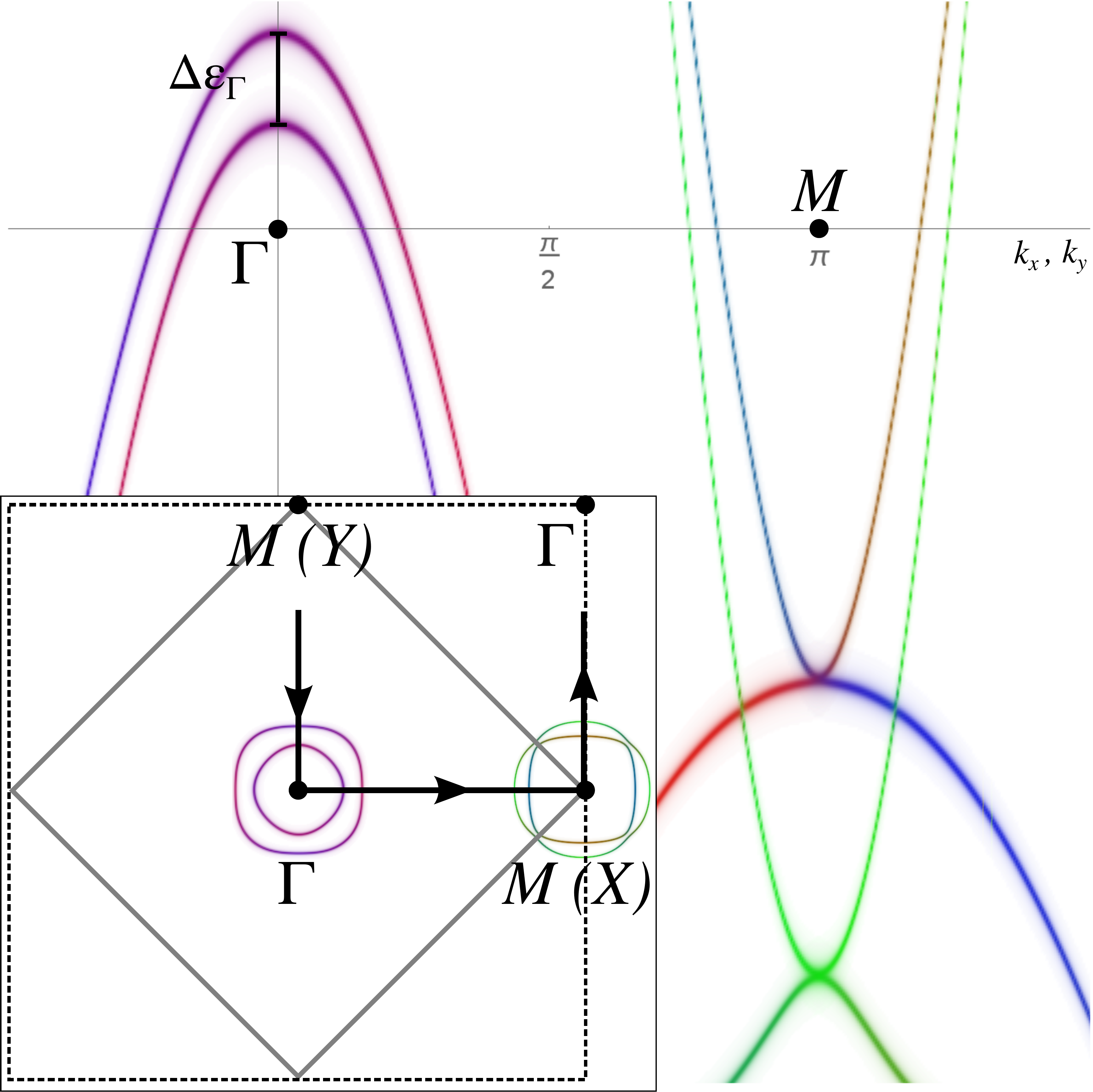}
	\caption{\label{fig:band-spectrum-tetragonal-phase}Electronic band dispersion in the tetragonal phase in the presence of spin-orbit
		coupling. The latter gives rise to a strong mixing of electronic states
		and a band splitting of $|\lambda|$ at the $\Gamma$ point, but does not affect the doublets at the $M$ point. The orbital content
		of the electronic states is represented by different colors: Blue and
		red correspond to $xz$ and $yz$ orbitals, respectively, while green,
		to the $xy$ orbital. 
		Here, the electronic states are represented
		in the two Fe Brillouin zone, where the $X$ and $Y$ points are both folded onto
		the $M$ point. The dispersions plotted correspond to the cuts indicated in the inset. The Fermi surface is also shown in the inset.
		}
\end{figure}

\begin{figure*}
\subfloat[SDW order for vanishing SDW-OLC coupling
]{\includegraphics[width=0.20\paperwidth]{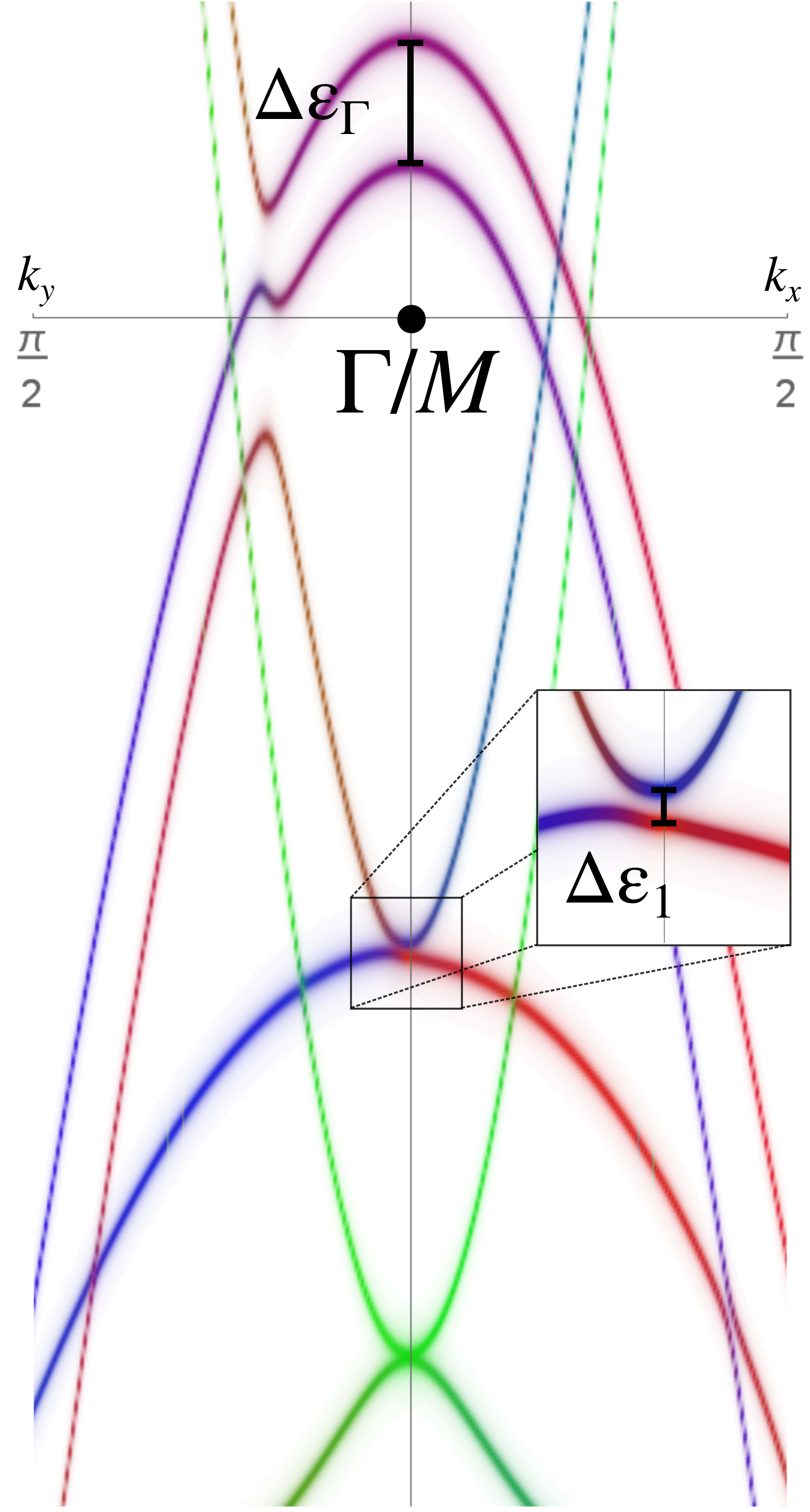}}
\hfill{}
\subfloat[
Coupled $ E_{M1}^{-} $ SDW and OLC orders
]{\includegraphics[width=0.20\paperwidth]{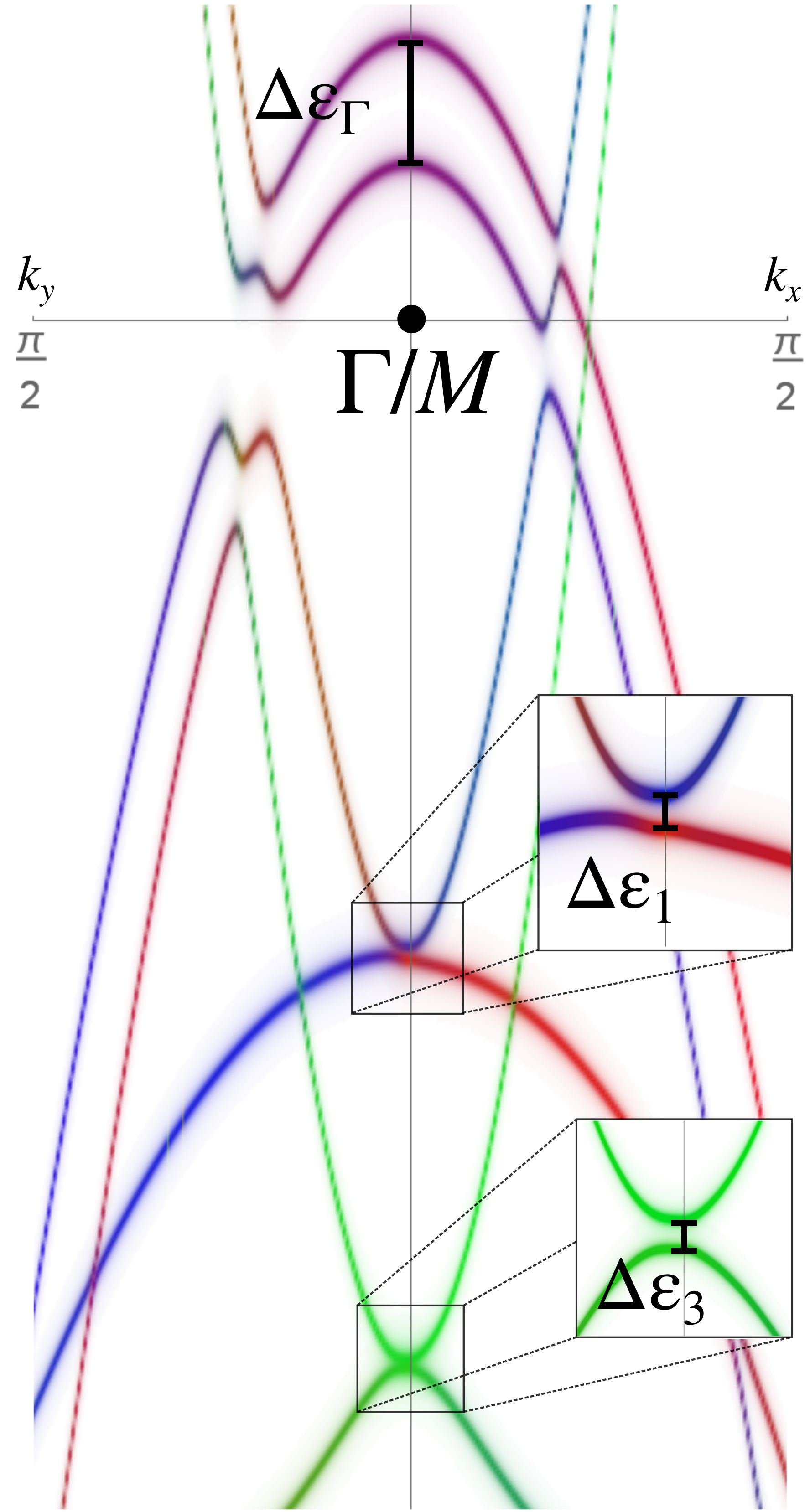}}
\hfill{}
\subfloat[
Coupled $ E_{M2}^{-} $ SDW and OLC orders
]{\includegraphics[width=0.20\paperwidth]{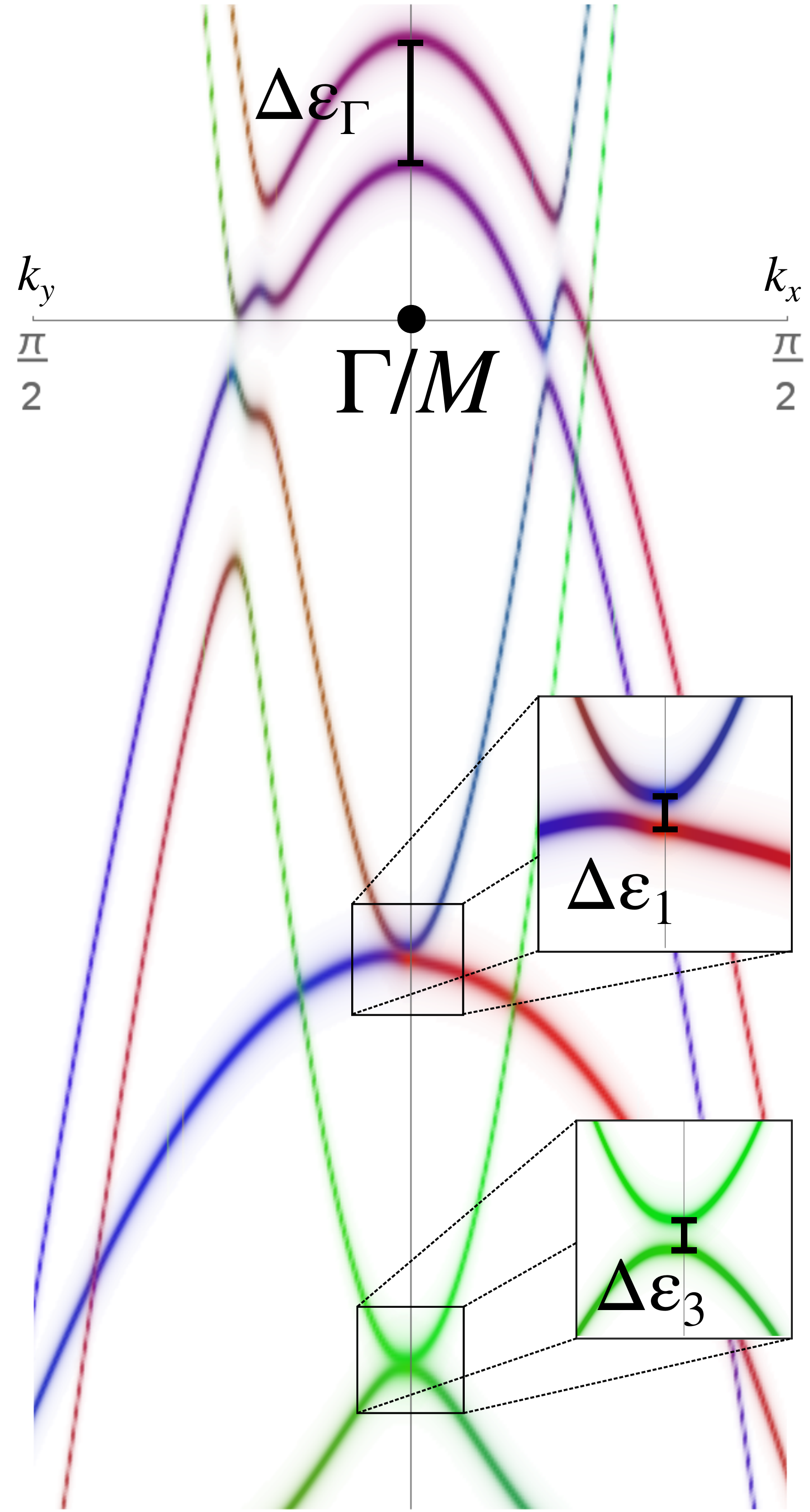}}
\hfill{}
\subfloat[
Coupled $E_{M3}^{+}$ SDW and OLC orders
]{\includegraphics[width=0.20\paperwidth]{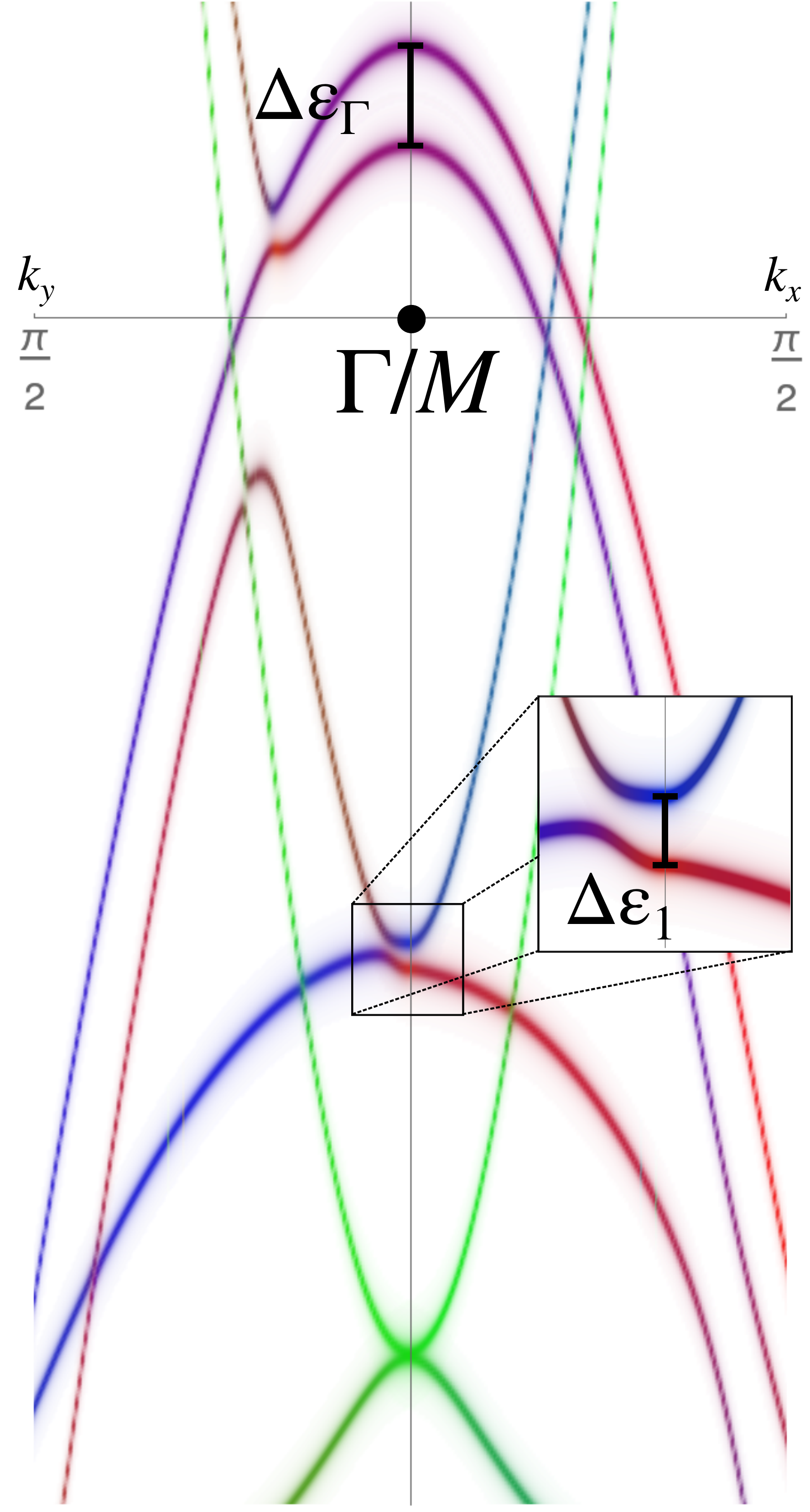}}
\protect
\caption{\label{fig:Effective-band-spectrum} Reconstructed electronic dispersions in the SDW phase for  (a) vanishing and (b)-(d) non-vanishing coupling to the OLC order parameter. In the absence of SDW-OLC coupling, the reconstructed electronic structures associated with each SDW irreducible representation are the same (a). However, for finite SDW-OLC coupling, the resulting band dispersion is different for the (b),(c) $E_{M1,M2}$ and (d) $E_{M3}$ representations. The dispersions plotted correspond to the cuts displayed in the inset of Fig. \ref{fig:band-spectrum-tetragonal-phase}. The insets zoom in on the splittings of the energy doublets at the $M$ point associated with each case. The electronic dispersion parameters are taken from Ref. \cite{Cvetkovic2013} (see also Appendix \ref{sec:Orbital-dispersion-relations}). To highlight the main features caused by each electronic reconstruction, the spin-orbit coupling is set to $ \lambda = 80 \text{ meV} $, the SDW order parameter to $ M=65\text{ meV} $, and the induced OLC order parameter, to $ \Phi = M $. 
} 
\end{figure*}
\begin{figure*}
	\subfloat[SDW order for vanishing SDW-OLC coupling 
	]
	{\includegraphics[width=0.20\paperwidth]{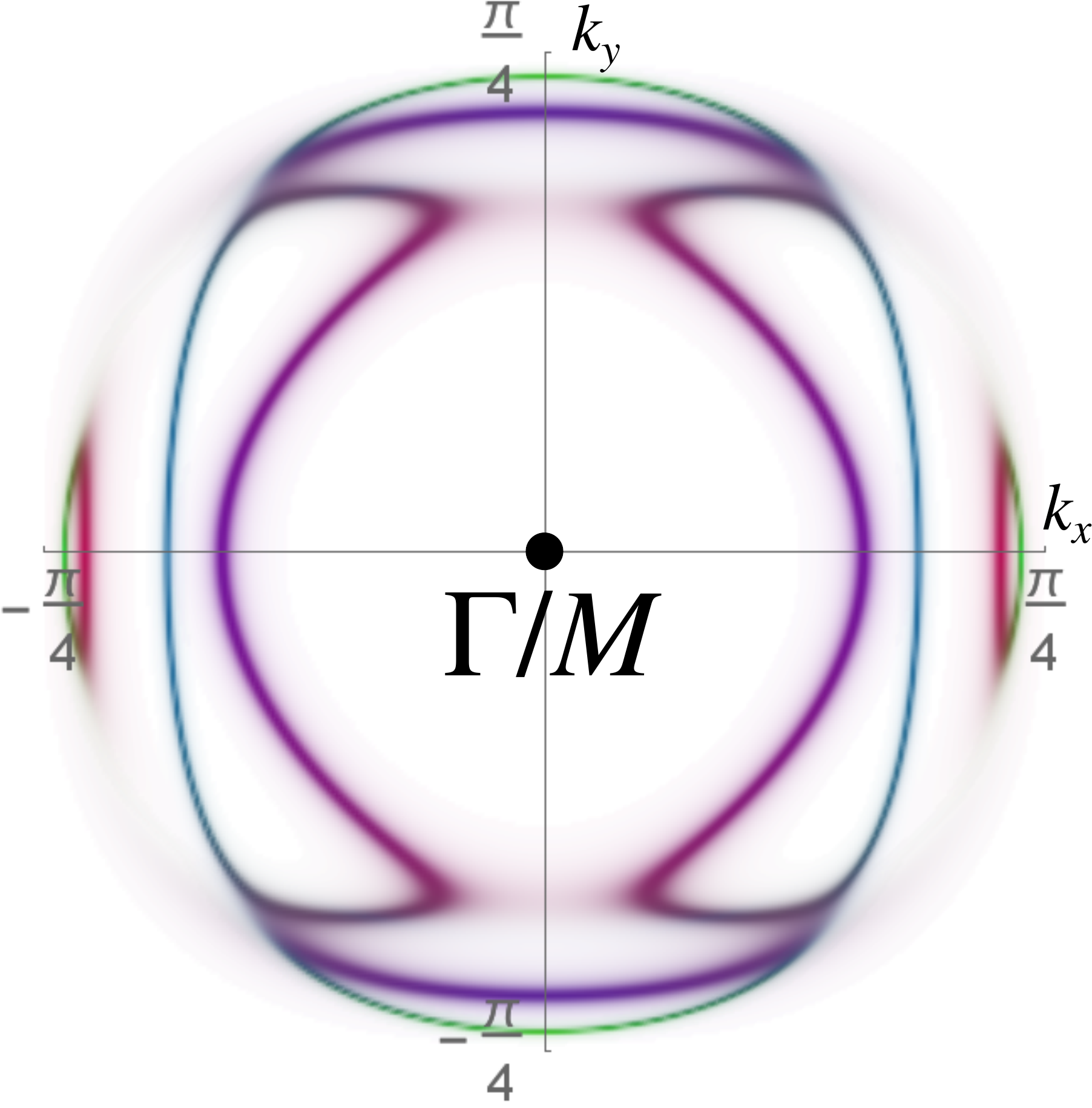}}
	\hfill{}
	\subfloat[Coupled $ E_{M1}^{-} $ SDW and OLC orders 
	]{\includegraphics[width=0.20\paperwidth]{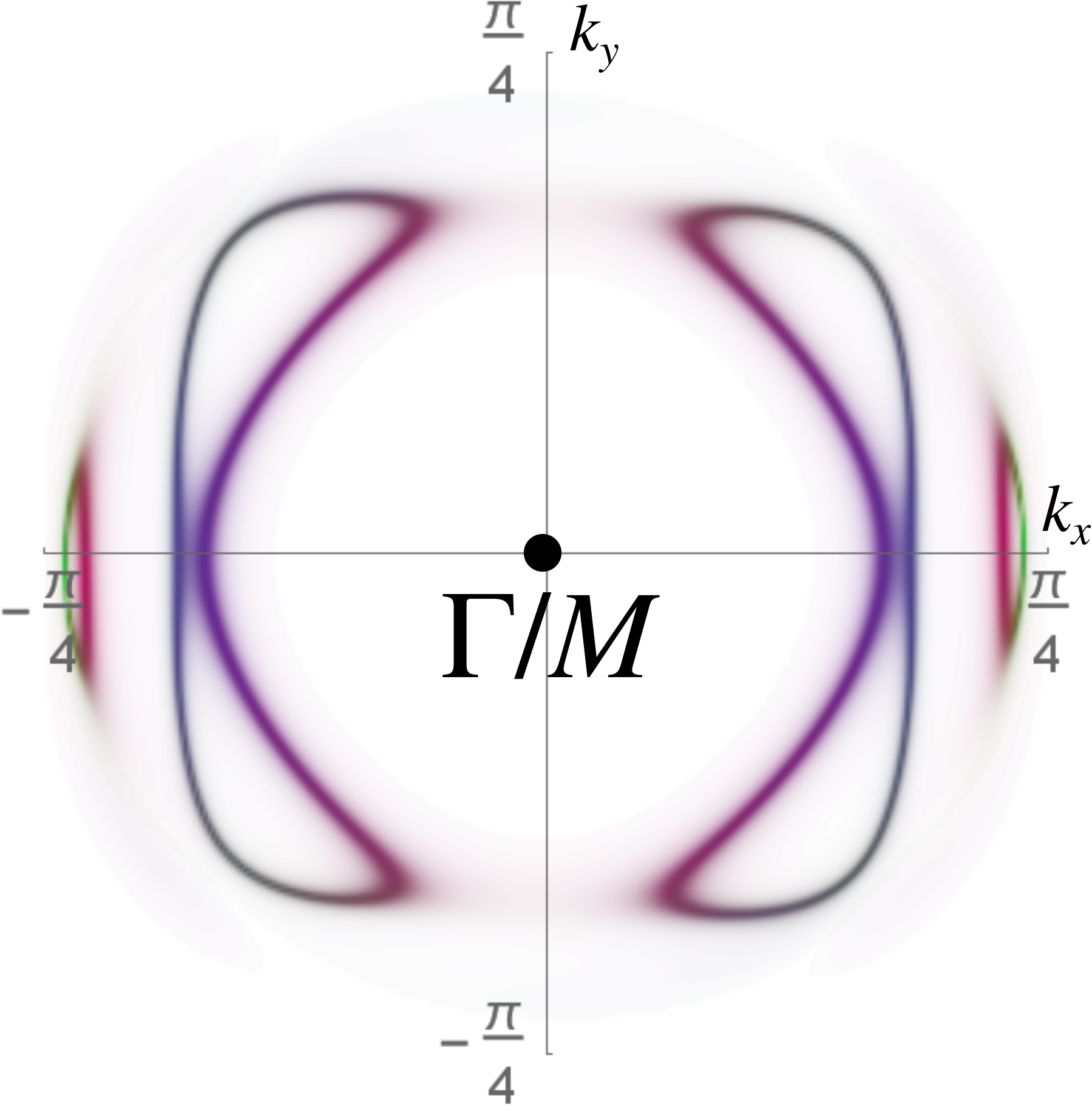}}
	\hfill{}
	\subfloat[Coupled $ E_{M2}^{-} $ SDW and OLC orders 
	]{\includegraphics[width=0.20\paperwidth]{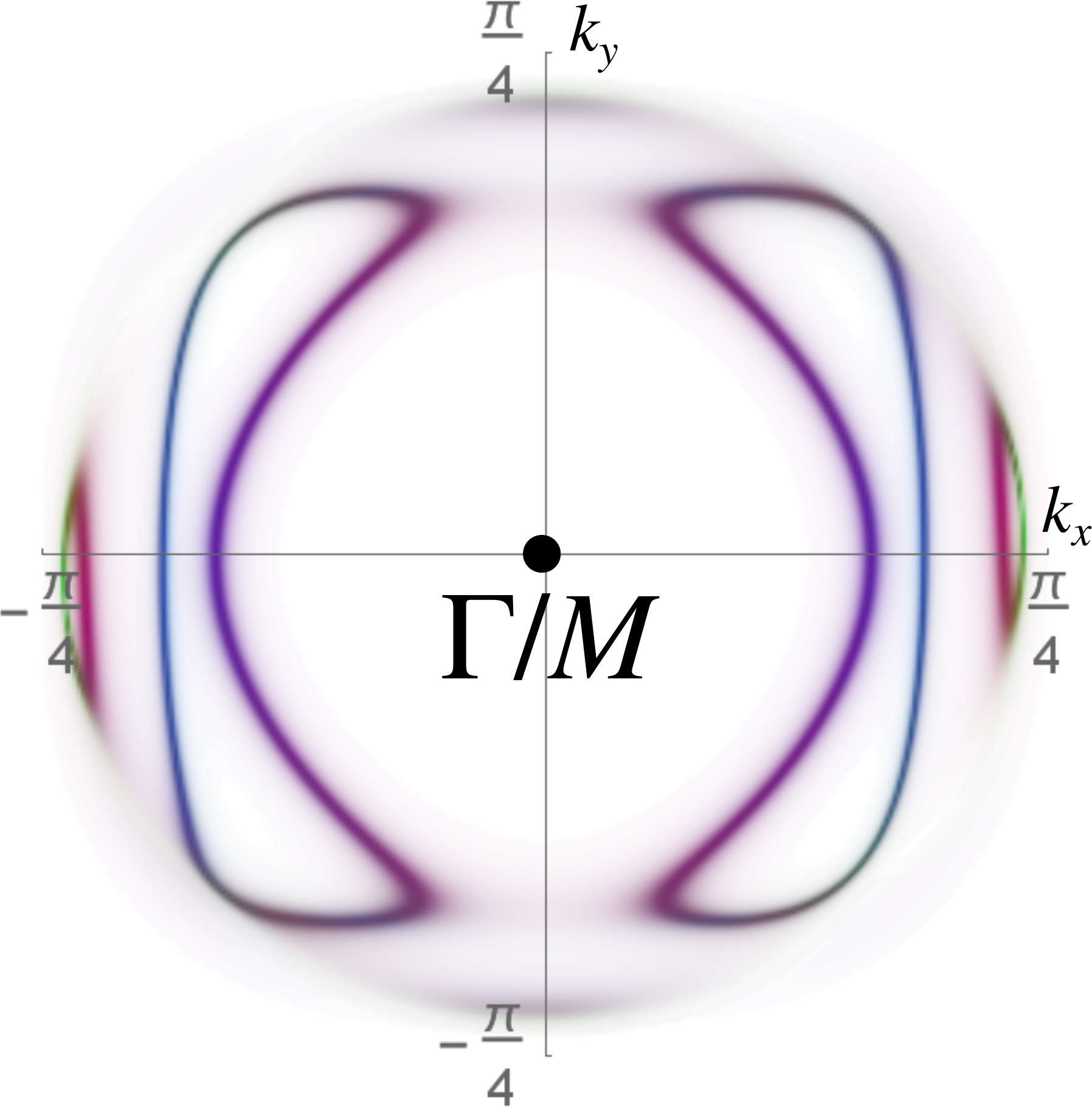}}
	\hfill{}
	\subfloat[Coupled $ E_{M3}^{+} $ SDW and OLC orders
	]{\includegraphics[width=0.20\paperwidth]{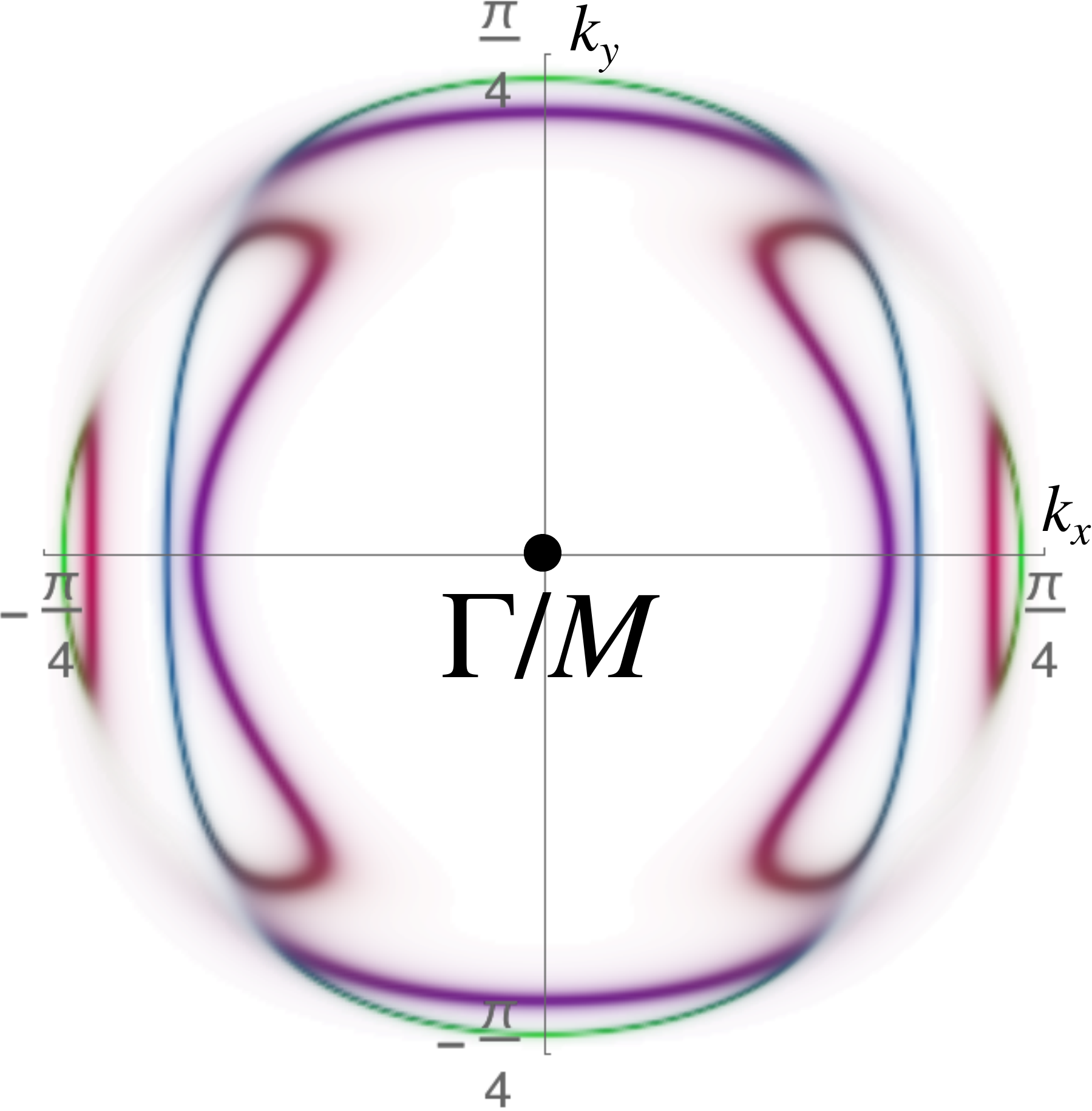}}
	\protect
	\caption{\label{fig:Effective-fermi-surfaces} Reconstructed Fermi surfaces corresponding to the band dispersions shown in Fig. \ref{fig:Effective-band-spectrum}. (a) corresponds to the SDW phase with vanishing SDW-OLC coupling, and is the same for all three irreducible representations of the SDW order parameter. (b)--(d) corresponds to a non-vanishing SDW-OLC coupling. Note that, for the case of the $E_{M1,M2}$ representations, there is additional Fermi surface gapping as compared to the case of the $E_{M3}$ representation and the case of vanishing SDW-OLC coupling. The parameters are the same as in Fig. \ref{fig:Effective-band-spectrum}.
}
\vspace{2cm}
\end{figure*}

A direct manifestation of OLC order is on the electronic spectrum
near the high-symmetry points, which can be probed experimentally
using ARPES \cite{ColdeaWatson2017,FedorovYareskoKimEtAl2016}. To
discuss this effect, it is convenient to refer to the coordinate system
corresponding to the 2-Fe Brillouin zone. Recall that both the $X$
and $Y$ points of the 1-Fe Brillouin zone are folded onto the $M$
point of the 2-Fe Brillouin zone (see inset of Fig. \ref{fig:band-spectrum-tetragonal-phase}).
Since the $M$ point only admits two-dimensional irreducible representations,
all electronic states at the $M$ point are doubly degenerate (on
top of the standard Kramers degeneracy). This degeneracy holds even
in the presence of spin-orbit coupling $\lambda$. In particular,
the two lowest-energy doublets at $M$ correspond to $\left[\epsilon_{\mathbf{k}=0}^{X}\right]_{11}=\left[\epsilon_{\mathbf{k}=0}^{Y}\right]_{22}=\epsilon_{1}$
and $\left[\epsilon_{\mathbf{k}=0}^{X}\right]_{22}=\left[\epsilon_{\mathbf{k}=0}^{Y}\right]_{11}=\epsilon_{3}$,
with $\left|\epsilon_{3}\right|>\left|\epsilon_{1}\right|$, and the
matrix elements corresponding to the non-interacting Hamiltonian of
Eq. (\ref{H0}). Note that the off-diagonal elements $\left[\epsilon_{\mathbf{k}=0}^{X}\right]_{12}=\left[\epsilon_{\mathbf{k}=0}^{Y}\right]_{12}=0$
vanish at the $M$ point. The $\Gamma$ point also has one energy
doublet, $\left[\epsilon_{\mathbf{k}=0}^{\Gamma}\right]_{11}=\left[\epsilon_{\mathbf{k}=0}^{\Gamma}\right]_{22}=\epsilon_{\Gamma}$,
associated with the two-dimensional irreducible representation $E_{g}$
at the zone center. In contrast to the doublets at $M$, however,
this doublet is split by the SOC (see Fig. \ref{fig:band-spectrum-tetragonal-phase}). 

In terms of the orbitals, which are good quantum numbers at the $M$
point, the existence of the doublet $\epsilon_{1}$ ($\epsilon_{3}$)
is a consequence of the fact that the $xz$ orbital ($xy$ orbital)
on-site energy at $X$ is equivalent to the $yz$ orbital ($xy$ orbital)
on-site energy energy at $Y$. Because long-range SDW and OLC orders
mix these orbitals, they lead to splittings of the corresponding doublets.
Now because the orbitals involved in the SDW order parameter $M_{E_{Mi}}^{\mu}$
are not necessarily the same as those involved in the OLC order parameter
$\Phi_{E_{Mi}}^{\mu}\neq0$, each order parameter will affect differently
each doublet. In particular, because intra-orbital magnetism can only
involve $xz$ and $yz$ orbitals (assuming, of course, that there
is not a $xy$ hole pocket at $\Gamma$, as it happens in a few iron
pnictides), the onset of long-range SDW order can only split the $\epsilon_{1}$
doublet. However, if the OLC order parameter belongs to the $E_{M1}$
or $E_{M2}$ irreducible representations, it involves also the $xy$
orbital, as shown in Table \ref{tab:Possible-intra-orbital-SDW}.
As a result, it leads to a splitting of the $\epsilon_{3}$ doublet,
but not of $\epsilon_{1}$.

To make this analysis quantitative, we assume non-zero values for
$M_{E_{Mi}}^{\mu}\neq0$ and $\Phi_{E_{Mi}}^{\mu}\neq0$ and diagonalize
the quadratic Hamiltonian $H_{0}+H_{\mathrm{SOC}}+H_{\mathrm{SDW}}+H_{\mathrm{OLC}}$
for each different irreducible representation. In Fig. \ref{fig:Effective-band-spectrum},
we show the corresponding band dispersions in the SDW+OLC state for
three different possible types of order. Focusing at the doublets
at the $\Gamma$ and $M$ points, we can obtain all induced splittings
analytically to leading order in the SOC $\lambda$, in the SDW order
parameter $M$, and in the OLC order parameter $\Phi$. When the latter
transform as components of the irreducible representations $E_{M1}$
and $E_{M2}$, we find the following leading-order band splittings
\begin{subequations}
\begin{align}
\Delta\epsilon_{\Gamma} & \approx|\lambda|\\
\Delta\epsilon_{1} & \approx\frac{M^{2}}{|\epsilon_{\Gamma}-\epsilon_{1}|}\\
\Delta\epsilon_{3} & \approx\frac{\Phi^{2}}{|\epsilon_{\Gamma}-\epsilon_{3}|}.
\end{align}
\end{subequations}

The fact that the orders that transform as $E_{M1}$ and $E_{M2}$
yield equivalent results is attributed to the fact that both orders
couple orbitals at $\Gamma$ to the $xy$ orbital at $X/Y$. A difference
in band shifts appear only in higher order in the order parameters.
Since $\Phi=-\chi_{\Phi}g(\lambda)M$, where $\chi_{\Phi}^{-1}=r_{\Phi}$
is the OLC susceptibility, the ratio between the doublet splittings
at the $M$ point provide an interesting way to estimate the coupling
between the SDW and OLC orders:
\begin{align}
\Delta\epsilon_{3}/\Delta\epsilon_{1} & \approx\chi_{\Phi}^{2}g^{2}(\lambda)\,\frac{|\epsilon_{\Gamma}-\epsilon_{1}|}{|\epsilon_{\Gamma}-\epsilon_{3}|}.
\end{align}

For coupled SDW-OLC orders that transform as $E_{M3}$, the band splittings
become, to leading order

\begin{subequations}
\begin{align}
\Delta\epsilon_{\Gamma} & \approx|\lambda|-2\frac{M\Phi}{|\epsilon_{\Gamma}-\epsilon_{1}|}\\
\Delta\epsilon_{1} & \approx\frac{M^{2}+\Phi^{2}}{|\epsilon_{\Gamma}-\epsilon_{1}|}\\
\Delta\epsilon_{3} & \approx0
\end{align}
\end{subequations}

In comparing these results with ARPES experiments, one has to keep
in mind that orbital order also leads to splittings of these doublets
\cite{Fernandes2014a}. However, for orders that transform as $ E_{M1} $ and $ E_{M2} $, the additional splitting of the doublet $ \epsilon_3 $ represents an unique signature since it is unaffected by orbital and intra-orbital spin  magnetic orders involving predominantly $ xz / yz $ orbitals. 
Thus, proper consideration of this effect
is needed to extract the doublet splittings produced only by SDW and
OLC order.

The onset of OLC order also impacts the reconstruction of the Fermi surface in the magnetically ordered state, which can also be probed by ARPES. Specifically, due to the inter-orbital nature of the OLC orders, 
additional parts of the Fermi surface can be gapped, as compared to the case in which only intra-orbital SDW is present, depicted in Fig. \ref{fig:Effective-fermi-surfaces}(a). Figures \ref{fig:Effective-fermi-surfaces}(b)--(d) depict our results for the reconstructed Fermi surface
assuming SDW and OLC order parameters of equal magnitudes, $M=\Phi$. Note that, in the case of order parameters belonging to the $ E_{M1,M2} $ representations, the onset of OLC order leads to an additional gapping of the Fermi surface along one direction. However, in the case of $ E_{M3} $ order, the Fermi surface remains very similar as the case without OLC order. 

It is important to point out that the expressions derived in this section are valid if intra-orbital magnetism involves only the $xz/yz$ orbitals. For the cases where intra-orbital magnetism involving also the $xy$ orbital appears, which is presumably the case when the additional $xy$ hole pocket is present, the situation changes. This is because the $xy$ SDW order parameter also leads to splitting of the $\epsilon_3$ doublet, which may mask or make it harder to distinguish the contribution arising from the OLC order parameter.

\section{Conclusions}

In this paper, we showed that the onset of long-range SDW order belonging
to one of two-dimensional irreducible representations $E_{M_{i}}$
of the space-group describing a generic iron-pnictide/chalcogenide
plane must trigger an orbital loop-current order belonging to the
same irreducible representation. Such a coupling between these two
types of order is mediated by the spin-orbit interaction, which has
been experimentally found to be sizable in these materials. Each irreducible
representation implies different patterns of loop-current order, as
they only allow very specific combination of charge flow between the
Fe orbitals. For instance, the most widely realized magnetic ground
state in the iron-based superconductors is the stripe SDW state belonging
to the $E_{M1}$ irreducible representation. It corresponds to stripes
of parallel spins whose magnetic moments are parallel to the ordering
vectors $\mathbf{Q}_{1}=\left(\pi,0\right)$ or $\mathbf{Q}_{2}=\left(0,\pi\right)$.
The corresponding OLC patterns have ordering vectors $\mathbf{Q}_{2}=\left(0,\pi\right)$
or $\mathbf{Q}_{1}=\left(\pi,0\right)$, respectively, and involve
charge transfer between neighboring Fe $yz/xz$ and $xy$ orbitals.
Thus, even though the intra-orbital SDW order affects directly only
the $xz$ and $yz$ orbitals, it indirectly impacts also the $xy$
orbital via the induced secondary inter-orbital OLC order. We note
that, by symmetry, OLC order is also induced by other types of magnetic
ground states, such as the $C_{4}$ double-\textbf{Q} magnetic phases
observed in certain hole-doped pnictides \cite{AvciChmaissemAllredEtAl2014,BoehmerHardyWangEtAl2015,AllredTaddeiBugarisEtAl2016,MeierDingKreyssigEtAl2017}.

In our considerations, we assumed that the microscopic electronic
interactions always promote SDW order, of which the OLC order is just
a byproduct. This seems to be the case according to Hartree-Fock \cite{GastiasoroAndersen2015}
and renormalization group calculations \cite{Chubukov2016}. It remains
to be seen, however, whether the OLC order can ever become the leading
instability of the system. In this case, besides OLC order, OLC fluctuations
would become important, which can possibly affect nematicity \cite{Chubukov2015}
and even superconductivity. Note that, since the SDW and OLC order
parameters share the same symmetry properties, both their fluctuations
promote sign-changing $s^{+-}$ superconductivity. Interestingly,
other types of loop-current orders have been proposed in different
correlated systems, most notably cuprates \cite{ChakravartyLaughlinMorrEtAl2001,Varma1997}, in which the order is however considered as leading instability, 
and iridates \cite{ZhaoTorchinskyChuEtAl2016}. Given that these are
also multi-orbital systems, it is interesting to explore whether any
of the effects unveiled here could be relevant in those compounds.

Our finding epitomizes the unique entanglement between spin and orbital
degrees of freedom in the iron-based superconductors, beyond the well-established
interconnection between nematicity and ferro-orbital order \cite{FanfarilloCortijoValenzuela2015,YamakawaOnariKontani2016}.
It also raises the important question of whether the induced OLC order
has a strong impact on the microscopic properties of these materials.
Up to date, to the best of our knowledge, all properties of the SDW
state have been interpreted in terms of a magnetic order parameter
and the accompanying ferro-orbital order. To elucidate the extent
to which the OLC order parameter affects different observables will
require a re-analysis of several of previous experimental results.
Here, we focused on two distinct manifestations of OLC order that
can be probed experimentally, namely, the magnetic field induced by
the loop currents on the pnictogen/chalcogen site, and the splitting
of the $xy$ doublet of the band structure at the $M$ point. Interestingly,
the latter has been recently observed in FeSe, where, however, the usual
SDW order is absent \cite{FedorovYareskoKimEtAl2016}. Another possible
manifestation of the coupling between OLC and SDW is in the spin-wave
spectrum, as spin excitations become mixed with loop-current excitations.
Obviously, such an effect will be stronger the closer the system is
to an spontaneous OLC instability. Finally, since superconductivity
is observed to coexist microscopically with SDW order in some iron-based
compounds, an interesting open issue is how OLC order can impact this
unique thermodynamic state.
\begin{acknowledgments}
We thank A. Chubukov and I. Eremin for fruitful discussions. J.K. and
R.M.F. are supported by the U.S. Department of Energy, Office of Science,
Basic Energy Sciences, under Award No. DE-SC0012336. 
J.S. acknowledges funding from the Deutsche Forschungsgemeinschaft (DFG)  SCHM 1031/7-1.
\end{acknowledgments}

\appendix
\begin{widetext}
\section{Notation guide\label{sec:Notation-guide}}
For convenience, here we provide a summary of the different quantities
defined throughout the paper. Table \ref{tab:orbital_space} displays
the objects defined in the Fe $3d$ orbital space. Table \ref{tab:irreps} presents the quantities defined in terms of
irreducible representations of the $P4/nmm$ space group, which describes
a generic iron pnictogen/chalcogen plane. Finally, Table \ref{tab:extended_spinor} displays the quantities
defined in the extended spinor space that describes the low-energy
electronic model used throughout the text.

\begin{table}[h]
	\protect\caption{Quantities defined in orbital space. Here, $a$ and $b$ are orbital
		indices referring to one of the three Fe $3d$ orbitals: $xz$, $yz$,
		and $xy$. \label{tab:orbital_space}}
	\begin{ruledtabular}
\begin{centering}
\begin{tabular}{l>{\raggedright}p{0.86\columnwidth}}
$d_{a,\mathbf{k}\sigma}$ & Annihilation operator for an electron at orbital $a$ with momentum
$\mathbf{k}$ and spin $\sigma$
\tabularnewline
\hline 
\noalign{\vskip\doublerulesep}
$t_{ij}^{a,b}$ & Amplitude for an electron hopping from site $i$, orbital $a$ to
site $j$, orbital $b$\tabularnewline
\hline 
\noalign{\vskip\doublerulesep}
$\mathbf{m}_{\mathbf{Q}_{i}}^{(a,b)}$ & Spin-density wave order parameter involving orbital $a$ at momentum
$\mathbf{k}$ and orbital $b$ at momentum $\mathbf{k}+\mathbf{Q}_{i}$.
Only two intra-orbitals combinations are considered in the main text,
$\mathbf{m}_{\mathbf{Q}_{1}}\equiv\mathbf{m}_{\mathbf{Q}_{1}}^{(yz,yz)}$
and $\mathbf{m}_{\mathbf{Q}_{2}}\equiv\mathbf{m}_{\mathbf{Q}_{1}}^{(xz,xz)}$\tabularnewline
\hline 
\noalign{\vskip\doublerulesep}
$\phi_{\mathbf{Q}_{i}}^{(a,b)}$ & Orbital loop-current order parameter involving orbital $a$ at momentum
$\mathbf{k}$ and orbital $b$ at momentum $\mathbf{k}+\mathbf{Q}_{i}$. 
\end{tabular}
\par\end{centering}
\end{ruledtabular}
\end{table}

\begin{table}[h]
	\protect\caption{Quantities defined in terms of irreducible representations of the
		$P4/nmm$ space group. \label{tab:irreps}}
	\begin{ruledtabular}
\begin{centering}
\begin{tabular}{l>{\raggedright}p{0.87\columnwidth}}
$E_{Mi}$ & Two-dimensional irreducible representations of the $P4/nmm$ group
at the $M=\left(\pi,\pi\right)$ point of the Brillouin zone corresponding
to the 2-Fe unit cell. In this paper we focus on $i=1,\,2,\,3$. \tabularnewline
\hline 
\noalign{\vskip\doublerulesep}
$M_{E_{Mi}}^{\mu}$ & Spin-density wave order parameter expressed in terms of the two-dimensional
irreducible representations $E_{Mi}$. Each doublet ($\mu=+,\,-$
) corresponds to combinations of the components of $\mathbf{m}_{\mathbf{Q}_{i}}$
introduced in Table \ref{tab:orbital_space}.\tabularnewline
\hline 
\noalign{\vskip\doublerulesep}
$\Phi_{E_{Mi}}^{\mu}$ & Orbital loop-current order parameter expressed in terms of the two-dimensional
irreducible representations $E_{Mi}$. Each doublet ($\mu=+,\,-$
) corresponds to combinations of $\phi_{\mathbf{Q}_{i}}^{(a,b)}$
introduced in Table \ref{tab:orbital_space}.
\end{tabular}
\par\end{centering}
\end{ruledtabular}
\end{table}
\pagebreak

\begin{table}[h]
	\protect\caption{Quantities defined in the space of the low-energy electronic state
		spinors. \label{tab:extended_spinor}}
\begin{ruledtabular}
\begin{centering}
\begin{tabular}{l>{\raggedright}p{0.7\columnwidth}}
$\Psi_{\mathbf{k}}=\left(\psi_{\Gamma,\mathbf{k}},\psi_{X,\mathbf{k}+\mathbf{Q}_{1}},\psi_{Y,\mathbf{k}+\mathbf{Q}_{2}}\right)^{\text{T}}$ & Extended $12$-dimensional electronic spinor. $\psi_{\Gamma}$, $\psi_{X}$,
and $\psi_{Y}$ are each four-dimensional spinors (two spins indices
and two orbital indices) describing the low-energy states near
the $\Gamma$, $X$, and $Y$ points of the 1-Fe unit cell Brillouin
zone. \tabularnewline
\hline 
\noalign{\vskip\doublerulesep}
$\Lambda_{\Gamma}^{\text{SOC}}$ and $\Lambda_{M}^{\text{SOC}}$ & Spin-orbit coupling vertices defined in the $\psi_{\Gamma}$ and in
the $\psi_{X}/\psi_{Y}$ subspaces according to Eqs. (\ref{eq:H-SOC})
and (\ref{eq:VSOC1})-(\ref{eq:VSOC2}).\tabularnewline
\hline 
\noalign{\vskip\doublerulesep}
$\varUpsilon_{E_{Mi},\mu}^{M}$ & Vertex describing the coupling between the spin density-wave order
parameter $M_{E_{Mi}}^{\mu}$ and the electronic states described
by the extended spinor $\Psi_{\mathbf{k}}$. The six $12\times12$
vertices are given in Appendix \ref{sec:Vertex-Parts}.\tabularnewline
\hline 
\noalign{\vskip\doublerulesep}
$\varUpsilon_{E_{Mi},\mu}^{\Phi}$ & Vertex describing the coupling between the orbital loop-current order
parameter $\Phi_{E_{Mi}}^{\mu}$ and the electronic states described
by the extended spinor $\Psi_{\mathbf{k}}$. The six $12\times12$
vertices are given in Appendix \ref{sec:Vertex-Parts}.
\end{tabular}
\par\end{centering}

\end{ruledtabular}
\end{table}

\section{Classification of SDW orders parameters involving $ d_{xy} $ orbitals}
Table \ref{tab:Possible-intra-orbital-SDW-XY}  shows how $xy$ intra-orbital SDW order parameters transform according to the irreducible representations $E_{Mi}:\,\left(E_{Mi}^{+},E_{Mi}^{-}\right)$. Comparison with Table  \ref{tab:Possible-intra-orbital-SDW} in the main text shows that, for a given magnetic order parameter with ordering vector $\mathbf{Q}_{i}$ and moment direction $m^{\alpha}$, the same types of OLC order are induced, regardless of whether the intra-orbital SDW arises from $xz/yz$ or $xy$ orbitals.
\begin{table}[h]
	\begin{minipage}{.5\textwidth}
		\protect\caption{\label{tab:Possible-intra-orbital-SDW-XY} Possible spin
		density-wave orders containing electronic $ xy $-orbital states and the corresponding OLC orders
		.}	
	\begin{ruledtabular}
		\begin{tabular}[b]{>{\raggedright}p{0.2\columnwidth}cccc}
			\noalign{\vskip\doublerulesep}
			\multirow{1}{0.2\columnwidth}{} & \multicolumn{2}{c}{%
				\begin{minipage}[t]{0.35\columnwidth}%
					\begin{center}
						intra-orbital \\
						spin order
						\par\end{center}%
				\end{minipage}} & \multicolumn{2}{c}{%
				\begin{minipage}[t]{0.35\columnwidth}%
					\begin{center}
						orbital loop\\
						current order 
						\par\end{center}%
				\end{minipage}}
				\tabularnewline[0.4cm]
				\cline{2-5} 
				\noalign{\vskip\doublerulesep}
				\noalign{\vskip\doublerulesep}
				irrep of $P4/nmm$ space group & %
				\begin{minipage}[t]{0.15\columnwidth}%
					field%
				\end{minipage}  & %
				\begin{minipage}[t]{0.15\columnwidth}%
					orbital and spin comp.%
				\end{minipage} & %
				\begin{minipage}[t]{0.1\columnwidth}%
					field%
				\end{minipage}  & %
				\begin{minipage}[t]{0.2\columnwidth}%
					orbital composition%
				\end{minipage}
				\tabularnewline[0.4cm]
				\hline 
				\noalign{\vskip\doublerulesep}
				\hline 
				\noalign{\vskip\doublerulesep}
				$\begin{pmatrix}E_{M1}^{+}\\
				E_{M1}^{-}
				\end{pmatrix}$ & $\begin{pmatrix}m_{\mathbf{Q}_{2}}^{y}\\
				m_{\mathbf{Q}_{1}}^{x}
				\end{pmatrix}$ & $\begin{pmatrix}xy,\,\sigma^{y}\\
				xy,\,\sigma^{x}
				\end{pmatrix}$ & $\begin{pmatrix}\phi_{\mathbf{Q}_{1}}^{\left(yz,xy\right)}\\
				\phi_{\mathbf{Q}_{2}}^{\left(xz,xy\right)}
				\end{pmatrix}$ & $\begin{pmatrix}yz,\, xy\\
				xz,\, xy
				\end{pmatrix}$
				\tabularnewline[0.4cm]
				\hline 
				\noalign{\vskip\doublerulesep}
				\noalign{\vskip\doublerulesep}
				$\begin{pmatrix}E_{M2}^{+}\\
				E_{M2}^{-}
				\end{pmatrix}$ & $\begin{pmatrix}m_{\mathbf{Q}_{2}}^{x}\\
				m_{\mathbf{Q}_{1}}^{y}
				\end{pmatrix}$ & $\begin{pmatrix}xy,\,\sigma^{x}\\
				xy,\,\sigma^{y}
				\end{pmatrix}$ & $\begin{pmatrix}\phi_{\mathbf{Q}_{1}}^{\left(xz,xy\right)}\\
				\phi_{\mathbf{Q}_{2}}^{\left(yz,xy\right)}
				\end{pmatrix}$ & $\begin{pmatrix}xz,\, xy\\
				yz,\, xy
				\end{pmatrix}$
				\tabularnewline[0.4cm]
				\hline 
				\noalign{\vskip\doublerulesep}
				\noalign{\vskip\doublerulesep}
				$\begin{pmatrix}E_{M3}^{+}\\
				E_{M3}^{-}
				\end{pmatrix}$ & $\begin{pmatrix}m_{\mathbf{Q}_{1}}^{z}\\
				m_{\mathbf{Q}_{2}}^{z}
				\end{pmatrix}$ & $\begin{pmatrix}xy,\,\sigma^{z}\\
				xy,\,\sigma^{z}
				\end{pmatrix}$ & $\begin{pmatrix}\phi_{\mathbf{Q}_{1}}^{\left(xz,yz\right)}\\
				\phi_{\mathbf{Q}_{2}}^{\left(yz,xz\right)}
				\end{pmatrix}$ & $\begin{pmatrix}xz,\, yz\\
				yz,\, xz
				\end{pmatrix}$
				\tabularnewline[0.4cm]
				\hline 
				\noalign{\vskip\doublerulesep}
				\noalign{\vskip\doublerulesep}
				$\begin{pmatrix}E_{M4}^{+}\\
				E_{M4}^{-}
				\end{pmatrix}$ & - & - & $\begin{pmatrix}\phi_{\mathbf{Q}_{1}}^{\left(xy,xy\right)}\\
				\phi_{\mathbf{Q}_{2}}^{\left(xy,xy\right)}
				\end{pmatrix}$ & $\begin{pmatrix}xy,\, xy\\
				xy,\, xy
				\end{pmatrix}$\tabularnewline[0.4cm]
			\end{tabular}
		\end{ruledtabular}
	\end{minipage}
	\end{table}

\section{\label{sec:Orbital-dispersion-relations}Orbital dispersion relations }

Here we give explicit expressions for the electronic dispersion relations introduced in
Eq. \ref{H0}:
\begin{eqnarray*}
\epsilon_{\mathbf{k}}^{\Gamma} & = & \begin{pmatrix}\epsilon_{\Gamma}+2\frac{\mathbf{k}^{2}}{2m_{\Gamma}}+b\left(k_{x}^{2}-k_{y}^{2}\right) & 4ck_{x}k_{y}\\
4ck_{x}k_{y} & \epsilon_{\Gamma}+2\frac{\mathbf{k}^{2}}{2m_{\Gamma}}-b\left(k_{x}^{2}-k_{y}^{2}\right)
\end{pmatrix},\\
\epsilon_{\mathbf{k}}^{X} & = & \begin{pmatrix}\epsilon_{1}+2\frac{\mathbf{k}^{2}}{2m_{1}}+a_{1}\left(k_{x}^{2}-k_{y}^{2}\right) & -iv(\mathbf{k})\\
iv(\mathbf{k}) & \epsilon_{3}+2\frac{\mathbf{k}^{2}}{2m_{3}}+a_{3}\left(k_{x}^{2}-k_{y}^{2}\right)
\end{pmatrix},\\
\epsilon_{\mathbf{k}}^{Y} & = & \epsilon_{\mathbf{k}}^{X}\big|_{k_{x}\rightarrow k_{y},k_{y}\rightarrow-k_{x}}
\end{eqnarray*}
with $v(\mathbf{k})=2vk_{y}+2p_{1}k_{y}\left(k_{y}^{2}+3k_{x}^{2}\right)-2p_{2}k_{y}\left(k_{x}^{2}-k_{y}^{2}\right)$.
The parameters $\epsilon_{\Gamma},m_{\Gamma},b,c$ for the dispersion
at $\Gamma$ and $\epsilon_{1},m_{1},a_{1},\epsilon_{3},m_{3},a_{3},v,p_{1},p_{2}$
at the $X$ and $Y$ points can be found in Ref. \cite{Cvetkovic2013}.

\section{\label{sec:Vertex-Parts}Vertex functions}
The vertices introduced in Eqs. \ref{eq:Mvertex}
and \ref{eq:phiVertex} for different orders are derived from their
orbital and spin composition as stated in Table \ref{tab:Possible-intra-orbital-SDW}.
They are represented as matrices in enlarged orbital space $\Psi=\left(\psi_{\Gamma},\psi_{X},\psi_{Y}\right)^{\text{T}}$
and are given by 
\begin{itemize}
\item $E_{M1}$:
\[
\begin{array}{ll}
\varUpsilon_{E_{M1},+}^{M}=-\begin{pmatrix}0 & 0 & \tau^{-}\sigma^{y}\\
0 & 0 & 0\\
\tau^{+}\sigma^{y} & 0 & 0
\end{pmatrix}, & \varUpsilon_{E_{M1},+}^{\Phi}=\begin{pmatrix}0 & i\tau^{+}\sigma^{0} & 0\\
-i\tau^{-}\sigma^{0} & 0 & 0\\
0 & 0 & 0
\end{pmatrix}\\
\varUpsilon_{E_{M1},-}^{M}=\begin{pmatrix}0 & \tau^{\uparrow}\sigma^{x} & 0\\
\tau^{\uparrow}\sigma^{x} & 0 & 0\\
0 & 0 & 0
\end{pmatrix}, & \varUpsilon_{E_{M1},-}^{\Phi}=\begin{pmatrix}0 & 0 & -i\tau^{\downarrow}\sigma^{0}\\
0 & 0 & 0\\
i\tau^{\downarrow}\sigma^{0} & 0 & 0
\end{pmatrix}
\end{array}
\]

\item $E_{M2}$:
\[
\begin{array}{ll}
\varUpsilon_{E_{M2},+}^{M}=-\begin{pmatrix}0 & 0 & \tau^{-}\sigma^{x}\\
0 & 0 & 0\\
\tau^{+}\sigma^{x} & 0 & 0
\end{pmatrix}, & \varUpsilon_{E_{M2},+}^{\Phi}=\begin{pmatrix}0 & -i\tau^{\downarrow}\sigma^{0} & 0\\
i\tau^{\downarrow}\sigma^{0} & 0 & 0\\
0 & 0 & 0
\end{pmatrix}\\
\varUpsilon_{E_{M2},-}^{M}=\begin{pmatrix}0 & \tau^{\uparrow}\sigma^{y} & 0\\
\tau^{\uparrow}\sigma^{y} & 0 & 0\\
0 & 0 & 0
\end{pmatrix}, & \varUpsilon_{E_{M2},-}^{\Phi}=\begin{pmatrix}0 & 0 & i\tau^{+}\sigma^{0}\\
0 & 0 & 0\\
-i\tau^{-}\sigma^{0} & 0 & 0
\end{pmatrix}
\end{array}
\]

\item $E_{M3}$:
\[
\begin{array}{ll}
\varUpsilon_{E_{M3},+}^{M}=\begin{pmatrix}0 & \tau^{\uparrow}\sigma^{z} & 0\\
\tau^{\uparrow}\sigma^{z} & 0 & 0\\
0 & 0 & 0
\end{pmatrix}, & \varUpsilon_{E_{M3},+}^{\Phi}=\begin{pmatrix}0 & -i\tau^{-}\sigma^{0} & 0\\
i\tau^{+}\sigma^{0} & 0 & 0\\
0 & 0 & 0
\end{pmatrix}\\
\varUpsilon_{E_{M3},-}^{M}=-\begin{pmatrix}0 & 0 & \tau^{-}\sigma^{z}\\
0 & 0 & 0\\
\tau^{+}\sigma^{z} & 0 & 0
\end{pmatrix}, & \varUpsilon_{E_{M3},-}^{\Phi}=\begin{pmatrix}0 & 0 & i\tau^{\uparrow}\sigma^{0}\\
0 & 0 & 0\\
-i\tau^{\uparrow}\sigma^{0} & 0 & 0
\end{pmatrix}
\end{array}
\]

\end{itemize}
with 
\begin{eqnarray*}
\tau^{\uparrow} & = & \begin{pmatrix}1 & 0\\
0 & 0
\end{pmatrix},\quad\tau^{\downarrow}=\begin{pmatrix}0 & 0\\
0 & 1
\end{pmatrix},\\
\tau^{+} & = & \begin{pmatrix}0 & 1\\
0 & 0
\end{pmatrix},\quad\tau^{-}=\begin{pmatrix}0 & 0\\
1 & 0
\end{pmatrix}
\end{eqnarray*}
acting in orbital doublet space and $\sigma^{i}$ acting in spin space. 

\end{widetext}

\bibliographystyle{apsrev4-1}
\bibliography{paperbase}

\begin{thebibliography}{40}%
\makeatletter
\providecommand \@ifxundefined [1]{%
 \@ifx{#1\undefined}
}%
\providecommand \@ifnum [1]{%
 \ifnum #1\expandafter \@firstoftwo
 \else \expandafter \@secondoftwo
 \fi
}%
\providecommand \@ifx [1]{%
 \ifx #1\expandafter \@firstoftwo
 \else \expandafter \@secondoftwo
 \fi
}%
\providecommand \natexlab [1]{#1}%
\providecommand \enquote  [1]{``#1''}%
\providecommand \bibnamefont  [1]{#1}%
\providecommand \bibfnamefont [1]{#1}%
\providecommand \citenamefont [1]{#1}%
\providecommand \href@noop [0]{\@secondoftwo}%
\providecommand \href [0]{\begingroup \@sanitize@url \@href}%
\providecommand \@href[1]{\@@startlink{#1}\@@href}%
\providecommand \@@href[1]{\endgroup#1\@@endlink}%
\providecommand \@sanitize@url [0]{\catcode `\\12\catcode `\$12\catcode
  `\&12\catcode `\#12\catcode `\^12\catcode `\_12\catcode `\%12\relax}%
\providecommand \@@startlink[1]{}%
\providecommand \@@endlink[0]{}%
\providecommand \url  [0]{\begingroup\@sanitize@url \@url }%
\providecommand \@url [1]{\endgroup\@href {#1}{\urlprefix }}%
\providecommand \urlprefix  [0]{URL }%
\providecommand \Eprint [0]{\href }%
\providecommand \doibase [0]{http://dx.doi.org/}%
\providecommand \selectlanguage [0]{\@gobble}%
\providecommand \bibinfo  [0]{\@secondoftwo}%
\providecommand \bibfield  [0]{\@secondoftwo}%
\providecommand \translation [1]{[#1]}%
\providecommand \BibitemOpen [0]{}%
\providecommand \bibitemStop [0]{}%
\providecommand \bibitemNoStop [0]{.\EOS\space}%
\providecommand \EOS [0]{\spacefactor3000\relax}%
\providecommand \BibitemShut  [1]{\csname bibitem#1\endcsname}%
\let\auto@bib@innerbib\@empty
\bibitem [{\citenamefont {Johnston}(2010)}]{Johnston2010}%
  \BibitemOpen
  \bibfield  {author} {\bibinfo {author} {\bibfnamefont {D.~C.}\ \bibnamefont
  {Johnston}},\ }\href {\doibase 10.1080/00018732.2010.513480} {\bibfield
  {journal} {\bibinfo  {journal} {Adv. Phys.}\ }\textbf {\bibinfo {volume}
  {59}},\ \bibinfo {pages} {803} (\bibinfo {year} {2010})}\BibitemShut
  {NoStop}%
\bibitem [{\citenamefont {Paglione}\ and\ \citenamefont
  {Greene}(2010)}]{Paglione2010}%
  \BibitemOpen
  \bibfield  {author} {\bibinfo {author} {\bibfnamefont {J.}~\bibnamefont
  {Paglione}}\ and\ \bibinfo {author} {\bibfnamefont {R.~L.}\ \bibnamefont
  {Greene}},\ }\href {\doibase 10.1038/nphys1759} {\bibfield  {journal}
  {\bibinfo  {journal} {Nat. Phys.}\ }\textbf {\bibinfo {volume} {6}},\
  \bibinfo {pages} {645} (\bibinfo {year} {2010})}\BibitemShut {NoStop}%
\bibitem [{\citenamefont {Stewart}(2011)}]{Stewart2011}%
  \BibitemOpen
  \bibfield  {author} {\bibinfo {author} {\bibfnamefont {G.~R.}\ \bibnamefont
  {Stewart}},\ }\href {\doibase 10.1103/RevModPhys.83.1589} {\bibfield
  {journal} {\bibinfo  {journal} {Rev. Mod. Phys.}\ }\textbf {\bibinfo {volume}
  {83}},\ \bibinfo {pages} {1589} (\bibinfo {year} {2011})}\BibitemShut
  {NoStop}%
\bibitem [{\citenamefont {Dai}\ \emph {et~al.}(2012)\citenamefont {Dai},
  \citenamefont {Hu},\ and\ \citenamefont {Dagotto}}]{Dai2012}%
  \BibitemOpen
  \bibfield  {author} {\bibinfo {author} {\bibfnamefont {P.}~\bibnamefont
  {Dai}}, \bibinfo {author} {\bibfnamefont {J.}~\bibnamefont {Hu}}, \ and\
  \bibinfo {author} {\bibfnamefont {E.}~\bibnamefont {Dagotto}},\ }\href
  {\doibase 10.1038/nphys2438} {\bibfield  {journal} {\bibinfo  {journal} {Nat.
  Phys.}\ }\textbf {\bibinfo {volume} {8}},\ \bibinfo {pages} {709} (\bibinfo
  {year} {2012})}\BibitemShut {NoStop}%
\bibitem [{\citenamefont {{Avci}}\ \emph {et~al.}(2014)\citenamefont {{Avci}},
  \citenamefont {{Chmaissem}}, \citenamefont {{Allred}}, \citenamefont
  {{Rosenkranz}}, \citenamefont {{Eremin}}, \citenamefont {{Chubukov}},
  \citenamefont {{Bugaris}}, \citenamefont {{Chung}}, \citenamefont
  {{Kanatzidis}}, \citenamefont {{Castellan}}, \citenamefont {{Schlueter}},
  \citenamefont {{Claus}}, \citenamefont {{Khalyavin}}, \citenamefont
  {{Manuel}}, \citenamefont {{Daoud-Aladine}},\ and\ \citenamefont
  {{Osborn}}}]{AvciChmaissemAllredEtAl2014}%
  \BibitemOpen
  \bibfield  {author} {\bibinfo {author} {\bibfnamefont {S.}~\bibnamefont
  {{Avci}}}, \bibinfo {author} {\bibfnamefont {O.}~\bibnamefont {{Chmaissem}}},
  \bibinfo {author} {\bibfnamefont {J.~M.}\ \bibnamefont {{Allred}}}, \bibinfo
  {author} {\bibfnamefont {S.}~\bibnamefont {{Rosenkranz}}}, \bibinfo {author}
  {\bibfnamefont {I.}~\bibnamefont {{Eremin}}}, \bibinfo {author}
  {\bibfnamefont {A.~V.}\ \bibnamefont {{Chubukov}}}, \bibinfo {author}
  {\bibfnamefont {D.~E.}\ \bibnamefont {{Bugaris}}}, \bibinfo {author}
  {\bibfnamefont {D.~Y.}\ \bibnamefont {{Chung}}}, \bibinfo {author}
  {\bibfnamefont {M.~G.}\ \bibnamefont {{Kanatzidis}}}, \bibinfo {author}
  {\bibfnamefont {J.-P.}\ \bibnamefont {{Castellan}}}, \bibinfo {author}
  {\bibfnamefont {J.~A.}\ \bibnamefont {{Schlueter}}}, \bibinfo {author}
  {\bibfnamefont {H.}~\bibnamefont {{Claus}}}, \bibinfo {author} {\bibfnamefont
  {D.~D.}\ \bibnamefont {{Khalyavin}}}, \bibinfo {author} {\bibfnamefont
  {P.}~\bibnamefont {{Manuel}}}, \bibinfo {author} {\bibfnamefont
  {A.}~\bibnamefont {{Daoud-Aladine}}}, \ and\ \bibinfo {author} {\bibfnamefont
  {R.}~\bibnamefont {{Osborn}}},\ }\href {\doibase 10.1038/ncomms4845}
  {\bibfield  {journal} {\bibinfo  {journal} {Nat. Commun.}\ }\textbf {\bibinfo
  {volume} {5}},\ \bibinfo {eid} {3845} (\bibinfo {year} {2014})}\BibitemShut
  {NoStop}%
\bibitem [{\citenamefont {{B{\"o}hmer}}\ \emph {et~al.}(2015)\citenamefont
  {{B{\"o}hmer}}, \citenamefont {{Hardy}}, \citenamefont {{Wang}},
  \citenamefont {{Wolf}}, \citenamefont {{Schweiss}},\ and\ \citenamefont
  {{Meingast}}}]{BoehmerHardyWangEtAl2015}%
  \BibitemOpen
  \bibfield  {author} {\bibinfo {author} {\bibfnamefont {A.~E.}\ \bibnamefont
  {{B{\"o}hmer}}}, \bibinfo {author} {\bibfnamefont {F.}~\bibnamefont
  {{Hardy}}}, \bibinfo {author} {\bibfnamefont {L.}~\bibnamefont {{Wang}}},
  \bibinfo {author} {\bibfnamefont {T.}~\bibnamefont {{Wolf}}}, \bibinfo
  {author} {\bibfnamefont {P.}~\bibnamefont {{Schweiss}}}, \ and\ \bibinfo
  {author} {\bibfnamefont {C.}~\bibnamefont {{Meingast}}},\ }\href {\doibase
  10.1038/ncomms8911} {\bibfield  {journal} {\bibinfo  {journal} {Nat.
  Commun.}\ }\textbf {\bibinfo {volume} {6}},\ \bibinfo {eid} {7911} (\bibinfo
  {year} {2015})}\BibitemShut {NoStop}%
\bibitem [{\citenamefont {{Allred}}\ \emph {et~al.}(2016)\citenamefont
  {{Allred}}, \citenamefont {{Taddei}}, \citenamefont {{Bugaris}},
  \citenamefont {{Krogstad}}, \citenamefont {{Lapidus}}, \citenamefont
  {{Chung}}, \citenamefont {{Claus}}, \citenamefont {{Kanatzidis}},
  \citenamefont {{Brown}}, \citenamefont {{Kang}}, \citenamefont {{Fernandes}},
  \citenamefont {{Eremin}}, \citenamefont {{Rosenkranz}}, \citenamefont
  {{Chmaissem}},\ and\ \citenamefont {{Osborn}}}]{AllredTaddeiBugarisEtAl2016}%
  \BibitemOpen
  \bibfield  {author} {\bibinfo {author} {\bibfnamefont {J.~M.}\ \bibnamefont
  {{Allred}}}, \bibinfo {author} {\bibfnamefont {K.~M.}\ \bibnamefont
  {{Taddei}}}, \bibinfo {author} {\bibfnamefont {D.~E.}\ \bibnamefont
  {{Bugaris}}}, \bibinfo {author} {\bibfnamefont {M.~J.}\ \bibnamefont
  {{Krogstad}}}, \bibinfo {author} {\bibfnamefont {S.~H.}\ \bibnamefont
  {{Lapidus}}}, \bibinfo {author} {\bibfnamefont {D.~Y.}\ \bibnamefont
  {{Chung}}}, \bibinfo {author} {\bibfnamefont {H.}~\bibnamefont {{Claus}}},
  \bibinfo {author} {\bibfnamefont {M.~G.}\ \bibnamefont {{Kanatzidis}}},
  \bibinfo {author} {\bibfnamefont {D.~E.}\ \bibnamefont {{Brown}}}, \bibinfo
  {author} {\bibfnamefont {J.}~\bibnamefont {{Kang}}}, \bibinfo {author}
  {\bibfnamefont {R.~M.}\ \bibnamefont {{Fernandes}}}, \bibinfo {author}
  {\bibfnamefont {I.}~\bibnamefont {{Eremin}}}, \bibinfo {author}
  {\bibfnamefont {S.}~\bibnamefont {{Rosenkranz}}}, \bibinfo {author}
  {\bibfnamefont {O.}~\bibnamefont {{Chmaissem}}}, \ and\ \bibinfo {author}
  {\bibfnamefont {R.}~\bibnamefont {{Osborn}}},\ }\href {\doibase
  10.1038/nphys3629} {\bibfield  {journal} {\bibinfo  {journal} {Nat. Phys.}\
  }\textbf {\bibinfo {volume} {12}},\ \bibinfo {pages} {493} (\bibinfo {year}
  {2016})}\BibitemShut {NoStop}%
\bibitem [{\citenamefont {Meier}\ \emph {et~al.}(2018)\citenamefont {Meier},
  \citenamefont {Ding}, \citenamefont {Kreyssig}, \citenamefont {Bud'ko},
  \citenamefont {Sapkota}, \citenamefont {Kothapalli}, \citenamefont {Borisov},
  \citenamefont {Valentí}, \citenamefont {Batista}, \citenamefont {Orth},
  \citenamefont {Fernandes}, \citenamefont {Goldman}, \citenamefont {Furukawa},
  \citenamefont {B{\"o}hmer},\ and\ \citenamefont
  {Canfield}}]{MeierDingKreyssigEtAl2017}%
  \BibitemOpen
  \bibfield  {author} {\bibinfo {author} {\bibfnamefont {W.~R.}\ \bibnamefont
  {Meier}}, \bibinfo {author} {\bibfnamefont {Q.~P.}\ \bibnamefont {Ding}},
  \bibinfo {author} {\bibfnamefont {A.}~\bibnamefont {Kreyssig}}, \bibinfo
  {author} {\bibfnamefont {S.~L.}\ \bibnamefont {Bud'ko}}, \bibinfo {author}
  {\bibfnamefont {A.}~\bibnamefont {Sapkota}}, \bibinfo {author} {\bibfnamefont
  {K.}~\bibnamefont {Kothapalli}}, \bibinfo {author} {\bibfnamefont
  {V.}~\bibnamefont {Borisov}}, \bibinfo {author} {\bibfnamefont
  {R.}~\bibnamefont {Valentí}}, \bibinfo {author} {\bibfnamefont {C.~D.}\
  \bibnamefont {Batista}}, \bibinfo {author} {\bibfnamefont {P.~P.}\
  \bibnamefont {Orth}}, \bibinfo {author} {\bibfnamefont {R.~M.}\ \bibnamefont
  {Fernandes}}, \bibinfo {author} {\bibfnamefont {A.~I.}\ \bibnamefont
  {Goldman}}, \bibinfo {author} {\bibfnamefont {Y.}~\bibnamefont {Furukawa}},
  \bibinfo {author} {\bibfnamefont {A.~E.}\ \bibnamefont {B{\"o}hmer}}, \ and\
  \bibinfo {author} {\bibfnamefont {P.~C.}\ \bibnamefont {Canfield}},\ }\href
  {\doibase 10.1038/s41535-017-0076-x} {\bibfield  {journal} {\bibinfo
  {journal} {npj Quantum Materials}\ }\textbf {\bibinfo {volume} {3}},\
  \bibinfo {pages} {5} (\bibinfo {year} {2018})}\BibitemShut {NoStop}%
\bibitem [{\citenamefont {Fernandes}\ \emph {et~al.}(2014)\citenamefont
  {Fernandes}, \citenamefont {Chubukov},\ and\ \citenamefont
  {Schmalian}}]{Fernandes2014}%
  \BibitemOpen
  \bibfield  {author} {\bibinfo {author} {\bibfnamefont {R.~M.}\ \bibnamefont
  {Fernandes}}, \bibinfo {author} {\bibfnamefont {A.~V.}\ \bibnamefont
  {Chubukov}}, \ and\ \bibinfo {author} {\bibfnamefont {J.}~\bibnamefont
  {Schmalian}},\ }\href {\doibase 10.1038/nphys2877} {\bibfield  {journal}
  {\bibinfo  {journal} {Nat. Phys.}\ }\textbf {\bibinfo {volume} {10}},\
  \bibinfo {pages} {97} (\bibinfo {year} {2014})}\BibitemShut {NoStop}%
\bibitem [{\citenamefont {{Yi}}\ \emph {et~al.}(2011)\citenamefont {{Yi}},
  \citenamefont {{Lu}}, \citenamefont {{Chu}}, \citenamefont {{Analytis}},
  \citenamefont {{Sorini}}, \citenamefont {{Kemper}}, \citenamefont {{Moritz}},
  \citenamefont {{Mo}}, \citenamefont {{Moore}}, \citenamefont {{Hashimoto}},
  \citenamefont {{Lee}}, \citenamefont {{Hussain}}, \citenamefont
  {{Devereaux}}, \citenamefont {{Fisher}},\ and\ \citenamefont
  {{Shen}}}]{YiLuChuEtAl2011}%
  \BibitemOpen
  \bibfield  {author} {\bibinfo {author} {\bibfnamefont {M.}~\bibnamefont
  {{Yi}}}, \bibinfo {author} {\bibfnamefont {D.}~\bibnamefont {{Lu}}}, \bibinfo
  {author} {\bibfnamefont {J.-H.}\ \bibnamefont {{Chu}}}, \bibinfo {author}
  {\bibfnamefont {J.~G.}\ \bibnamefont {{Analytis}}}, \bibinfo {author}
  {\bibfnamefont {A.~P.}\ \bibnamefont {{Sorini}}}, \bibinfo {author}
  {\bibfnamefont {A.~F.}\ \bibnamefont {{Kemper}}}, \bibinfo {author}
  {\bibfnamefont {B.}~\bibnamefont {{Moritz}}}, \bibinfo {author}
  {\bibfnamefont {S.-K.}\ \bibnamefont {{Mo}}}, \bibinfo {author}
  {\bibfnamefont {R.~G.}\ \bibnamefont {{Moore}}}, \bibinfo {author}
  {\bibfnamefont {M.}~\bibnamefont {{Hashimoto}}}, \bibinfo {author}
  {\bibfnamefont {W.-S.}\ \bibnamefont {{Lee}}}, \bibinfo {author}
  {\bibfnamefont {Z.}~\bibnamefont {{Hussain}}}, \bibinfo {author}
  {\bibfnamefont {T.~P.}\ \bibnamefont {{Devereaux}}}, \bibinfo {author}
  {\bibfnamefont {I.~R.}\ \bibnamefont {{Fisher}}}, \ and\ \bibinfo {author}
  {\bibfnamefont {Z.-X.}\ \bibnamefont {{Shen}}},\ }\href {\doibase
  10.1073/pnas.1015572108} {\bibfield  {journal} {\bibinfo  {journal} {Proc.
  Natl. Acad. Sci.}\ }\textbf {\bibinfo {volume} {108}},\ \bibinfo {pages}
  {6878} (\bibinfo {year} {2011})}\BibitemShut {NoStop}%
\bibitem [{\citenamefont {{Applegate}}\ \emph {et~al.}(2012)\citenamefont
  {{Applegate}}, \citenamefont {{Singh}}, \citenamefont {{Chen}},\ and\
  \citenamefont {{Devereaux}}}]{ApplegateSinghChenEtAl2012}%
  \BibitemOpen
  \bibfield  {author} {\bibinfo {author} {\bibfnamefont {R.}~\bibnamefont
  {{Applegate}}}, \bibinfo {author} {\bibfnamefont {R.~R.~P.}\ \bibnamefont
  {{Singh}}}, \bibinfo {author} {\bibfnamefont {C.-C.}\ \bibnamefont {{Chen}}},
  \ and\ \bibinfo {author} {\bibfnamefont {T.~P.}\ \bibnamefont
  {{Devereaux}}},\ }\href {\doibase 10.1103/PhysRevB.85.054411} {\bibfield
  {journal} {\bibinfo  {journal} {\prb}\ }\textbf {\bibinfo {volume} {85}},\
  \bibinfo {eid} {054411} (\bibinfo {year} {2012})}\BibitemShut {NoStop}%
\bibitem [{\citenamefont {Fernandes}\ \emph {et~al.}(2013)\citenamefont
  {Fernandes}, \citenamefont {B{\"o}hmer}, \citenamefont {Meingast},\ and\
  \citenamefont {Schmalian}}]{Fernandes2013a}%
  \BibitemOpen
  \bibfield  {author} {\bibinfo {author} {\bibfnamefont {R.~M.}\ \bibnamefont
  {Fernandes}}, \bibinfo {author} {\bibfnamefont {A.~E.}\ \bibnamefont
  {B{\"o}hmer}}, \bibinfo {author} {\bibfnamefont {C.}~\bibnamefont
  {Meingast}}, \ and\ \bibinfo {author} {\bibfnamefont {J.}~\bibnamefont
  {Schmalian}},\ }\href {http://dx.doi.org/10.1103/PhysRevLett.111.137001}
  {\bibfield  {journal} {\bibinfo  {journal} {Phys. Rev. Lett.}\ }\textbf
  {\bibinfo {volume} {111}},\ \bibinfo {pages} {137001} (\bibinfo {year}
  {2013})}\BibitemShut {NoStop}%
\bibitem [{\citenamefont {Borisenko}\ \emph {et~al.}(2015)\citenamefont
  {Borisenko}, \citenamefont {Evtushinsky}, \citenamefont {Liu}, \citenamefont
  {Morozov}, \citenamefont {Kappenberger}, \citenamefont {Wurmehl},
  \citenamefont {Büchner}, \citenamefont {Yaresko}, \citenamefont {Kim},
  \citenamefont {Hoesch},\ and\ \citenamefont {Zhigadlo}}]{Borisenko2015}%
  \BibitemOpen
  \bibfield  {author} {\bibinfo {author} {\bibfnamefont {S.~V.}\ \bibnamefont
  {Borisenko}}, \bibinfo {author} {\bibfnamefont {D.~V.}\ \bibnamefont
  {Evtushinsky}}, \bibinfo {author} {\bibfnamefont {Z.-H.}\ \bibnamefont
  {Liu}}, \bibinfo {author} {\bibfnamefont {I.}~\bibnamefont {Morozov}},
  \bibinfo {author} {\bibfnamefont {R.}~\bibnamefont {Kappenberger}}, \bibinfo
  {author} {\bibfnamefont {S.}~\bibnamefont {Wurmehl}}, \bibinfo {author}
  {\bibfnamefont {B.}~\bibnamefont {Büchner}}, \bibinfo {author} {\bibfnamefont
  {A.~N.}\ \bibnamefont {Yaresko}}, \bibinfo {author} {\bibfnamefont {T.~K.}\
  \bibnamefont {Kim}}, \bibinfo {author} {\bibfnamefont {M.}~\bibnamefont
  {Hoesch}}, \ and\ \bibinfo {author} {\bibfnamefont {N.~D.}\ \bibnamefont
  {Zhigadlo}},\ }\href {\doibase 10.1038/nphys3594} {\bibfield  {journal}
  {\bibinfo  {journal} {Nat. Phys.}\ }\textbf {\bibinfo {volume} {12}},\
  \bibinfo {pages} {311} (\bibinfo {year} {2015})}\BibitemShut {NoStop}%
\bibitem [{\citenamefont {Fernandes}\ and\ \citenamefont
  {Vafek}(2014)}]{Fernandes2014a}%
  \BibitemOpen
  \bibfield  {author} {\bibinfo {author} {\bibfnamefont {R.~M.}\ \bibnamefont
  {Fernandes}}\ and\ \bibinfo {author} {\bibfnamefont {O.}~\bibnamefont
  {Vafek}},\ }\href {\doibase 10.1103/physrevb.90.214514} {\bibfield  {journal}
  {\bibinfo  {journal} {Phys. Rev. B}\ }\textbf {\bibinfo {volume} {90}},\
  \bibinfo {pages} {214514} (\bibinfo {year} {2014})}\BibitemShut {NoStop}%
\bibitem [{\citenamefont {Cvetkovic}\ and\ \citenamefont
  {Vafek}(2013)}]{Cvetkovic2013}%
  \BibitemOpen
  \bibfield  {author} {\bibinfo {author} {\bibfnamefont {V.}~\bibnamefont
  {Cvetkovic}}\ and\ \bibinfo {author} {\bibfnamefont {O.}~\bibnamefont
  {Vafek}},\ }\href {\doibase 10.1103/physrevb.88.134510} {\bibfield  {journal}
  {\bibinfo  {journal} {Phys. Rev. B}\ }\textbf {\bibinfo {volume} {88}},\
  \bibinfo {pages} {134510} (\bibinfo {year} {2013})}\BibitemShut {NoStop}%
\bibitem [{\citenamefont {Christensen}\ \emph {et~al.}(2015)\citenamefont
  {Christensen}, \citenamefont {Kang}, \citenamefont {Andersen}, \citenamefont
  {Eremin},\ and\ \citenamefont {Fernandes}}]{Christensen2015}%
  \BibitemOpen
  \bibfield  {author} {\bibinfo {author} {\bibfnamefont {M.~H.}\ \bibnamefont
  {Christensen}}, \bibinfo {author} {\bibfnamefont {J.}~\bibnamefont {Kang}},
  \bibinfo {author} {\bibfnamefont {B.~M.}\ \bibnamefont {Andersen}}, \bibinfo
  {author} {\bibfnamefont {I.}~\bibnamefont {Eremin}}, \ and\ \bibinfo {author}
  {\bibfnamefont {R.~M.}\ \bibnamefont {Fernandes}},\ }\href {\doibase
  10.1103/physrevb.92.214509} {\bibfield  {journal} {\bibinfo  {journal} {Phys.
  Rev. B}\ }\textbf {\bibinfo {volume} {92}},\ \bibinfo {pages} {214509}
  (\bibinfo {year} {2015})}\BibitemShut {NoStop}%
\bibitem [{\citenamefont {{Fernandes}}\ and\ \citenamefont
  {{Chubukov}}(2017)}]{FernandesChubukov2017}%
  \BibitemOpen
  \bibfield  {author} {\bibinfo {author} {\bibfnamefont {R.~M.}\ \bibnamefont
  {{Fernandes}}}\ and\ \bibinfo {author} {\bibfnamefont {A.~V.}\ \bibnamefont
  {{Chubukov}}},\ }\href {\doibase 10.1088/1361-6633/80/1/014503} {\bibfield
  {journal} {\bibinfo  {journal} {Rep. Prog. Phys.}\ }\textbf {\bibinfo
  {volume} {80}},\ \bibinfo {eid} {014503} (\bibinfo {year}
  {2017})}\BibitemShut {NoStop}%
\bibitem [{\citenamefont {{Tomi{\'c}}}\ \emph {et~al.}(2014)\citenamefont
  {{Tomi{\'c}}}, \citenamefont {{Jeschke}},\ and\ \citenamefont
  {{Valent{\'{\i}}}}}]{TomicJeschkeValenti2014}%
  \BibitemOpen
  \bibfield  {author} {\bibinfo {author} {\bibfnamefont {M.}~\bibnamefont
  {{Tomi{\'c}}}}, \bibinfo {author} {\bibfnamefont {H.~O.}\ \bibnamefont
  {{Jeschke}}}, \ and\ \bibinfo {author} {\bibfnamefont {R.}~\bibnamefont
  {{Valent{\'{\i}}}}},\ }\href {\doibase 10.1103/PhysRevB.90.195121} {\bibfield
   {journal} {\bibinfo  {journal} {\prb}\ }\textbf {\bibinfo {volume} {90}},\
  \bibinfo {eid} {195121} (\bibinfo {year} {2014})}\BibitemShut {NoStop}%
\bibitem [{\citenamefont {{Lin}}\ \emph {et~al.}(2011)\citenamefont {{Lin}},
  \citenamefont {{Berlijn}}, \citenamefont {{Wang}}, \citenamefont {{Lee}},
  \citenamefont {{Yin}},\ and\ \citenamefont {{Ku}}}]{LinBerlijnWangEtAl2011}%
  \BibitemOpen
  \bibfield  {author} {\bibinfo {author} {\bibfnamefont {C.-H.}\ \bibnamefont
  {{Lin}}}, \bibinfo {author} {\bibfnamefont {T.}~\bibnamefont {{Berlijn}}},
  \bibinfo {author} {\bibfnamefont {L.}~\bibnamefont {{Wang}}}, \bibinfo
  {author} {\bibfnamefont {C.-C.}\ \bibnamefont {{Lee}}}, \bibinfo {author}
  {\bibfnamefont {W.-G.}\ \bibnamefont {{Yin}}}, \ and\ \bibinfo {author}
  {\bibfnamefont {W.}~\bibnamefont {{Ku}}},\ }\href {\doibase
  10.1103/PhysRevLett.107.257001} {\bibfield  {journal} {\bibinfo  {journal}
  {Phys. Rev. Lett.}\ }\textbf {\bibinfo {volume} {107}},\ \bibinfo {eid}
  {257001} (\bibinfo {year} {2011})}\BibitemShut {NoStop}%
\bibitem [{\citenamefont {Coldea}\ and\ \citenamefont
  {Watson}(2018)}]{ColdeaWatson2017}%
  \BibitemOpen
  \bibfield  {author} {\bibinfo {author} {\bibfnamefont {A.~I.}\ \bibnamefont
  {Coldea}}\ and\ \bibinfo {author} {\bibfnamefont {M.~D.}\ \bibnamefont
  {Watson}},\ }\href {\doibase 10.1146/annurev-conmatphys-033117-054137}
  {\bibfield  {journal} {\bibinfo  {journal} {Annu. Rev. Condens. Matter
  Phys.}\ }\textbf {\bibinfo {volume} {9}},\ \bibinfo {pages} {125} (\bibinfo
  {year} {2018})}\BibitemShut {NoStop}%
\bibitem [{\citenamefont {Bohm}(1949)}]{Bohm49}%
  \BibitemOpen
  \bibfield  {author} {\bibinfo {author} {\bibfnamefont {D.}~\bibnamefont
  {Bohm}},\ }\href {\doibase 10.1103/PhysRev.75.502} {\bibfield  {journal}
  {\bibinfo  {journal} {Phys. Rev.}\ }\textbf {\bibinfo {volume} {75}},\
  \bibinfo {pages} {502} (\bibinfo {year} {1949})}\BibitemShut {NoStop}%
\bibitem [{\citenamefont {Kiselev}\ \emph {et~al.}(2017)\citenamefont
  {Kiselev}, \citenamefont {Scheurer}, \citenamefont {W\"olfle},\ and\
  \citenamefont {Schmalian}}]{Kiselev17}%
  \BibitemOpen
  \bibfield  {author} {\bibinfo {author} {\bibfnamefont {E.~I.}\ \bibnamefont
  {Kiselev}}, \bibinfo {author} {\bibfnamefont {M.~S.}\ \bibnamefont
  {Scheurer}}, \bibinfo {author} {\bibfnamefont {P.}~\bibnamefont {W\"olfle}},
  \ and\ \bibinfo {author} {\bibfnamefont {J.}~\bibnamefont {Schmalian}},\
  }\href {https://link.aps.org/doi/10.1103/PhysRevB.95.125122} {\bibfield
  {journal} {\bibinfo  {journal} {Phys. Rev. B}\ }\textbf {\bibinfo {volume}
  {95}},\ \bibinfo {pages} {125122} (\bibinfo {year} {2017})}\BibitemShut
  {NoStop}%
\bibitem [{\citenamefont {Kang}\ and\ \citenamefont
  {Tesanovic}(2011)}]{Kang2011}%
  \BibitemOpen
  \bibfield  {author} {\bibinfo {author} {\bibfnamefont {J.}~\bibnamefont
  {Kang}}\ and\ \bibinfo {author} {\bibfnamefont {Z.}~\bibnamefont
  {Tesanovic}},\ }\href {http://dx.doi.org/10.1103/PhysRevB.83.020505}
  {\bibfield  {journal} {\bibinfo  {journal} {Phys. Rev. B}\ }\textbf {\bibinfo
  {volume} {83}},\ \bibinfo {pages} {020505(R)} (\bibinfo {year}
  {2011})}\BibitemShut {NoStop}%
\bibitem [{\citenamefont {Podolsky}\ \emph {et~al.}(2009)\citenamefont
  {Podolsky}, \citenamefont {Kee},\ and\ \citenamefont {Kim}}]{Podolsky2009}%
  \BibitemOpen
  \bibfield  {author} {\bibinfo {author} {\bibfnamefont {D.}~\bibnamefont
  {Podolsky}}, \bibinfo {author} {\bibfnamefont {H.-Y.}\ \bibnamefont {Kee}}, \
  and\ \bibinfo {author} {\bibfnamefont {Y.~B.}\ \bibnamefont {Kim}},\ }\href
  {\doibase 10.1209/0295-5075/88/17004} {\bibfield  {journal} {\bibinfo
  {journal} {Europhysics Letters}\ }\textbf {\bibinfo {volume} {88}},\ \bibinfo
  {pages} {17004} (\bibinfo {year} {2009})}\BibitemShut {NoStop}%
\bibitem [{\citenamefont {Chubukov}\ \emph {et~al.}(2015)\citenamefont
  {Chubukov}, \citenamefont {Fernandes},\ and\ \citenamefont
  {Schmalian}}]{Chubukov2015}%
  \BibitemOpen
  \bibfield  {author} {\bibinfo {author} {\bibfnamefont {A.~V.}\ \bibnamefont
  {Chubukov}}, \bibinfo {author} {\bibfnamefont {R.~M.}\ \bibnamefont
  {Fernandes}}, \ and\ \bibinfo {author} {\bibfnamefont {J.}~\bibnamefont
  {Schmalian}},\ }\href {https://doi.org/10.1103/PhysRevB.91.201105} {\bibfield
   {journal} {\bibinfo  {journal} {Phys. Rev. B}\ }\textbf {\bibinfo {volume}
  {91}},\ \bibinfo {pages} {201105} (\bibinfo {year} {2015})}\BibitemShut
  {NoStop}%
\bibitem [{\citenamefont {Chubukov}\ \emph {et~al.}(2008)\citenamefont
  {Chubukov}, \citenamefont {Efremov},\ and\ \citenamefont
  {Eremin}}]{Chubukov2008}%
  \BibitemOpen
  \bibfield  {author} {\bibinfo {author} {\bibfnamefont {A.~V.}\ \bibnamefont
  {Chubukov}}, \bibinfo {author} {\bibfnamefont {D.~V.}\ \bibnamefont
  {Efremov}}, \ and\ \bibinfo {author} {\bibfnamefont {I.}~\bibnamefont
  {Eremin}},\ }\href {\doibase 10.1103/physrevb.78.134512} {\bibfield
  {journal} {\bibinfo  {journal} {Phys. Rev. B}\ }\textbf {\bibinfo {volume}
  {78}},\ \bibinfo {pages} {134512} (\bibinfo {year} {2008})}\BibitemShut
  {NoStop}%
\bibitem [{\citenamefont {Varma}(1997)}]{Varma1997}%
  \BibitemOpen
  \bibfield  {author} {\bibinfo {author} {\bibfnamefont {C.~M.}\ \bibnamefont
  {Varma}},\ }\href {\doibase 10.1103/physrevb.55.14554} {\bibfield  {journal}
  {\bibinfo  {journal} {Phys. Rev. B}\ }\textbf {\bibinfo {volume} {55}},\
  \bibinfo {pages} {14554} (\bibinfo {year} {1997})}\BibitemShut {NoStop}%
\bibitem [{\citenamefont {{Chakravarty}}\ \emph {et~al.}(2001)\citenamefont
  {{Chakravarty}}, \citenamefont {{Laughlin}}, \citenamefont {{Morr}},\ and\
  \citenamefont {{Nayak}}}]{ChakravartyLaughlinMorrEtAl2001}%
  \BibitemOpen
  \bibfield  {author} {\bibinfo {author} {\bibfnamefont {S.}~\bibnamefont
  {{Chakravarty}}}, \bibinfo {author} {\bibfnamefont {R.~B.}\ \bibnamefont
  {{Laughlin}}}, \bibinfo {author} {\bibfnamefont {D.~K.}\ \bibnamefont
  {{Morr}}}, \ and\ \bibinfo {author} {\bibfnamefont {C.}~\bibnamefont
  {{Nayak}}},\ }\href {\doibase 10.1103/PhysRevB.63.094503} {\bibfield
  {journal} {\bibinfo  {journal} {\prb}\ }\textbf {\bibinfo {volume} {63}},\
  \bibinfo {eid} {094503} (\bibinfo {year} {2001})}\BibitemShut {NoStop}%
\bibitem [{\citenamefont {Lee}\ and\ \citenamefont {Wen}(2008)}]{Lee2008}%
  \BibitemOpen
  \bibfield  {author} {\bibinfo {author} {\bibfnamefont {P.~A.}\ \bibnamefont
  {Lee}}\ and\ \bibinfo {author} {\bibfnamefont {X.-G.}\ \bibnamefont {Wen}},\
  }\href {\doibase 10.1103/physrevb.78.144517} {\bibfield  {journal} {\bibinfo
  {journal} {Phys. Rev. B}\ }\textbf {\bibinfo {volume} {78}},\ \bibinfo
  {pages} {144517} (\bibinfo {year} {2008})}\BibitemShut {NoStop}%
\bibitem [{\citenamefont {{Gastiasoro}}\ and\ \citenamefont
  {{Andersen}}(2015)}]{GastiasoroAndersen2015}%
  \BibitemOpen
  \bibfield  {author} {\bibinfo {author} {\bibfnamefont {M.~N.}\ \bibnamefont
  {{Gastiasoro}}}\ and\ \bibinfo {author} {\bibfnamefont {B.~M.}\ \bibnamefont
  {{Andersen}}},\ }\href {\doibase 10.1103/PhysRevB.92.140506} {\bibfield
  {journal} {\bibinfo  {journal} {\prb}\ }\textbf {\bibinfo {volume} {92}},\
  \bibinfo {eid} {140506} (\bibinfo {year} {2015})}\BibitemShut {NoStop}%
\bibitem [{\citenamefont {Graser}\ \emph {et~al.}(2009)\citenamefont {Graser},
  \citenamefont {Maier}, \citenamefont {Hirschfeld},\ and\ \citenamefont
  {Scalapino}}]{Graser2009}%
  \BibitemOpen
  \bibfield  {author} {\bibinfo {author} {\bibfnamefont {S.}~\bibnamefont
  {Graser}}, \bibinfo {author} {\bibfnamefont {T.~A.}\ \bibnamefont {Maier}},
  \bibinfo {author} {\bibfnamefont {P.~J.}\ \bibnamefont {Hirschfeld}}, \ and\
  \bibinfo {author} {\bibfnamefont {D.~J.}\ \bibnamefont {Scalapino}},\ }\href
  {\doibase 10.1088/1367-2630/11/2/025016} {\bibfield  {journal} {\bibinfo
  {journal} {New J. Phys.}\ }\textbf {\bibinfo {volume} {11}},\ \bibinfo
  {pages} {025016} (\bibinfo {year} {2009})}\BibitemShut {NoStop}%
\bibitem [{\citenamefont {Fernandes}\ \emph {et~al.}(2012)\citenamefont
  {Fernandes}, \citenamefont {Chubukov}, \citenamefont {Knolle}, \citenamefont
  {Eremin},\ and\ \citenamefont {Schmalian}}]{Fernandes2012a}%
  \BibitemOpen
  \bibfield  {author} {\bibinfo {author} {\bibfnamefont {R.~M.}\ \bibnamefont
  {Fernandes}}, \bibinfo {author} {\bibfnamefont {A.~V.}\ \bibnamefont
  {Chubukov}}, \bibinfo {author} {\bibfnamefont {J.}~\bibnamefont {Knolle}},
  \bibinfo {author} {\bibfnamefont {I.}~\bibnamefont {Eremin}}, \ and\ \bibinfo
  {author} {\bibfnamefont {J.}~\bibnamefont {Schmalian}},\ }\href {\doibase
  10.1103/physrevb.85.024534} {\bibfield  {journal} {\bibinfo  {journal} {Phys.
  Rev. B}\ }\textbf {\bibinfo {volume} {85}},\ \bibinfo {pages} {024534}
  (\bibinfo {year} {2012})}\BibitemShut {NoStop}%
\bibitem [{\citenamefont {Lederer}\ and\ \citenamefont
  {Kivelson}(2012)}]{Lederer2012}%
  \BibitemOpen
  \bibfield  {author} {\bibinfo {author} {\bibfnamefont {S.}~\bibnamefont
  {Lederer}}\ and\ \bibinfo {author} {\bibfnamefont {S.~A.}\ \bibnamefont
  {Kivelson}},\ }\href {\doibase 10.1103/physrevb.85.155130} {\bibfield
  {journal} {\bibinfo  {journal} {Phys. Rev. B}\ }\textbf {\bibinfo {volume}
  {85}},\ \bibinfo {pages} {155130} (\bibinfo {year} {2012})}\BibitemShut
  {NoStop}%
\bibitem [{\citenamefont {Kissikov}\ \emph {et~al.}(2018)\citenamefont
  {Kissikov}, \citenamefont {Sarkar}, \citenamefont {Lawson}, \citenamefont
  {Bush}, \citenamefont {Timmons}, \citenamefont {Tanatar}, \citenamefont
  {Prozorov}, \citenamefont {Bud'ko}, \citenamefont {Canfield}, \citenamefont
  {Fernandes},\ and\ \citenamefont {Curro}}]{KissikovSarkarLawsonEtAl2017}%
  \BibitemOpen
  \bibfield  {author} {\bibinfo {author} {\bibfnamefont {T.}~\bibnamefont
  {Kissikov}}, \bibinfo {author} {\bibfnamefont {R.}~\bibnamefont {Sarkar}},
  \bibinfo {author} {\bibfnamefont {M.}~\bibnamefont {Lawson}}, \bibinfo
  {author} {\bibfnamefont {B.~T.}\ \bibnamefont {Bush}}, \bibinfo {author}
  {\bibfnamefont {E.~I.}\ \bibnamefont {Timmons}}, \bibinfo {author}
  {\bibfnamefont {M.~A.}\ \bibnamefont {Tanatar}}, \bibinfo {author}
  {\bibfnamefont {R.}~\bibnamefont {Prozorov}}, \bibinfo {author}
  {\bibfnamefont {S.~L.}\ \bibnamefont {Bud'ko}}, \bibinfo {author}
  {\bibfnamefont {P.~C.}\ \bibnamefont {Canfield}}, \bibinfo {author}
  {\bibfnamefont {R.~M.}\ \bibnamefont {Fernandes}}, \ and\ \bibinfo {author}
  {\bibfnamefont {N.~J.}\ \bibnamefont {Curro}},\ }\href
  {https://doi.org/10.1038/s41467-018-03377-8} {\bibfield  {journal} {\bibinfo
  {journal} {Nat. Commun.}\ }\textbf {\bibinfo {volume} {9}},\ \bibinfo {pages}
  {1058} (\bibinfo {year} {2018})}\BibitemShut {NoStop}%
\bibitem [{\citenamefont {{Ueland}}\ \emph {et~al.}(2015)\citenamefont
  {{Ueland}}, \citenamefont {{Pandey}}, \citenamefont {{Lee}}, \citenamefont
  {{Sapkota}}, \citenamefont {{Choi}}, \citenamefont {{Haskel}}, \citenamefont
  {{Rosenberg}}, \citenamefont {{Lang}}, \citenamefont {{Harmon}},
  \citenamefont {{Johnston}}, \citenamefont {{Kreyssig}},\ and\ \citenamefont
  {{Goldman}}}]{UelandPandeyLeeEtAl2015}%
  \BibitemOpen
  \bibfield  {author} {\bibinfo {author} {\bibfnamefont {B.~G.}\ \bibnamefont
  {{Ueland}}}, \bibinfo {author} {\bibfnamefont {A.}~\bibnamefont {{Pandey}}},
  \bibinfo {author} {\bibfnamefont {Y.}~\bibnamefont {{Lee}}}, \bibinfo
  {author} {\bibfnamefont {A.}~\bibnamefont {{Sapkota}}}, \bibinfo {author}
  {\bibfnamefont {Y.}~\bibnamefont {{Choi}}}, \bibinfo {author} {\bibfnamefont
  {D.}~\bibnamefont {{Haskel}}}, \bibinfo {author} {\bibfnamefont {R.~A.}\
  \bibnamefont {{Rosenberg}}}, \bibinfo {author} {\bibfnamefont {J.~C.}\
  \bibnamefont {{Lang}}}, \bibinfo {author} {\bibfnamefont {B.~N.}\
  \bibnamefont {{Harmon}}}, \bibinfo {author} {\bibfnamefont {D.~C.}\
  \bibnamefont {{Johnston}}}, \bibinfo {author} {\bibfnamefont
  {A.}~\bibnamefont {{Kreyssig}}}, \ and\ \bibinfo {author} {\bibfnamefont
  {A.~I.}\ \bibnamefont {{Goldman}}},\ }\href {\doibase
  10.1103/PhysRevLett.114.217001} {\bibfield  {journal} {\bibinfo  {journal}
  {Phys. Rev. Lett.}\ }\textbf {\bibinfo {volume} {114}},\ \bibinfo {eid}
  {217001} (\bibinfo {year} {2015})}\BibitemShut {NoStop}%
\bibitem [{\citenamefont {{Fedorov}}\ \emph {et~al.}(2016)\citenamefont
  {{Fedorov}}, \citenamefont {{Yaresko}}, \citenamefont {{Kim}}, \citenamefont
  {{Kushnirenko}}, \citenamefont {{Haubold}}, \citenamefont {{Wolf}},
  \citenamefont {{Hoesch}}, \citenamefont {{Gr{\"u}neis}}, \citenamefont
  {{B{\"u}chner}},\ and\ \citenamefont
  {{Borisenko}}}]{FedorovYareskoKimEtAl2016}%
  \BibitemOpen
  \bibfield  {author} {\bibinfo {author} {\bibfnamefont {A.}~\bibnamefont
  {{Fedorov}}}, \bibinfo {author} {\bibfnamefont {A.}~\bibnamefont
  {{Yaresko}}}, \bibinfo {author} {\bibfnamefont {T.~K.}\ \bibnamefont
  {{Kim}}}, \bibinfo {author} {\bibfnamefont {Y.}~\bibnamefont
  {{Kushnirenko}}}, \bibinfo {author} {\bibfnamefont {E.}~\bibnamefont
  {{Haubold}}}, \bibinfo {author} {\bibfnamefont {T.}~\bibnamefont {{Wolf}}},
  \bibinfo {author} {\bibfnamefont {M.}~\bibnamefont {{Hoesch}}}, \bibinfo
  {author} {\bibfnamefont {A.}~\bibnamefont {{Gr{\"u}neis}}}, \bibinfo {author}
  {\bibfnamefont {B.}~\bibnamefont {{B{\"u}chner}}}, \ and\ \bibinfo {author}
  {\bibfnamefont {S.~V.}\ \bibnamefont {{Borisenko}}},\ }\href {\doibase
  10.1038/srep36834} {\bibfield  {journal} {\bibinfo  {journal} {Sci. Rep.}\
  }\textbf {\bibinfo {volume} {6}},\ \bibinfo {eid} {36834} (\bibinfo {year}
  {2016})}\BibitemShut {NoStop}%
\bibitem [{\citenamefont {Chubukov}\ \emph {et~al.}(2016)\citenamefont
  {Chubukov}, \citenamefont {Khodas},\ and\ \citenamefont
  {Fernandes}}]{Chubukov2016}%
  \BibitemOpen
  \bibfield  {author} {\bibinfo {author} {\bibfnamefont {A.~V.}\ \bibnamefont
  {Chubukov}}, \bibinfo {author} {\bibfnamefont {M.}~\bibnamefont {Khodas}}, \
  and\ \bibinfo {author} {\bibfnamefont {R.~M.}\ \bibnamefont {Fernandes}},\
  }\href {https://doi.org/10.1103/PhysRevX.6.041045} {\bibfield  {journal}
  {\bibinfo  {journal} {Phys. Rev. X}\ }\textbf {\bibinfo {volume} {6}},\
  \bibinfo {pages} {041045} (\bibinfo {year} {2016})}\BibitemShut {NoStop}%
\bibitem [{\citenamefont {{Zhao}}\ \emph {et~al.}(2016)\citenamefont {{Zhao}},
  \citenamefont {{Torchinsky}}, \citenamefont {{Chu}}, \citenamefont
  {{Ivanov}}, \citenamefont {{Lifshitz}}, \citenamefont {{Flint}},
  \citenamefont {{Qi}}, \citenamefont {{Cao}},\ and\ \citenamefont
  {{Hsieh}}}]{ZhaoTorchinskyChuEtAl2016}%
  \BibitemOpen
  \bibfield  {author} {\bibinfo {author} {\bibfnamefont {L.}~\bibnamefont
  {{Zhao}}}, \bibinfo {author} {\bibfnamefont {D.~H.}\ \bibnamefont
  {{Torchinsky}}}, \bibinfo {author} {\bibfnamefont {H.}~\bibnamefont {{Chu}}},
  \bibinfo {author} {\bibfnamefont {V.}~\bibnamefont {{Ivanov}}}, \bibinfo
  {author} {\bibfnamefont {R.}~\bibnamefont {{Lifshitz}}}, \bibinfo {author}
  {\bibfnamefont {R.}~\bibnamefont {{Flint}}}, \bibinfo {author} {\bibfnamefont
  {T.}~\bibnamefont {{Qi}}}, \bibinfo {author} {\bibfnamefont {G.}~\bibnamefont
  {{Cao}}}, \ and\ \bibinfo {author} {\bibfnamefont {D.}~\bibnamefont
  {{Hsieh}}},\ }\href {\doibase 10.1038/nphys3517} {\bibfield  {journal}
  {\bibinfo  {journal} {Nat. Phys.}\ }\textbf {\bibinfo {volume} {12}},\
  \bibinfo {pages} {32} (\bibinfo {year} {2016})}\BibitemShut {NoStop}%
\bibitem [{\citenamefont {{Fanfarillo}}\ \emph {et~al.}(2015)\citenamefont
  {{Fanfarillo}}, \citenamefont {{Cortijo}},\ and\ \citenamefont
  {{Valenzuela}}}]{FanfarilloCortijoValenzuela2015}%
  \BibitemOpen
  \bibfield  {author} {\bibinfo {author} {\bibfnamefont {L.}~\bibnamefont
  {{Fanfarillo}}}, \bibinfo {author} {\bibfnamefont {A.}~\bibnamefont
  {{Cortijo}}}, \ and\ \bibinfo {author} {\bibfnamefont {B.}~\bibnamefont
  {{Valenzuela}}},\ }\href {\doibase 10.1103/PhysRevB.91.214515} {\bibfield
  {journal} {\bibinfo  {journal} {\prb}\ }\textbf {\bibinfo {volume} {91}},\
  \bibinfo {eid} {214515} (\bibinfo {year} {2015})}\BibitemShut {NoStop}%
\bibitem [{\citenamefont {{Yamakawa}}\ \emph {et~al.}(2016)\citenamefont
  {{Yamakawa}}, \citenamefont {{Onari}},\ and\ \citenamefont
  {{Kontani}}}]{YamakawaOnariKontani2016}%
  \BibitemOpen
  \bibfield  {author} {\bibinfo {author} {\bibfnamefont {Y.}~\bibnamefont
  {{Yamakawa}}}, \bibinfo {author} {\bibfnamefont {S.}~\bibnamefont {{Onari}}},
  \ and\ \bibinfo {author} {\bibfnamefont {H.}~\bibnamefont {{Kontani}}},\
  }\href {\doibase 10.1103/PhysRevX.6.021032} {\bibfield  {journal} {\bibinfo
  {journal} {Phys. Rev. X}\ }\textbf {\bibinfo {volume} {6}},\ \bibinfo {eid}
  {021032} (\bibinfo {year} {2016})}\BibitemShut {NoStop}%
\end{thebibliography}%

\end{document}